\documentclass[5p,twocolumn,10pt,times]{elsarticle}
\usepackage{amsmath}
\usepackage{graphicx}
\usepackage{xcolor}
\usepackage{amsmath}
\usepackage{algorithm}
\usepackage[noend]{algpseudocode}
\usepackage{soul}
\usepackage{ulem} 
\usepackage{wrapfig}
\usepackage{dirtytalk}
\usepackage{float}
\newcommand{\name}{TopoKnit}
\usepackage{color}

\usepackage{subcaption}

\usepackage{geometry}
\begin{document}
\baselineskip11pt

\begin{frontmatter}

\title{\name\/ : A Process-Oriented Representation for Modeling the Topology\\ of Yarns in Weft-Knitted Textiles }

\author{Levi Kapllani$^{1,2}$}
\author{Chelsea Amanatides$^{2}$}
\author{Genevieve Dion$^{2,3}$}
\author{Vadim Shapiro$^4$}
\author{David E. Breen$^{1,2}$}
\address{$^1$Computer Science Department, $^2$Center for Functional Fabrics,
$^3$Design Department, Drexel University}
\address{$^4$University of Wisconsin - Madison, and International Computer Science Institute}
\cortext[David Breen]{Corresponding author}
\ead{david@cs.drexel.edu}

\begin{abstract} 
Machine knitted textiles are complex multi-scale material structures increasingly important in many industries, including consumer products, architecture, composites, medical, and military. Computational modeling, simulation, and design of industrial fabrics require efficient representations of the spatial, material, and physical properties of such structures. We propose a process-oriented representation, \name\/, that defines a foundational data structure for representing the topology of weft-knitted textiles at the yarn scale. Process space serves as an intermediary between the machine and fabric spaces, and supports a concise, computationally efficient evaluation approach based on on-demand, near constant-time queries. In this paper, we define the properties of the process space, and design a data structure to represent it and algorithms to evaluate it. We demonstrate the effectiveness of the representation scheme by providing results of evaluations of the data structure in support of common topological operations in the fabric space.
\end{abstract}


\end{frontmatter}


\section{Introduction}
Knitting has been a technique for producing versatile textiles for over a millennium, and the knitting process was first automated with machinery in the 16th century \cite{spencer1983knitting}.
Knitting produces fabrics with varied mechanical properties that can be
shaped into many forms. Knitted textiles are increasingly important in a number of industries, including consumer products, architecture, composites, medical, and military. In order for these textiles to be widely deployed and reach their full industrial potential, computer-based modeling and simulation tools must be developed to support the design and optimization of knitted structures.

With this need in mind, we have developed a process-oriented representation
for modeling the topology of yarns in weft-knitted textiles. Our initial focus has been on representing fabrics that can be manufactured by weft-knitting machines consisting of two flat beds of needles and a single yarn.
Since it has been shown that the structures formed by yarns and their
interactions dominate the mechanical behavior of knitted textiles \cite{Liu2017:ORM,Liu2018:CAM,Tekerek2018:IIR,Knittel2015:SFT}, \name\/ represents these types of fabrics with yarns
and the neighborhoods where they contact each other.
The goal of our work is to define a low-level representation of knitted fabrics that has the properties of both completeness (i.e.~capable of storing all knitted fabrics that can be manufactured from a specific set
of stitches) and validity (i.e.~ guaranteed to represent only 
physically valid states of the fabric). Additionally, the representation should support efficient query algorithms for evaluation and generation of many types of knitted structure models.

To date a variety of models have been proposed for knitted textiles, covering a spectrum of length scales from loop/stitch structures, to fabric swatches and
complete garments.  The focus of these models range from idealized purely geometric models of yarn paths, to mechanical models of stitches, to simulation models of
fabrics and clothing. There has also been abundant work on visual models for accurately rendering textiles. This assortment of models have not always been compatible with each other nor have they provided a conceptual/theoretical foundation on which to build the consistent, robust analysis tools that are essential for evaluating manufacturability and providing predictive simulation capabilities. \name\/ is a major step towards an approach capable of representing the
topological structures needed for all such models.

Our ultimate goal is to develop a multiscale data structure that can capture the topological, geometric and mechanical properties/behaviors of knitted textiles. This yarn-level representation, where mechanical behaviors are
derived from robust geometric models, which in turn are based on accurate topological structures, should support the simulation and optimization techniques that are essential for design operations that ensure manufacturability of the material. The representation should capture relationships, structures and phenomena at a variety of spatial and temporal scales. These features include yarn geometry, frictional contacts, and deformations occurring at the loop, stitch, pattern, swatch and garment levels.

As a first step toward achieving this goal we describe here a low-level
representation, and associated data structure, that can be used to derive
the topological relationships of
the yarns in the various stitches that make up a knitted fabric manufactured on a 2-flatbed weft-knitting machine. For a given set of stitch operations, the data structure is capable of representing all fabrics produced by this class of knitting machines; thus endowing it with the property of
completeness.
We see this process-oriented representation, which we call
\name\/, as providing the necessary foundation on which to build more detailed models of knitted textiles that include topology, geometry and mechanics. \name\/ not only provides a foundational data structure that captures
the fundamental primitives of knitted textiles, but it also includes a rich set of access functions that allows for queries of the features of the topological structure of the fabric.
This alleviates the requirement for an explicit representation of what could be a very complex topology by performing on-demand evaluation as certain information is needed by higher-level applications. Examples of applications that require this type of efficient
topological representation include simulations of electrical current,
water, and heat flow through textiles
\cite{vallett2017digital,hong2016model,shen2019analysis}. 

\section{Related Work}
The early work on modeling knitted textiles focused primarily on defining and analyzing the geometric structure of knit stitches \cite{peirce1947geometrical,leaf1955,munden1959}. This work was remarkably done without the mathematical infrastructure of splines \cite{SplinesBook01}, which was not widely available until the 1970s. Much later work did utilize splines to describe the centerlines of yarn geometry in knitted materials \cite{demiroz2000studyI,kurbak2008basicI,kurbak2008basicII,kurbak2009geometrical}. Follow-on research applied minimum energy analysis to determine the shape of relaxed yarn loops in individual stitches \cite{shanahan1970theoretical,hepworth1976,dejong1977energyI,semnani2003new}, as well as larger bulk properties of plain-knitted fabrics \cite{dejong1977energyII,choi2003energy,choi2006shape}. This work was extended by Kyosev et al.\ \cite{kyosev20053d} to include the compression of the yarns in the loop.
Sherburn, Lin, et al.~\cite{Sherburn2007,Lin:2012:AGM} developed a modeling approach/system working on microscopic, mesoscopic and macroscopic scales to predict the mechanical properties of textiles.
Duhovic and Bhattacharyya \cite{duhovic2006simulating} simulated the knitting process in order to understand how each of a yarn's deformation mechanisms contribute to the overall deformation energy/behavior of a yarn in a knitted fabric.
In recent work, Knittel, Wadekar et al.~\cite{ChelseaThesis2019,Knittel2020,Wadekar2020GMK,Wadekar2020GMC} investigated
helicoid scaffolds as a framework within which to study the structure
of knitted fabrics. 

The first system to model and visualize complete knitted fabrics was developed by Meissner, Eberhardt and Strasser \cite{Meissner:1998:AKF,Eberhardt:2000:KF}.
Their system (KnitSim) accepts Stoll knitting machine commands (knit a stitch, transfer loops between beds, and rack the beds) and simulates the knitting process to produce an explicit topological representation of a knitted material. The topology consists of Bonding Points (BPs), where yarns wrap around each other, and edges, representing yarns, that connect the BPs.  Yarns are defined as a linked list of BPs.
By making assumptions about the length of yarns between BPs, a relaxation process is run to produce a 2D geometric layout for the fabric.
Eberhardt and Weber \cite{eberhardt1999particle} then applied Breen et al.'s particle approach \cite{Breen94b} to simulate the draping behavior of knitted fabrics.

In ground-breaking work Kaldor et al.\ \cite{Kaldor:2008:SKC,Kaldor:2010:EYC} simulated complete swatches and articles of clothing consisting of knitted fabrics by modeling the geometry and physics of individual yarns in these items. The yarns in the model are defined with cubic B-spline curves surrounded by a constant radius to produce a swept surface with a circular cross-section. Yarn dynamics are dictated by both energy terms (kinetic and bending) and hard constraints to prevent yarn extension and collisions, while friction interactions, a critical component of correct yarn behavior, are approximated using a velocity filter that penalizes locally non-rigid motion. While the topological structure of the fabric is implied by yarn contacts, it is not explicitly represented in their model.

The Kaldor et al.\ work was extended by Yuksel et al.\ \cite{Yuksel2012,Wu2018} to produce Stitch Meshes, an approach to generating Kaldor-style knitted material models of clothing from polygonal models that represent the clothing's surface. A shortcoming of this approach is that it utilizes stitches that cannot be created on a knitting machine, thus limiting its usefulness in a manufacturing setting. This project has recently been enhanced to generate graphical instructions for hand knitters \cite{Wu:2019:KSM}. Aspects of this work were utilized by Leaf et al.~\cite{leaf2018interactive} to produce an interactive design tool for simulating swatch-level patterns for knit and woven textiles. To improve relaxation time, they employ a GPU implementation using periodic boundary conditions to simulate modeling unit cells. Stitch Meshes have also been used to model crocheted fabrics
\cite{Guo:2020:SKC}.

Igarashi et al.~\cite{Igarashi:2008:K3Ma,Igarashi:2008K3Mb} developed a technique for generating hand-knitting patterns
from 3D models, along with an interactive design and visualization system. In related work, McCann, et al.~\cite{McCann:2016:CMK,Vidya:2018:AMK,Lin:2018:ETP,Narayanan2019} have developed algorithms that can generate knitting machine commands based on a variety of polygonal models as input. These algorithms allow for the interactive design of 3D shapes,which can then be manufactured on a commercial knitting machine.
In related work, Popescu et al. ~\cite{popescu2018automated} describe
an approach for generating an automated knitting pattern given a 3D
model, without being constrained to developable surfaces. 
Motivated by Narayanan et al.~\cite{Vidya:2018:AMK}, Kaspar et al.~ \cite{kaspar2019knitting} introduce an interactive system which allows users of different skill levels to create and customize machine-knitted textiles.

Cirio et al.\ \cite{cirio2017yarn} define a topological representation of knits consisting of a limited set of stitches, in contrast to the yarn-geometry-based approach of Kaldor et al. As with the Yuksel et al.\ work, they incorporated stitches that are not manufacturable on knitting machines. They are able to generate geometric models for rendering from their representation and have defined simplified mechanical relationships on the topological structure that supports efficient simulation of the approximate physical
behavior of their virtual knitted fabrics.
Aspects of this work has recently incorporated yarn-level models to
provide a more efficient computational method for detailed cloth
simulations \cite{CCRMO20}.
This work was extended by S\'{a}nchez et al. \cite{sanchez2020robust} to include simulation of stacked layers with implicit contact handling using a Eulerian-on-Lagrangian (EoL) discretization.

Similar to the Meissner et al.~work, Counts \cite{counts2018knitting} presents a graph-based representation aiming to approximate the 2D
layout of knitted textiles.
The developed algorithms use information from a simulation to generate a
topological graph from a limited set of hand and machine knitting
instructions.  Using loops as its fundamental primitive, as well as the
small set of supported stitches (knit and transfer), limits the yarn
topology it can represent. The approach does offer the ability to 
generate machine knitting instructions from the graph, but its
disadvantages include not supporting increases/decreases and more
complex stitches, and requiring a simulation of the knitting process
to generate the graph.

Finite Element Modeling (FEM) has been used to analyze the mechanical properties of knitted fabrics. Liu et
al.~\cite{Liu2017:ORM,Liu2018:CAM,Liu2019:PFE} perform simulations with
solid yarn-level geometric models \cite{Wadekar2020OYG}, while Dinh et
al.~\cite{Dinh:2018:PMP},
Poincloux et al.~\cite{Poincloux:2018:GEK}
Liu et al.~\cite{Liu2019:MHA} and
Sperl et al.~\cite{Sperl2020:HYL} 
base simulations
on a homogenized unit cell.
All of these efforts simulate simple weft-knitted textiles, i.e.~their
virtual fabrics only consist of Knit and Purl stitches.
Their approaches do not rely on an underlying topological model, which will make it difficult to extend them to more complicated knitted fabrics constructed from different types of stitches.

\name\ improves upon previous work by providing a general topological representation of
fabrics that can be produced by flatbed weft-knitting machines with a
standard set of stitches. Our work stands apart from the Kaldor
et al.\ work, which simply modeled the geometry of the yarns in Knit and Purl stitches, without representing the underlying topology of the knitted
structures. Our focus on incorporating process-based features in our model provides critical manufacturability properties not found in the Yuksel et al.~and Cirio et al.~work, which describe materials that cannot be manufactured on knitting machines. The McCann et al.~work does not create a
model of a knitted material, but instead
makes assumptions about the physical size of a single stitch in order
to generate knitting machine instructions that produce actual fabric.

While Meissner et al.'s topological model utilizes similar primitives
as \name\/, with Bonding Points (BPs) (our Contact Neighborhoods (CNs))
connected by yarn edges,
there are significant differences between the methods and
capabilities of the two approaches.
They generate a topological data structure from a limited set of
low-level machine instructions, an approach that requires a simulation of
the knitting process.
We utilize a process-oriented approach based on higher-level stitch
commands, which captures how the stitches modify an all-Knit-stitch
structure and does not require a knitting simulation.
Their approach is more limited than ours in that they assume that
a row of BPs only interacts with the BPs from the previous row.
This is clearly not the case with more complicated stitch patterns,
as can been seen in some of our
examples, e.g.~Figures \ref{fig:pattern3}, \ref{fig:pattern8} and
\ref{fig:patternVertical}.

Meissner et al.'s results are presented in low resolution and are
difficult to interpret.
Some of their results appear to be topologically incorrect, unphysical
and generated from unexplained machine commands and stitching scenarios.
They do not provide sufficient evidence to evaluate the correctness of
their results. For example, they do not offer the input instructions and
resulting topology graphs for their geometric results.
Additionally, they do not provide the technical information needed to
evaluate the generality and robustness of their approach, e.g.
the details on how to generate the BP data structure
from machine instructions.
\\ 

\section{Fabrication of Weft-Knitted Textiles}

\begin{wrapfigure}{L}{0.35\linewidth}
  \centering
  \includegraphics[width=0.75\linewidth]{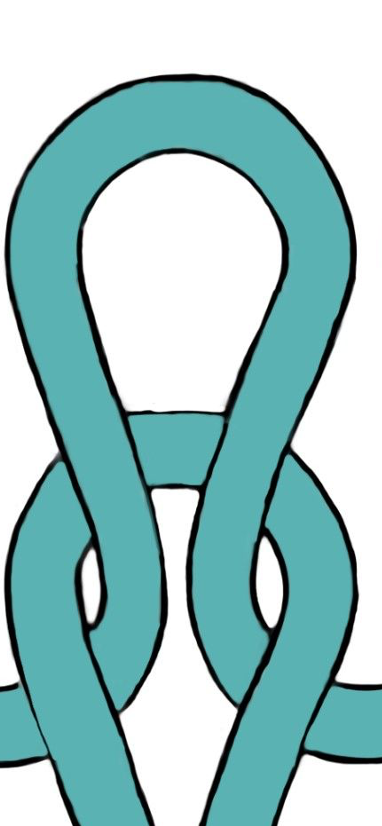}
 \caption{A single stitch, with its ``legs'' holding the ``head'' (upper loop) of the stitch below.}
   \label{fig:singleLoop}
\end{wrapfigure}

\begin{figure}[p]
  \includegraphics[width=1\linewidth]{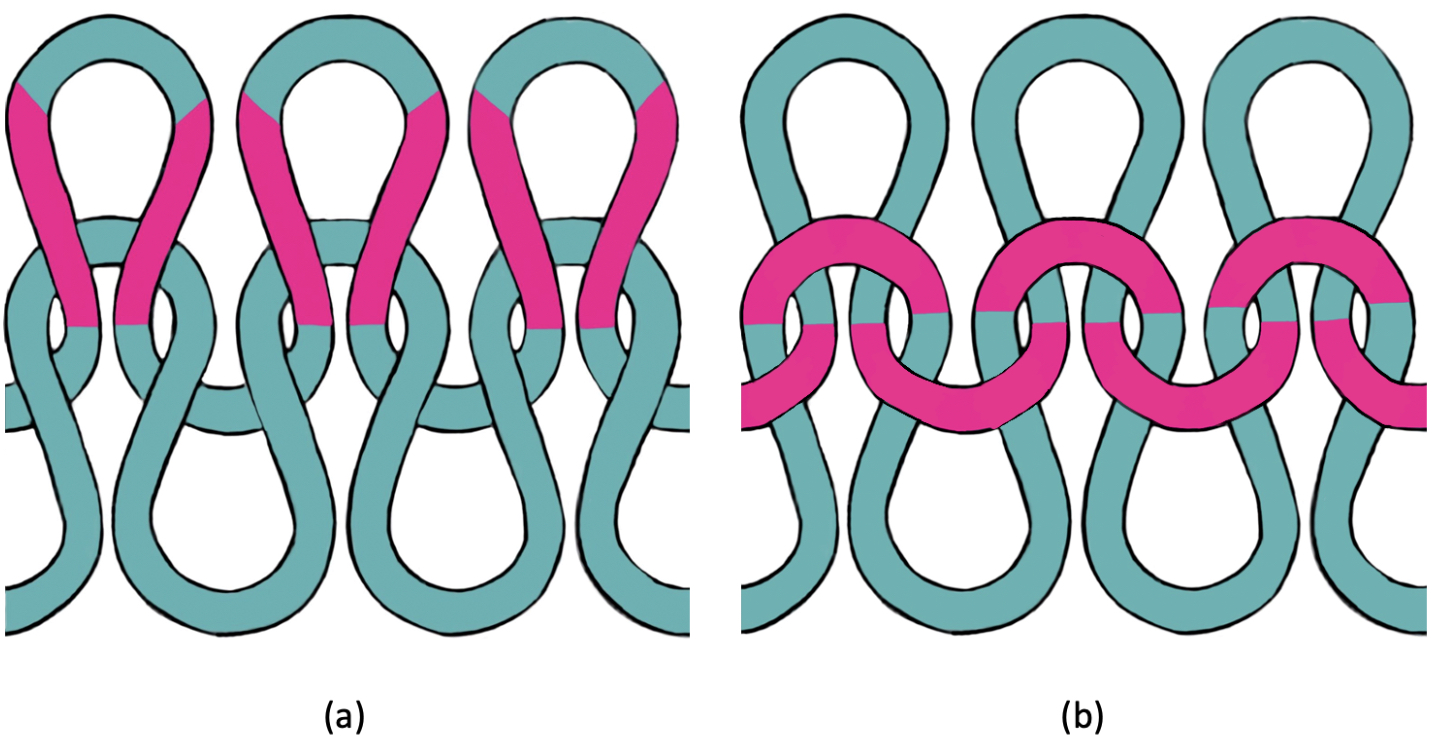}
  \vspace{-5mm}
  \caption{(a) All-Knit pattern. (b) All-Purl pattern.}
  \label{fig:pattern}
\vspace{6mm}
  \includegraphics[width=1\linewidth]{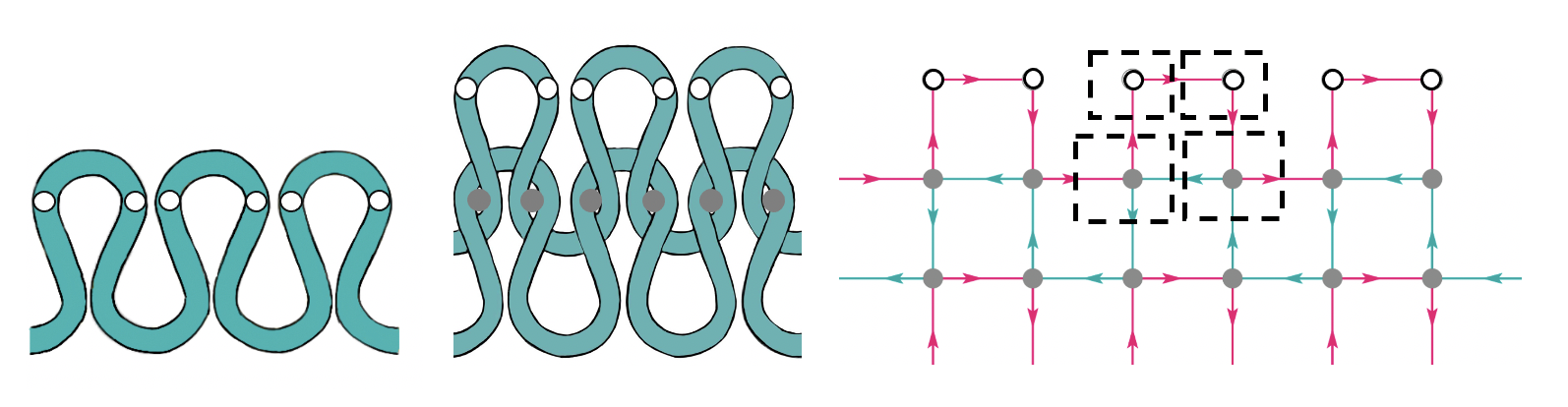}\\
    \mbox{\hspace{11mm}(a)\hspace{16mm} (b)\hspace{24mm} (c)}
    \caption{A Knit stitch is created by pulling a loop of yarn through a loop held from the previous row. (a) Row of loops. (b) Knit stitches produced from another row of stitches. (c) Topological representation.}
  \label{fig:knitStitch}
 \end{figure}

\begin{figure}[p]
\vspace{6mm}
  \includegraphics[width=1\linewidth]{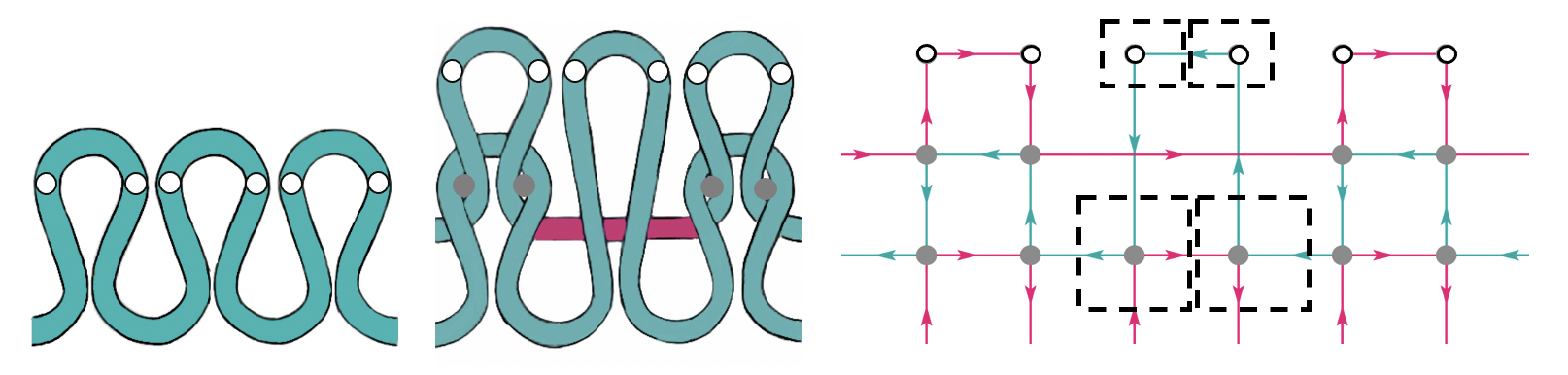}
   \mbox{\hspace{10mm}(a)\hspace{16mm} (b)\hspace{24mm} (c)} 
  \caption{A Miss stitch is created when the needle holds a loop from a previous row as the yarn passes by, without knitting a new stitch.}
  \label{fig:missStitch}
\vspace{5mm}
\end{figure}

The loop is the fundamental structural element of knitted textiles.
A new loop is formed when a yarn is drawn through a previously existing
loop, as seen in
Figure \ref{fig:singleLoop}\footnote{Figures \ref{fig:singleLoop},\ref{fig:pattern}(a)(b) to \ref{fig:missStitch}(a)(b), \ref{fig:singleLoopCNs}(a) and \ref{fig:allKnit}(b) were partially produced with the Shima Seiki SDS-One APEX3 KnitPaint system.}. 
When
this process is repeated across a row, and then subsequently again in other rows, the fabric is formed \cite{spencer1983knitting}. When the yarn is drawn through the loop(s) held on the needle from back to front, a \textit{Knit} stitch is created, as shown in Figure \ref{fig:pattern}(a). A Knit stitch can be distinguished by the small ``v'' shapes visible on the fabric, formed by the vertical yarns of the stitches. When the yarn is drawn through the loop(s) held on the needle from front to back, a \textit{Purl} stitch is created (Figure \ref{fig:pattern}(b)), which is distinguishable by the horizontal
sequence of small curves visible on the fabric, formed by the heads and
tails of the stitches. It is important to note that Knit and Purl stitches are structurally the same, however the side from which they are viewed determines their labeling in the knitted structure. 

\begin{figure}[p]
\vspace{6mm}
  \includegraphics[width=1.0\linewidth]{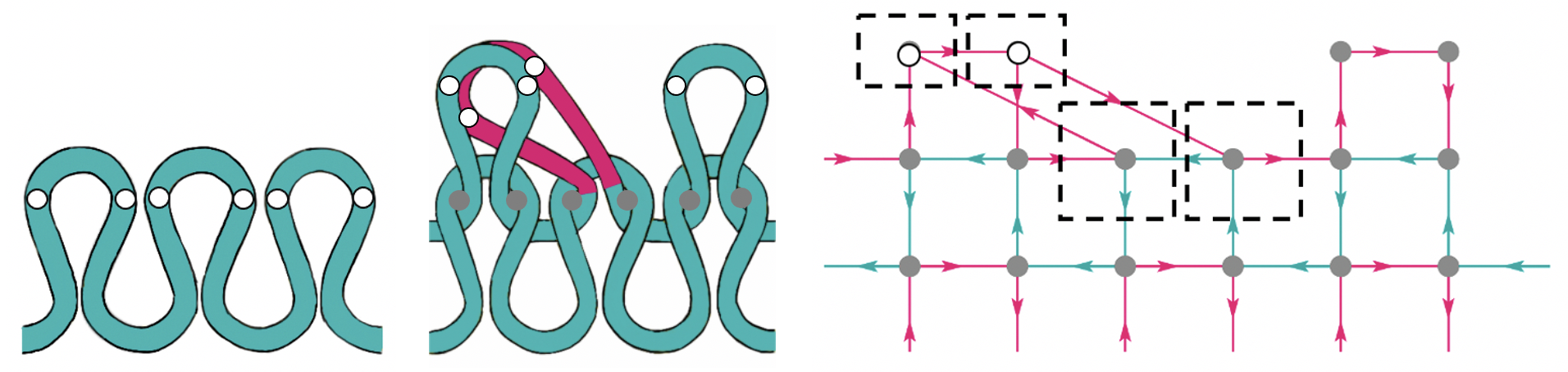}\\
    \mbox{\hspace{9mm}(a)\hspace{16mm} (b)\hspace{22mm} (c)}
  \caption{A Transfer stitch is created when the loop of a Knit or Purl 
  stitch is transferred up to three needle positions away to the left or
  right.}
  \label{fig:transferStitch}
 \end{figure}
 
 \begin{figure}[p]
\vspace{6mm}
  \includegraphics[width=1\linewidth]{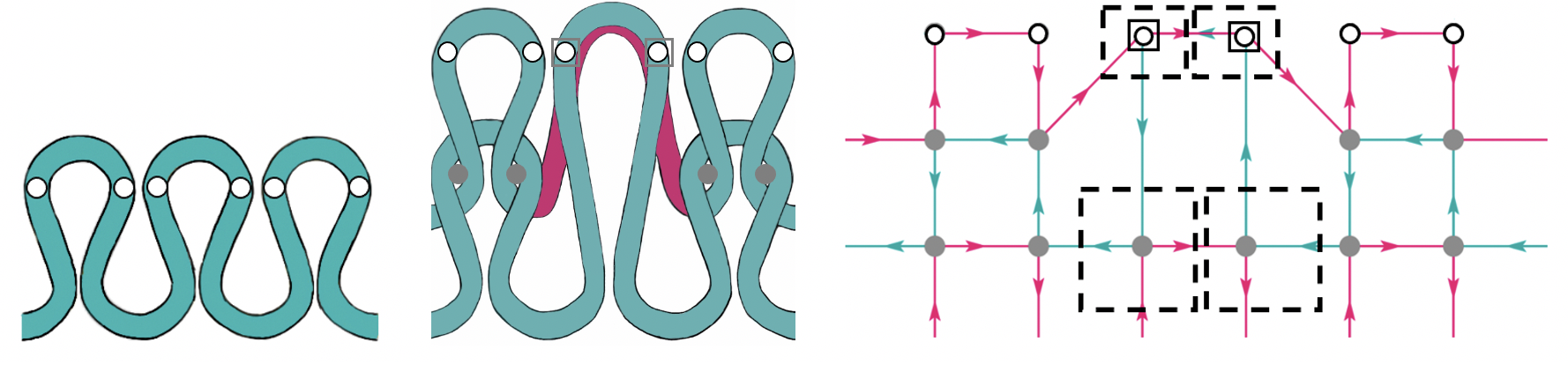}
   \mbox{\hspace{10mm}(a)\hspace{16mm} (b)\hspace{24mm} (c)}   \caption{A Tuck stitch is created by tucking a yarn onto a loop held from the previous row, instead of creating a new stitch.}
  \label{fig:tuckStitch}
   \end{figure}
   
Additionally, a number of other stitches can be formed, which may be combined to produce a vast variety of knitted textiles. These stitches include: 
\newline
 \textbf{Transfer stitch} \newline
A Transfer stitch is produced when a Knit or Purl stitch is created and
then its head loop is transferred, via a sequence of needle bed transfers
and rackings, to another needle location.
The Transfer stitch presented in Figure \ref{fig:transferStitch}(a) 
transfers the loop one needle to the left, however, depending on the type of Transfer stitch there can be up to three loop movements to the left or right. When a new yarn comes to the needle holding the transferred loop, both overlapping loops will be knit together.
The tranferred loop is highlighted in magenta in Figure
\ref{fig:transferStitch}(b). 
\newline
\textbf{Tuck stitch} \newline
A Tuck stitch is created when a yarn is tucked onto the needle and pulled
up, instead of being pulled through the held loop. The tucked loop
is held on the needle together with the loop from the previous row, as shown in Figure \ref{fig:tuckStitch}(a). The needle holds both loops, which will
be knitted together when a stitch instruction is executed above it on
the next row.
The tucked loop is highlighted in magenta in Figure
\ref{fig:tuckStitch}(b). 
\newline
 \textbf{Miss stitch}\newline
Similarly to the Tuck stitch, during the execution of a Miss stitch the
needle holds the loop from the previous row, but the new yarn is not hooked by the needle. Instead the yarn passes by, creating a horizontal
segment of yarn across the front or the back of the held loop.
The resulting horizontal yarn is highlighted in magenta in Figure
\ref{fig:missStitch}(b). 
\newline
\newline  
\newline  
 \textbf{Empty stitch} \newline
An Empty stitch specifies that no machine operation will be executed
for a specific needle. Therefore, no yarn is looped on the needle at that location.

These six stitches (Knit, Purl, Tuck, Miss, Transfer and Empty) are the
fundamental stitches needed to create most knitted textiles, and can be
combined to generate complicated knitted patterns.
\name\/ supports all of these stitches;
thus allowing for a broad representation of knitted fabrics.

\section{Modeling Spaces, Primitives and Mappings}

When developing a model that supports the evaluation of the
validity, manufacturability and structural integrity of a particular
material, it is critical to define and assess the configuration spaces associated with the material and the expressiveness of the primitives that exist within these spaces.

When modeling  knitted  textiles,  three  spaces  are  available for primitive definition. The first is the configuration space of the knitting machine. This would include modeling the needles, racking beds, carriages and yarn carriers of the machine, their motion and their interaction with each other and with yarns. While the advantage of this space is that it can provide information about and insight into the manufacturing process, it requires a simulation process to produce a virtual version of the material. While this approach/space can be employed to produce knitting commands for manufacturing a knitted
fabric \cite{McCann:2016:CMK,Vidya:2018:AMK,Lin:2018:ETP}, it does not
provide a representation of the material itself. Since we wish to create a representation that is machine independent and able to model materials produced on all flatbed weft-knitting machines, this space was deemed too restrictive for our purposes.

The second is the fabric space of the material itself, which is a static fully evaluated space where the structures of the manufactured fabric are directly represented. For example, a plausible topological model of any fabric could be a cell complex where 0-cells represent all contacts between yarns, oriented 1-cells correspond to yarn segments connecting adjacent contact points,  and oriented 2-cells could be used to represent the yarn loops.   The advantage of this space is that allows for validity checking of the
fabric,
for example topological correctness, but the disadvantage is that the indirect property of manufacturability is more difficult to assess and enforce. Additionally, without modeling the manufacturing process, determining the relationship between machine parameters and material characteristics is problematic. Given the nearly countless possible combinations of stitches and the complex low-level structures that these combinations can produce, it is difficult to create a single static data structure that is guaranteed to accommodate the topology and geometry of all possible fabrics. For example, representing topology of a Knit or Purl stitch may be straightforward, but the topological structure of the combination of other stitches, such as
Transfer, Miss or Tuck stitches, is far less obvious. 

\subsection{Process Space}
Our work focuses on modeling knitted textiles within a \textit{process space}, an intermediate space between machine space and fabric space. This space captures aspects of both machine and fabric space, as it includes some structures of the material, as well as how those structures are manipulated by the machine during the knitting process.
Process space models the abstract processes that lead to the formation of the material.
In our work we focus on the processes that locally manipulate yarn loops
at the stitch command level.
It is the interlocking of these loops that form the central structures that hold a knitted fabric together.  A contiguous block of stitches may be grouped together into swatches that can be repeated to produce a whole fabric. Additionally, the loops may be moved to other needles, both on the front and back beds of a knitting machine in order to create even more complex stitches.

The two critical components of the material that play a central role in this process are the yarn and the yarn crossings where the stitches connect with each other. Our process-oriented representation has the 
yarn  crossing as its primary primitive, with the yarn being implicitly defined as the primitive that connects yarn crossings. The knitting process is then described as the creation and manipulation of these yarn crossings. Since the interaction of yarns at a crossing and the connections between yarn crossings may be complex, we encapsulate this complexity in a primitive we call a Contact Neighborhood (CN). This is a primitive which does not explicitly represent the geometry of the crossing yarns or the adjacencies between crossing yarns. The manipulations/movement of the CNs, for example via a Transfer stitch, are defined via 
mappings that change their location and possibly their state in the fabric.
Thus, the topological structure (a cell complex) underlying the fabric is constructed by specifying, transforming and interconnecting CNs as explained below. 

The process of knitting defines a natural bivariate parameterization of the resulting fabric. Machine weft knitting uses a row of $k$ needles. A single yarn is carried across these needles in a back-and-forth fashion. Each pass of the yarn may be labeled with integer $l$.  Therefore each stitch lies in a coordinate grid and has a unique $(k,l)$ grid location. The two indices $(k,l)$ produce a parameterization over the surface of the knitted material.  The definition of CNs and the way that they interconnect capture the inherent 2D topological structure of knitted textiles.

\subsection{Primitives}
   \label{primitives}
The fabric can be described in terms of loops, but its geometric and
mechanical properties are determined by lower-level components,
yarns and their ``contacts'', which is why we chose Contact Neighborhoods
(CNs) and the yarns that connect them to be the fundamental primitives
of our knitted textile data structure.
Loops and stitches were deemed to be too coarse as primitives because certain stitch operations can create structures that span several needle
locations and yarn rows or produce non-symmetric loops. While most loops are well-defined symmetric structures, there are cases where that symmetry is broken and the four CNs that define the loop are significantly different. Additionally, certain stitch combinations can produce complex structures that require more representational resolution than is available at the loop/stitch level.

While yarns come in many types, e.g.~staple (twisted short fibers),
monofilament (a single extruded material), and multifilament (twisted
monofilaments), our approach assumes that the yarn, as compared to its
constituent fibers, is the smallest physical element that needs to be
directly modeled when simulating knitted materials. Yarns may have
complex, non-linear properties, but these may be lumped without the
need for modeling the individual fibers that make up the yarn.
Previous work has shown that the shapes of the yarns in stitches play a
significant role in the overall mechanical behavior of knitted
materials \cite{Liu2017:ORM,Liu2018:CAM}, irrespective of the deformation
properties of the yarns. Therefore we chose the yarn as the lowest-level
primitive to provide the connection between CNs.

\begin{figure}[t]
  \centering
  \includegraphics[width=0.35\linewidth]{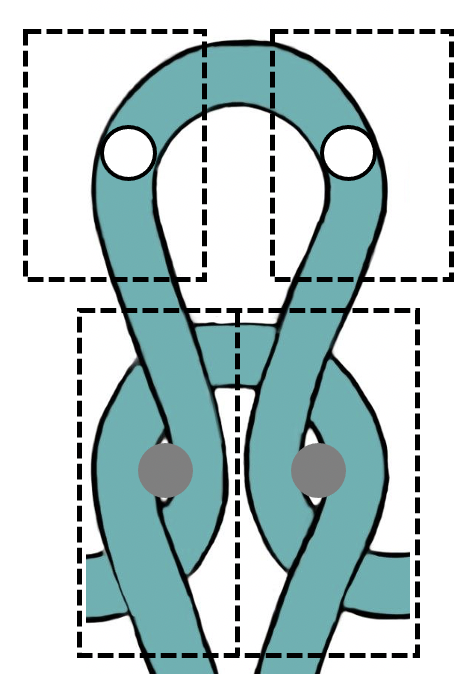}
  \hspace{3mm}
  \includegraphics[width=0.55\linewidth]{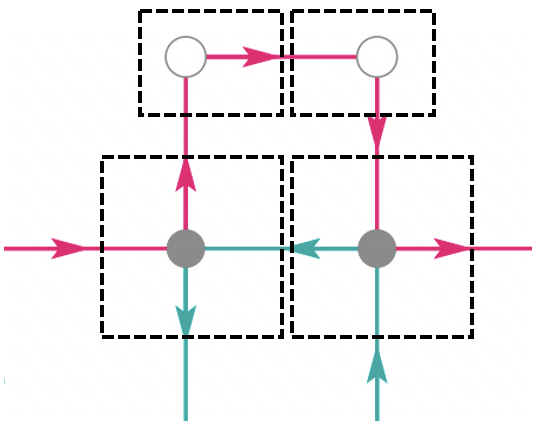}
  \mbox{\hspace{-14mm}(a)\hspace{38mm}(b)}
 \caption{(a) A single stitch is represented by the potential CNs (white disks) of its upper loop and the actualized CNs (grey disks) where its legs cross with the previous loop. (b) Topological representation of the stitch showing the CNs and the edges (yarns) that connect them. The actualized CNs
 consist of the contact point between yarns and
 four incident, directed edges. The potential CNs consist
 of two incident edges and a potential contact point.}
   \label{fig:singleLoopCNs}
\end{figure}

A yarn's direction is defined by the direction of the carrier movement at the time of the CN instantiation. In general the direction
of a yarn alternates left and right as each row of stitches is formed. The labeling/indexing of CNs is slightly different
from the labeling of the needles and yarns. Since the loop formation process generates two CNs for each stitch, each needle
$k$ is associated with two CNs ($2k$ and $2k+1$) in the CN's $i$ coordinate. A yarn's $l$ index maps directly to the CN's $j$ 
coordinate.

\subsubsection{CN Types}
There are three types of CNs, with each CN having a unique identifier
$(i,j)$, the location in the CN grid where it is first instantiated.
The two main CNs are the potential CN (PCN) and the actualized CN (ACN). PCNs are created when a yarn passes over a needle and it hooks the yarn, creating a loop. The PCNs are defined at the head of the loop and mark a location on a yarn where an ACN could be created by intertwining with another yarn. A PCN is defined by a potential contact point and two directed incident edges, as seen by the white disks and two edges surrounded by dashed rectangles in Figure \ref{fig:singleLoopCNs}.

A PCN is actualized into an ACN when the associated loop is used to create a stitch with another yarn,
i.e.\ when an actual yarn crossing is created. An ACN is defined by a contact point and four directed incident edges, as seen by the grey disks and four edges surrounded by dashed rectangles in
Figure \ref{fig:singleLoopCNs}. CN actualization does not have to occur at the location where the PCN is created. The PCN may be transferred to another location and converted into an ACN by a stitch at the new location.

\begin{figure}[t]
  \includegraphics[width=1\linewidth]{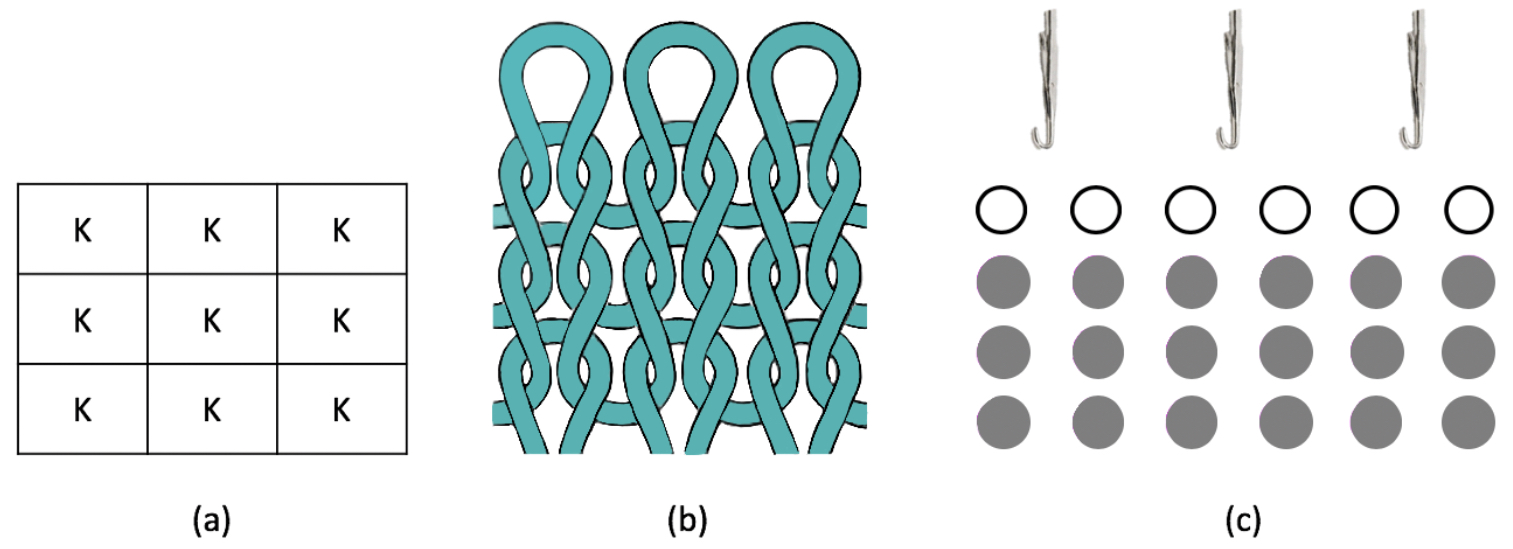}
  \caption{A 3x3 all-Knit stitch pattern: (a) Stitch pattern.
  (b) Resulting loops. (c) Corresponding Contact Neighborhoods.}
  \label{fig:allKnit}
\end{figure}

The CNs for a single Knit stitch are shown in Figure \ref{fig:singleLoopCNs}. It should be noted when a Knit stitch is executed at location $(i,j)$ it actualizes the PCN present at $(i,j)$ and creates a PCN at $(i,j+1)$. Figure \ref{fig:singleLoopCNs}(b) is a topological representation of the stitch, with the edges between the CN nodes representing the two yarns (one teal and the other magenta) connecting them. 
The directions of the yarns into and out of the CNs are signified with
arrows. The gray color of the ACNs signifies that this is a Knit stitch that
produces the specific yarn crossing relationships seen in
Figure \ref{fig:singleLoopCNs}(a). Figure \ref{fig:allKnit}(c) presents the
PCNs and ACNs that define a $3 \times 3$ swatch of Knit stitches
(Figure \ref{fig:allKnit}(a)). A schematic of the resulting stitches is given in Figure \ref{fig:allKnit}(b). Note that, given the mapping from
needle-yarn $(k, l)$ indexing to CN $(i, j)$ indexing, an $M \times N$ block of stitches produces a $2M \times (N + 1)$ block of CNs, since the loop formation process generates two PCNs for each stitch and the top row of stitches produces both ACNs at its legs and PCNs at its heads.

\begin{figure}
  \centering
  \includegraphics[width=1.0\linewidth]{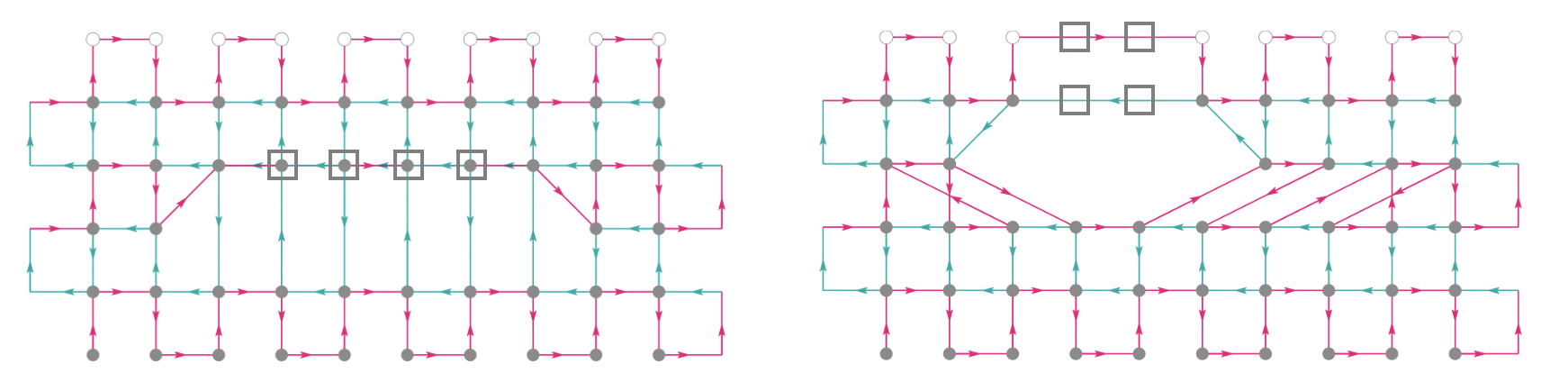}
  \mbox{\hspace{-0mm}(a)\hspace{40mm}(b)}
 \caption{(a) Multiple consecutive Tuck stitches produce unanchored CNs
 (UACNs), displayed as squares. (b) Knit stitches executed on needles with no yarn loops also
 produce UACNs.}
   \label{fig:createUACNs}
\end{figure}

The third CN type is an unanchored CN (UACN). An UACN is created when the yarn is pulled up by the needle, but the loop's legs are not anchored,
i.e.~not twisted or held down by another yarn. This occurs with two
or more consecutive Tuck stitches in a row and when attempting to make a
Knit or Purl stitch when no yarn loop is previously held on the needle,
as shown in Figure \ref{fig:createUACNs}.

When no CN is instantiated at index $(i,j)$ its state is defined as
Empty (E). This may occur
when the needle did not hook the yarn during the associated stitch
operation, e.g.~the Miss stitch, or when the yarn was not pulled over this
location, e.g. outside the boundary of the fabric.

\begin{table*}[]
\centering
\textbf{Knit, Purl and Transfer Stitch\\ \tiny \textcolor{white}{foo}\\}
\begin{tabular}{|c|c|c|c|}
\hline
j  CN value before stitch execution & j  CN value after stitch execution & j + 1 value (Knit $|$ Purl) & j + 1 value (Transfer) \\ \hline
null. PCN, [0,0] & K$|$P, ACN, [0,0] & null, PCN, [0,0] & null, PCN, [${\Delta}i$,0] \\ \hline
null, PCN, [${\Delta}i$,0] & K$|$P, PCN, [${\Delta}i$,0] & null, UACN$|$PCN, [0,0] & null, UACN$|$PCN, [${\Delta}i$,0] \\ \hline
null, UACN, [${\Delta}i$,0] & K$|$P, UACN$|$ACN, [${\Delta}i$,0] & null, UACN$|$PCN, [0,0] & null, UACN$|$PCN, [${\Delta}i$,0] \\ \hline
null, E, [0,-1] &{\color{black} K$|$P, E, [0,-1]} & null, UACN$|$PCN, [0,0] & null, UACN$|$PCN, [${\Delta}i$,0] \\ \hline
\end{tabular}
\caption{Values set at data structure elements $(i,j)$ and $(i,j+1)$ when
processing Knit, Purl and Transfer stitches.}
\label{tab: populatingDSKnitTransfer}
\end{table*}

\subsection{Mappings}

Since knitted textiles consist of more complicated stitches than just Knit and Purl, additional information is required to specify how the knitting process manipulates the CNs after they have been created from the loop formation process. For example, a Transfer stitch first produces a Knit or Purl stitch (thus
creating ACNs at its legs and PCNs at its head), but then transfers, i.e. moves, the head of the stitch to a
another nearby needle to its left or right.
See Figure \ref{fig:transferStitch}.
These manipulations can be represented by mappings that capture the
movement of the CNs and their state as they are transformed.
The mappings that may be applied to CNs include two pieces of information, 
\begin{enumerate}
    \item an actualization value that specifies the state of the CN $(i,j)$ at its final location,
        \item a vector that defines the CNs relative movement from its
	initial location in the CN $(i,j)$ grid.
\end{enumerate}
These mappings specify an incremental, local transformation, i.e.\ the first immediate step in what could be a multi-step movement of the CN through the fabric.  For example, a loop may be transferred to another needle, the loops at that needle could be tucked up to the next row of stitches, then another loop could be transferred onto the tucked destination. 
The incremental form of the mappings supports efficient, on-demand queries
of the model, which is an unevaluated representation of the local
processes that aggregate to form the final knitted textile.

Tuck and Miss stitches apply different manipulations to the PCNs, moving them vertically, rather than horizontally. The Tuck stitch (Figure \ref{fig:tuckStitch}) shifts the PCNs $(i,j)$ and $(i+1,j)$ up to the next row of stitches and overlays them on PCNs $(i,j+1)$ and $(i+1,j+1)$.  
The Miss stitch (Figure \ref{fig:missStitch}) also shifts up PCNs to the next row, but by not hooking the $j'th$ yarn the $j+1'th$ CNs defined for
this stitch (which are Empty) shift down one unit into the $j'th$ row.
Given the horizontal and vertical shifts that
may occur, PCNs may end up at a different location and their actualization
state might change or remain the same.

\section{Data Structure}
\label{section:dataStructure}

We have developed a data structure based on the process-oriented representation of Contact Neighborhoods (CNs) and associated mappings. The data structure can be populated with data derived from an $M \times N$ set of stitch instructions (e.g. Knit, Purl, Tuck, etc.) and produces a $2M \times (N + 1)$ array of CNs. The $(i,j)$ index into the array effectively provides a
unique ID  for the CN created at this position in the fabric and also
specifies a location in the $(i,j)$ grid for other CNs that move through
the fabric. There is a direct relationship between the $(m, n)$ stitch instruction and
the values stored at the four associated $(2m, n)$, $(2m + 1, n)$, $(2m, n + 1)$,
$(2m + 1, n + 1)$ CNs. See Figure \ref{fig:singleLoopCNs}. At each $(i,j)$ element in the CN array three parameters are stored, which
fully specify the CN that is created by the associated stitch instruction at
that parameterized location in the knitted fabric, as well as information
about the stitch itself.

The three parameters fall into two categories, location-based and CN-based. Location-based parameters, of which there is only one, store information that is specific to that $(i,j)$ location. CN-based parameters provide details about the CN that is created at location $(i,j)$, but which may be shifted to another location. The three parameters are: stitch type ST,
actualization value  AV and the movement vector $[{\Delta}i,{\Delta}j]$ MV.
Each element in the data structure therefore takes the form of (ST, AV, MV).

\subsection{Location-Based Parameter}

The one location-based parameter (ST) gives information about the type of stitch to be formed when actualizing CNs that end up at location $(i,j)$. For example, if a Knit stitch is to be knitted at $(i,j)$, this would be specified with a `K'  in the corresponding element in the data structure. 
A 'P' is stored for a Purl stitch.
The stitch type parameter value
determines how a CN  at this location is actualized, whether the yarn is drawn through the loop(s) from back to front (Knit) or from front to back (Purl). In both of these cases firm contacts are formed between yarns via the twisting of the legs of a loop with the head of another loop, as seen in Figure \ref{fig:singleLoopCNs}(a). Note that since two CNs are created when a loop is formed, the stitch type for locations $(i,j)$ and $(i+1,j)$ will be the same.

\subsection{CN-Based Parameters}

The AV parameter indicates if a CN has been defined and, if so, it
specifies the state of the CN as a result of the stitches executed during the knitting
process. This parameter can take four values: PCN, ACN, UACN and E, as explained
in Section \ref{primitives}.

The MV parameter $[{\Delta}i,{\Delta}j]$ indicates whether the CN is being
shifted in either the horizontal or vertical direction. In the case of a
Knit or Purl stitch, ${\Delta}i$ and ${\Delta}j$ are both 0, but when
using other stitches, such as the Transfer stitch, their values would depend
on the amount and direction of the movement needed for that specific
stitch.  A \say{one  needle movement to the left} Transfer stitch (See
Figure \ref{fig:transferStitch}), has ${\Delta}i$ equal to -2 and ${\Delta}j$
equal to 0. Note that CNs are moved and not loops, which is why the
CN is shifted two locations to the left rather than one. For a Tuck stitch,
the vector is $[0,1]$, since the CN is shifted up to the next row.
See Figure \ref{fig:tuckStitch}.
The Miss stitch involves the movement of two sets of CNs.
When Miss stitch data is written
at location $(i,j)$, a CN may already exist at this location
from the processing of a stitch instruction at $(i,j-1)$. The CN is shifted
up to the next row using MV parameter value $[0,1]$. Instead of defining
a CN at $(i,j+1)$ the yarn passes through and the CN that was supposed to
be defined at $(i,j+1)$  is now shifted downward with MV $[0,-1]$.
See Figure \ref{fig:missStitch} for clarification.

\begin{table*}[]
\centering
\textbf{Tuck and Miss Stitch\\ \tiny \textcolor{white}{oo}\\}
\begin{tabular}{|c|c|c|c|}
\hline
j  CN value before stitch execution & j  CN value after stitch execution & j + 1 value (Tuck) & j + 1 value (Miss) \\ \hline
null. PCN, [0,0] & null, PCN, [0,1] & null, UACN, [0,0] & null, E, [0,-1] \\ \hline
null, PCN, [${\Delta}i$,0] & null, PCN, [${\Delta}i$,1] & null, UACN, [0,0] & null, E, [0,-1] \\ \hline
null, UACN, [${\Delta}i$,0] & null, UACN, [${\Delta}i$,1] & null, UACN, [0,0] & null, E, [0,-1] \\ \hline
null, E, [0,-1] & {\color{black} null, E, [0,-1]}*& null, UACN, [0,0] & null, E, [0,-1] \\ \hline
\end{tabular}
\caption{Values set at data structure elements $(i,j)$ and $(i,j+1)$ when
processing Tuck and Miss stitches. *Writing a Tuck and Miss stitch above a
Miss stitch requires special processing away from the $j$ cell, as described
below.}
\label{tab: populatingDSTuckMiss}
\end{table*}

\subsection{Populating the Data Structure}
Given an $M \times N$ array of stitch instructions, a CN grid of size
$(2M, N+1)$ is allocated.
Each element in the grid is initialized as an
Empty stitch, with the values (null, E, [null,null]),
except the first row which we assume contain PCNs due to ``cast-on''
stitches that stabilize the bottom of the fabric.
The stitch instructions are processed and CN values are assigned in a
row-by-row,
alternating left-right then right-left order, similar to the knitting
process they represent.
In general the data is written to these CNs in pairs, the lower pair, then the upper pair.
The pre-condition state of the lower pair of CNs depends on the stitch
instruction executed below them in the $n-1$ row.
This state, along with the $(m,n)$ stitch instruction currently being
executed, determine if and how the lower and upper pairs of CNs will be
modified.

\begin{figure}[b!]
  \centering
  \includegraphics[width=1.0\linewidth]{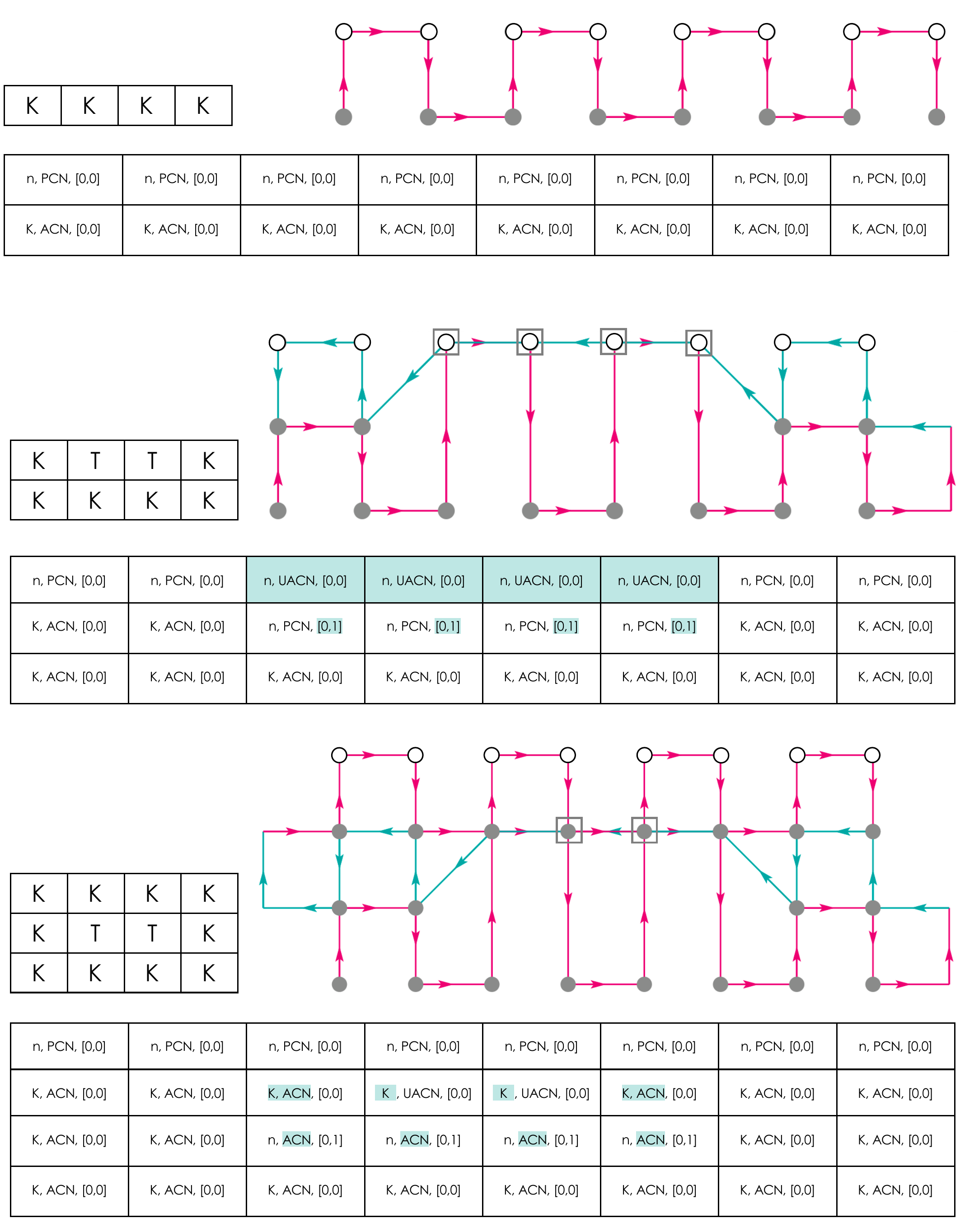}
  \begin{center}
  \vspace{0mm}
   \includegraphics[width=1.0\linewidth]{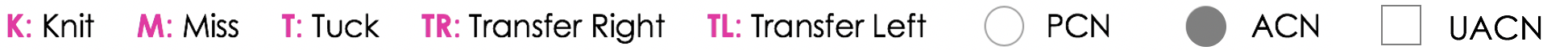}\\
  \end{center}
  \vspace{-3mm}
  \caption{Determining if a UACN is anchored. (top) A row
  of Knit (K) stitches is written to the data structure. The resulting
  topology graph is on the upper right. (middle) Two Tuck (T) stitches
  are executed in the next row. This tucks up the CNs in the second row
  and creates four UACNs in the third row. The changes to the data 
  structure are highlighted. (bottom) Another row of Knit stitches are written
  to the data structure. The outer UACNs are changed to ACNs, because they
  are connected to ACNs one row down and one column over. The tucked up
  loops are actualized (converted to ACNs) by the Knit stitches.}
  \label{fig:twoTucks}
\end{figure}

\subsubsection{Knit, Purl and Transfer Stitches}
Table \ref{tab: populatingDSKnitTransfer} details how information related
to Knit, Purl and Transfer stitch instructions is written into the CN data
structure, depending on the pre-condition parameter
values for the associated  $(i,j)$ element.  When modifying the lower CNs
(the $j$ cells) two actions are consistently applied to the
data structure values: A Knit or Purl stitch type is written in the ST
parameter and the MV parameter is left unchanged.
The AV parameter remains unchanged for a transferred PCN (a PCN with a 
non-zero ${\Delta}i$) and an Empty (E) CN state. AV is modified to ACN for a
stationary PCN (a PCN where ${\Delta}i = 0$). The AV parameter for transferred PCNs
will be modified when actualized by a Knit, Purl or Transfer at their final location.
A local operation is performed to determine if the UACN is truly
unanchored. The UACN is changed to an ACN if the location down and over
from the UACN holds an ACN, i.e.~the UACN is actually anchored.


For the upper CN pair ($j+1$ cells), the ST parameter remains null.
The AV parameter will be set to PCN
if the $j$ cell is an ACN or if another CN is moved to the $j$ cell.
Otherwise, no CN exists at the $j$ cell, which is one of the conditions
for being unanchored; thus the $j+1$ cell's AV parameter is set to UACN.
The MV parameter has the value [0,0] for Knit and Purl stitches,
since the loops of these stitches are not moved.
See Figure \ref{fig:knitStitch}.
For the Transfer stitch
the MV parameter is set to $[{\Delta}i,0]$, where ${\Delta}i$ can have
the values $\pm2, \pm4, \pm6$, depending on the direction and magnitude of
the specified horizontal shift.
See Figure \ref{fig:transferStitch}.


Figure \ref{fig:twoTucks} details how UACNs are written and
possibly modified while the CN data structure is being populated with
Knit and Purl stitches. It 
consists of a stitch pattern, the associated topology graph and data
structure elements. The top
diagram shows the state of the data structure and the topology graph after
four Knit stitches are processed.  Given that the data structure is
initialized with a row of PCNs, the Knit stitches actualize them and PCNs
are written in the next row up. The Knit stitches in the second row produce
the same result. The two Tuck stitches write `1' to the ${\Delta}j$
component of the associated CNs and create UACNs above them.

Note at this point there are two sets of CNs at the four interior CN
locations along the top of the structure. Four UACNs (designated with
squares)
from the teal yarn going right-to-left and four PCNs (designated with
circles with white centers) from the two magenta loops that were tucked up
one row.
When the top row of Knit stitches are written to the data structure,
the UACNs in the $j$ row are examined to determine if they might be
anchored. The UACNs in the 3rd and 6th columns are anchored because they
are connected to a local ACN one row down and one column over. The
Knit stitch
therefore actualizes them and their state is changed to ACN. The interior
UACNs are not directly attached to ACNs in a lower row and therefore do
not change their state.  The PCNs tucked up from a lower row intertwine
with the top yarn and are actualized to ACNs.

For more complex stitch patterns changing the state of an UACN to
anchored and then actualized can require non-local information.
For example, an UACN may be connected to an ACN in a lower
yarn row, but several needle positions away, which can be seen in
Figure \ref{fig:patternVertical}. The lower ACN does anchor the upper
CN. Determining this configuration and change of status requires
tracing the yarn from the UACN to a possibly faraway CN, a non-local
operation. Wanting the process of initially populating the data structure
to be based on local information only, we decided to determine the
final state of these UACNs at a later time, during the yarn tracing
procedure.

\subsubsection{Tuck and Miss Stitches}
Table \ref{tab: populatingDSTuckMiss} details the parameter values that
are written to the CN data structure for Tuck and Miss stitches.
In the lower CNs ($j$ cells), the ${\Delta}j$ component of the MV parameter
is set to 1, except for the Miss stitch precondition (${\Delta}j = -1$),
which is left unchanged.
See Figures \ref{fig:tuckStitch} and \ref{fig:missStitch}.
As signified by an asterisk* in the table, writing Tuck and Miss parameters
values above a Miss stitch requires special processing away from the four
cells that are normally affected by each stitch instruction. Since Tuck
and  Miss stitches effectively pull up the yarns held on the needle, a
column of these stitches will pull the yarn up multiple rows. In order to
capture this behavior, when writing a Tuck or Miss stitch above a
Miss stitch, we move down the column looking for the first cell with a
positive ${\Delta}j$ value. Once found, this cell's ${\Delta}j$ value is
then 
incremented by 1, indicating that associated CN has been pulled up one
more row.

\begin{figure}[]
  \centering
  \includegraphics[width=0.9\linewidth]{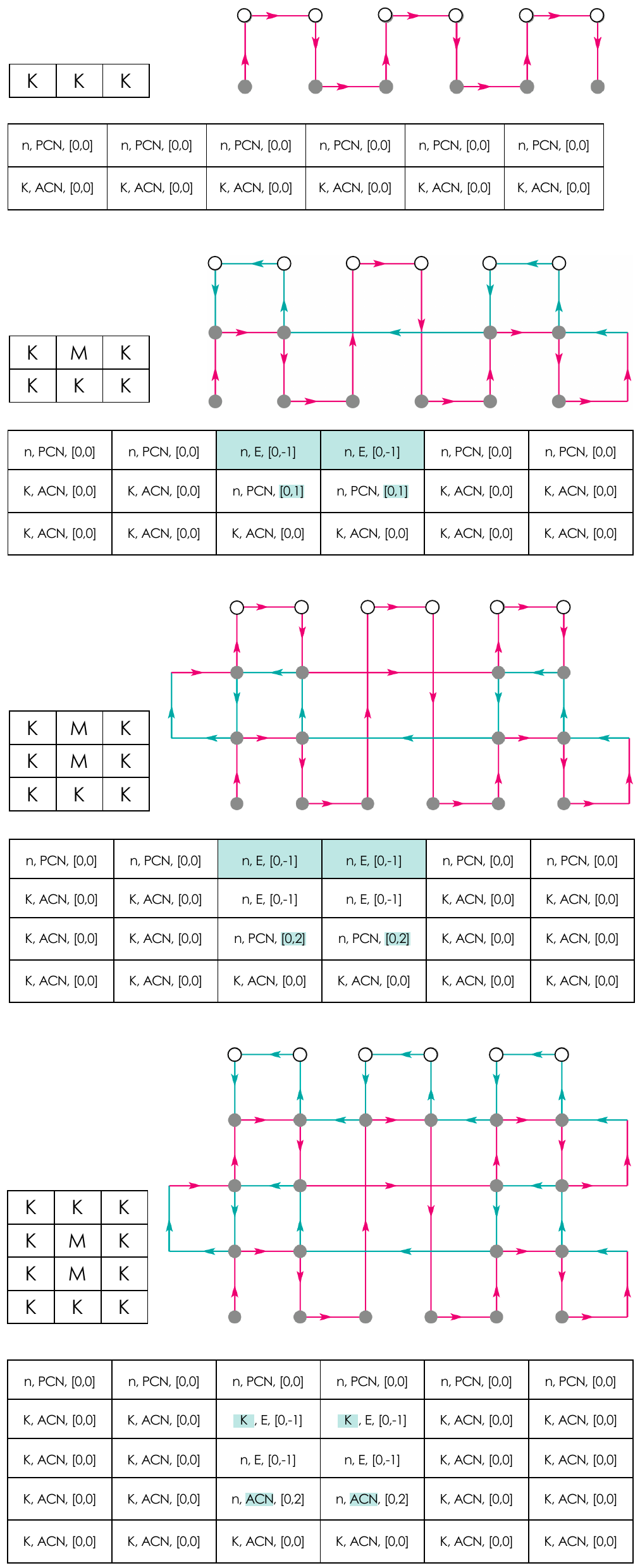}
  \caption{Updating ${\Delta}j$ of a Miss stitch.  (top) A row
  of Knit (K) stitches is written to the data structure. The resulting
  topology graph is on the upper right. (middle top) A Miss stitch is
  executed in the middle of the pattern in the next row. This pulls up
  the held yarn and the traversing yarn crosses to the next stitch. An Empty
  stitch is written at $j+1$. (middle bottom) Another Miss stitch is
  executed above the previous one. The held loop is pulled up another
  row. Note that its ${\Delta}j$ has been incremented to 2. (bottom)
  A final row of Knit stitches actualizes the held loop.}
  \label{fig:twoMiss}
\end{figure}

The $j+1$ cells of all Tuck stitches are defined as unanchored, because
a Tuck stitch does not make a yarn intertwining at location $(i,j)$.
Thus the $j+1$ parameter values are set as (null, UACN, [0,0]). For the
Miss stitch, no CNs are created at $j+1$, since the needle does not hook
the traversing yarn. Thus the AV value is set to E (Empty) and the 
${\Delta}j$ is set to -1, signifying that the yarn that should have
passed through this location can be found in the next row down.

The special processing needed for Miss stitches is further explained in
Figure \ref{fig:twoMiss}. First, a row of Knit stitches are written to the
data structure. The next row has a Miss stitch surrounded by Knit stitches.
The Miss stitch changes the PCN's ${\Delta}j$ to `1' and writes
``(null, E, [0,-1])'' to the cell above. The next row has the same
stitches.  In this case, the cell values of the $j$ cell are left
unchanged, and the CNs above the Miss stitch are also set to
``(null, E, [0,-1])''. Note though that the ${\Delta}j$ values of the CNs
that are pulled up another row have been set to `2'. The top row of
Knit stitches actualize the two exterior sets of PCNs from the third row
Knit stitches and the PCNs created by the Knit stitch at the center of
the first row.

\section{Topology Graph Algorithms}
\label{section:algorithms}

Given a stitch pattern, the \name\/ data structure and associated
algorithms can generate a topology graph of a knitted textile created
with a single yarn. See Figures
\ref{fig:pattern1} to \ref{fig:pattern7} for examples.
The graph specifies how the yarn flows
through the fabric and how points on the yarn contact other points
on the yarn as it travels through the various stitches.
The nodes in the graph, Contact Neighborhoods,
represent yarn crossings/intertwinings and the edges of the graph
represent the yarn segments that connect the yarn crossings.
The nodes of the graph are located at discrete $(i,j)$ locations in a
2D grid, which are related to the $k'th$ needle that created the associated
stitch with the $l'th$ row of the yarn.  
Zero, one or multiple nodes can be located at a single $(i,j)$ grid point.

\begin{figure}[t]
  \includegraphics[width=1\linewidth]{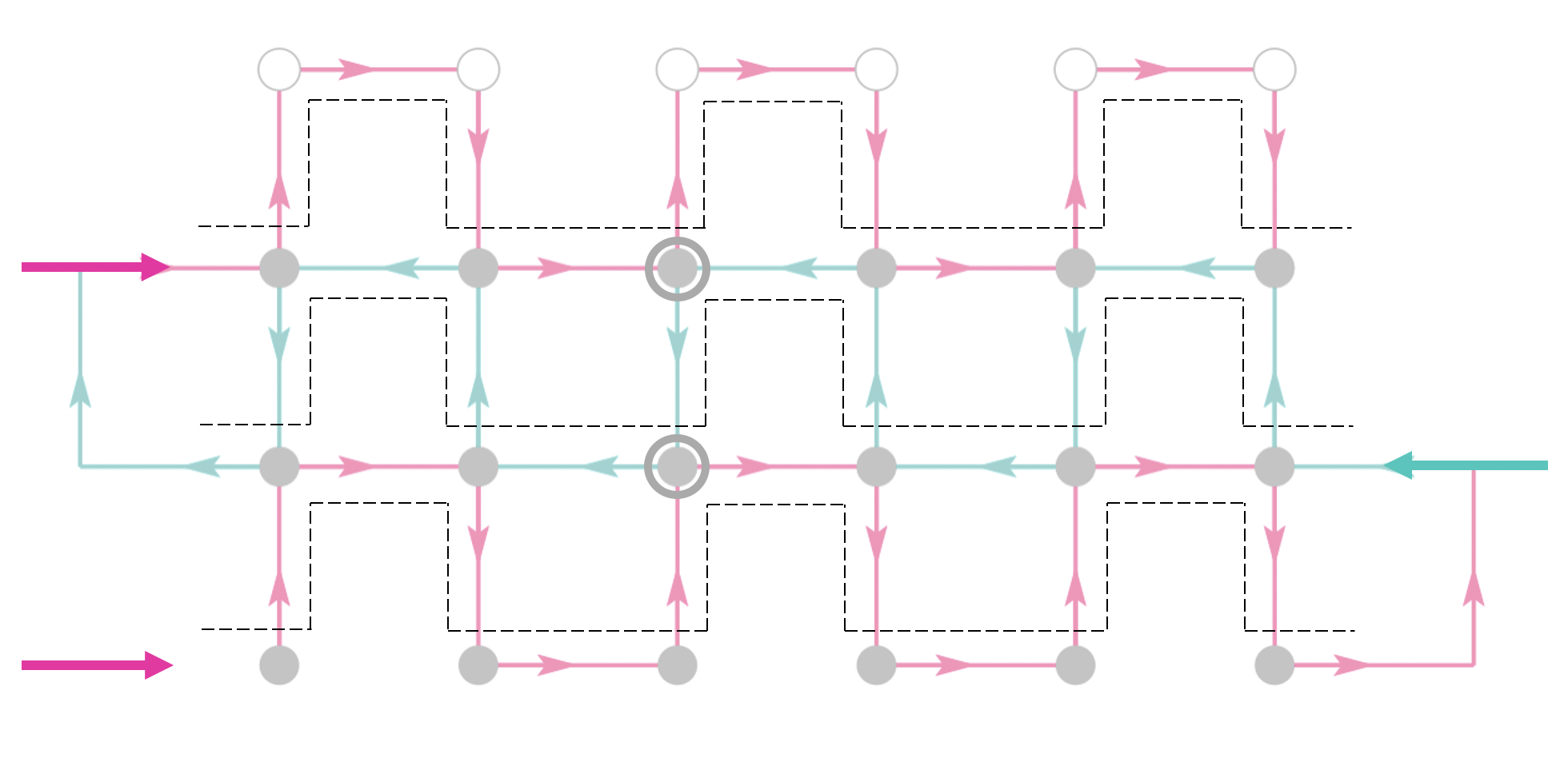}
  \caption{CNs processed in a square wave order for a 3x3 all-Knit pattern. The highlighted CNs are described Section \ref{sec:add-to-list}.}
  \label{fig:squareWave}
\end{figure}

\subsection{Yarn Path Algorithms}
\label{subsection:yarn_path_algorithms}

Once populated with local stitch information, 
algorithms may be applied to the \name\/ data structure
to compute a variety of yarn-level topological data about the fabric
produced by a given stitch pattern.

\subsubsection{Follow-The-Yarn Algorithm}
The main algorithm for extracting the yarn
topology is called FOLLOW\_THE\_YARN and produces a sequential list of
$(i,j)$ locations that the yarn passes through as it courses throughout
the fabric.  More specifically, each element in the yarn path list
is a location that contains an ACN, i.e.~this is a location where
two or more yarns intertwine, with these ACN locations being sequentially
ordered along the length of the yarn.
An integer that identifies the current stitch row number, an important
piece of information needed when determining the connectivity of an ACN,
is also stored with the grid location.

The algorithm starts at location $(0,0)$ (bottom left) in the data
structure and processes the CNs in a square-wave pattern as seen in
Figure \ref{fig:squareWave}.
The locations and associated CNs are processed along this path because it
captures the order in
which CNs are created and modified by the knitting process.
The arrows on the sides of the figure indicate the direction that the yarn
is carried for that row of stitches, first left-to-right (magenta arrow),
then right-to-left (teal arrow), and so on.
The square-wave pattern defines what we call \textit{leg} nodes and \textit{head}
nodes.  In this pattern a leg node CN is first processed (node $(i,j)$,
assuming that the
yarn direction is from left-to-right), followed by two head nodes
($(i,j+1)$ and $(i+1,j+1)$), and finally the other leg node
CN ($(i+1,j)$).
Note that with this pattern, two yarns flow into and out of each CN
(except at the top and 
bottom rows, which can be stabilized with ``cast-on'' and ``bind-off''
stitches), once as a head node and once as a leg node.
When representing an all-Knit pattern, all grid locations
will be added to the yarn path, which makes its topology match the square
wave (See Figure \ref{fig:squareWave}). Therefore the mappings for the
other stitches effectively encode the local topological modifications that
make the  associated loops  deviate from the default all-Knit structure.

\begin{algorithm}
\small
\caption{$FOLLOW\_THE\_YARN(DS)$}
\begin{algorithmic}[]
\State $i, j ,legNode, currentStitchRow \gets 0, 0,True, 0$
\State $yarnPath \gets []$
\While {There are CNs to process}
\If {$ADD\_TO\_LIST(i,j,legNode,yarnPath,DS)$}
\If {$legNode$}
\State $locInfo \gets (i,j,currentStitchRow)$
\Else  \Comment{Head node}
 \State $locInfo \gets \newline  \mbox{\hspace{16mm} (FINAL\_LOCATION(i,j,DS),currentStitchRow)}$
\EndIf

\State $yarnPath.append(locInfo)$
\EndIf
\State $i,j,legNode,currentStitchRow \gets \newline \mbox{\hspace{20mm}} NEXT\_CN(i,j,legNode,currentStitchRow)$
\EndWhile
\Return yarnPath
\end{algorithmic}
\label{alg:follow_yarn}
\end{algorithm}

As seen in Algorithm \ref{alg:follow_yarn},
the ADD\_TO\_LIST algorithm is invoked
for each CN processed in the square wave order, and determines if the CN
should be added to the yarn path.
If the ADD\_TO\_LIST algorithm returns True, the legNode status of
$(i,j)$ is checked.  If it is a leg node, the $(i,j)$ grid location of the
CN is added to the yarn path list.
Since only head nodes are moved in the fabric during the knitting process,
a leg node's final location matches the location $(i,j)$ being processed.
The FINAL\_LOCATION algorithm is called to determine a
head CN's final location in the $(i,j)$ grid.

Once a CN is processed, the NEXT\_CN algorithm computes the next
$(i,j)$ location in the square wave, along with the leg node status of CN
$(i,j)$, and updates the $currentStitchRow$ index if necessary.

\begin{algorithm}[t]
\small
\caption{$ADD\_TO\_LIST(i, j, legNode, yarnPath,DS)$}
\begin{algorithmic}[]
\If {$legNode$}
\State $numberOfACNs \gets LENGTH(ACNS\_AT(i, j,DS))$
\If {$numberOfACNs > 0$} \Comment{Anchored CN is here}\\
\hspace{9mm}\Return True
\Else\\
\hspace{9mm}\Return False
\EndIf
\Else \Comment{Head node}
\State $actualizationValue \gets DS[i][j].AV$
\If {$actualizationValue == ``E"$}\\
\hspace{9mm}\Return False
\ElsIf {$actualizationValue ==  ``UACN" $}
\If {$EVEN(i) == ODD(j)$} \Comment{Look backward}
\State {$(m,n) \gets Last \hspace{1mm} ACN\hspace{1mm} in\hspace{1mm} yarn \hspace{1mm} path$}
\Else \Comment{Look forward}
\State {$(m,n) \gets Next $\hspace{1mm}$ACN$\hspace{1mm}$ to$\hspace{1mm}$ be$\hspace{1mm}$ added$\hspace{1mm}$ to $\hspace{1mm}$yarn$\hspace{1mm}$path$}
\EndIf
\State $finalI, finalJ =  FINAL\_LOCATION(i, j,DS)$
\If {$n < finalJ$} \Comment{This CN is anchored}
\State $DS[i][j].AV \gets ``ACN"$ \Comment{Update CN's state}\\
\hspace{14.5mm}\Return True
\Else \\
\hspace{14.5mm}\Return False
\EndIf
\Else \Comment{An ACN or PCN} \\
\hspace{9mm}\Return True
\EndIf
\EndIf
\end{algorithmic}
\label{alg:add_to_list}
\end{algorithm}

\subsubsection{Add-To-List Algorithm}
\label{sec:add-to-list}
The two factors that determine whether a CN$(i,j)$ should be added to
the yarn path list are the CN's leg node status and its AV parameter.
See Algorithm \ref{alg:add_to_list}.
Leg node CNs are processed first.  The algorithm returns $True$ if
there exists at least one actualized CN at location $(i,j)$. This signifies
that the upper yarn was intertwined with another lower yarn at this
location.
Head  nodes require examining the CN's AV parameter. Empty CNs (E),
which can be part of a Miss stitch or outside the boundary of the fabric,
are not included in the yarn path list.

UACN head nodes require additional inspection to determine if they have
been anchored by another CN further away along the yarn. This determination
is performed by either looking back to the previous element in the yarn
path or looking forward to find the next ACN to be inserted in the yarn
path. The direction of the inspection is based on the structure of the
square wave pattern and the parities of the $i$ and $j$ variables.
The parity of $j$ indicates if the yarn is going from right to left
(even) or left to right (odd) across the head of the loop, thus determining
if the previous CN is at a lesser or greater $i$ value.  For example, the
first processed head node of a stitch is connected to a leg node
``behind'' it along the yarn.  The second head node is connected to a
leg node ``in front of'' it along the yarn. The parity of $i$ and $j$
values captures this first/second relationship, as seen in Figure
\ref{fig:squareWave}. The lower circled CN has an $(i,j)$ value of
$(2,1)$.  When the parities are different the anchored check looks
backwards. The upper circled CN has value $(2,2)$. When the parities are
the same, the check looks forward along the yarn for a CN that might be
at a lower $j$ value.

The final block of the AV check returns True, because all ACN and PCN
head nodes are added to the yarn path list.

\begin{algorithm}[t]
\small
\caption{$FINAL\_LOCATION(i, j,DS)$}
\begin{algorithmic}[]
\If {$j == lastRow$} \Comment{CNs in the last row don't move}\\
\hspace{3mm}\Return i, j
\ElsIf {$DS[i][j].{\Delta}i\hspace{1mm} != 0$}\Comment{Move CN horizontally}\\
\hspace{3mm}\Return FINAL\_LOC\_RECURSIVE(i + DS[i][j].${\Delta}i$, j,DS)
\Else \Comment{Move CN vertically} \\
\hspace{3mm}\Return FINAL\_LOC\_RECURSIVE(i, j  + DS[i][j].${\Delta}j$,DS)
\EndIf
\end{algorithmic}
\label{alg:final_loc}
\end{algorithm}
\begin{algorithm}[t]
\small
\caption{$FINAL\_LOC\_RECURSIVE(i, j,DS)$}
\begin{algorithmic}[]
\If {$ DS[i][j].ST ==  (K|P)$}\Comment{CN is actualized here}\\
\hspace{3mm}\Return i, j
\Else \Comment{Recursively accumulate vertical movements}\\
\hspace{3mm}\Return FINAL\_LOC\_RECURSIVE(i, j  + DS[i][j].${\Delta}j,DS$)
\EndIf
\end{algorithmic}
\label{alg:final_loc_rec}
\end{algorithm}

\subsubsection{Final-Location Algorithm}
Head node CNs can move within the $(i,j)$ grid, with their local
movements encoded in the MV parameter.
The FINAL\_LOCATION function aggregates the movement information to
determine the final location of the CN$(i,j)$  in the $(i,j)$ grid.
Since a Transfer stitch actualizes two leg nodes (by creating a Knit or
a Purl stitch) and then shifts its
newly created head nodes left or right, a held loop (represented by two
CNs) cannot be shifted vertically and then shifted horizontally;
nor do horizontal CN shifts accumulate.  Therefore, as seen in
Algorithms \ref{alg:final_loc} and \ref{alg:final_loc_rec},
FINAL\_LOCATION first applies horizontal movements, if any exist, before
recursively calling FINAL\_LOC\_RECURSIVE to aggregate any possible
vertical movements. See Figure \ref{fig:patternVertical}. The final location is reached at a location where a Knit or Purl stitch
is executed, which actualizes the CN moved to this location.

\subsubsection{ACNS-At Algorithm}
ACNS\_AT (Algorithm \ref{alg:acns_at}) determines
which ACNs are ultimately positioned at location $(i,j)$ in the CN grid,
after the knitting process is complete.
It computes
in near constant time, since it executes the FINAL\_LOCATION algorithm
on a small finite set of CNs in a local neighborhood of $(i,j)$.
It is only necessary to examine the local CNs because a loop of yarn can
only be stretched a limited amount before it or the needle holding it
breaks. Conventional knitting wisdom holds that a loop of yarn can only be
stretched/moved away three needle positions horizontally or three yarn
rows vertically.
Additionally, since knitting is a sequential process, a loop at row
$j+1$ cannot be shifted downwards to a previous, lower row. The physical
properties of yarns therefore limit the distance
over which they can be manipulated and leads to only requiring the
examination of a $13 \times 4$ neighborhood
($i - 6 \rightarrow i + 6$, $j \rightarrow j - 3$) around location $(i,j)$.
FINAL\_LOCATION is called for each CN$(i',j')$ in the neighborhood,
and if this CN's final location is $(i,j)$ and its AV state is $``ACN"$,
it is added to the returned ACN list.

\begin{algorithm}[t]
\small
\caption{$ACNS\_AT(i, j,DS)$}
\begin{algorithmic}[]
\State{$ACNList \gets []$}
\For{all CN $(i',j')$ in the local 13 $\times$ 4 neighborhood}
\If {$(FINAL\_LOCATION (i',j',DS) == (i,j))\hspace{1mm}\&\&$ \newline\mbox{\hspace{30mm} $(DS[i'][j'].AV == ``ACN")$}}
\State $ACNList.append((i',j'))$
\EndIf \\
\Return ACNList
\EndFor
\end{algorithmic}
\label{alg:acns_at}
\end{algorithm}

\begin{algorithm}[b]
\small
\caption{$NEXT\_CN(i,j,legNode,currentStitchRow)$}
\begin{algorithmic}[]
\If {$BORDER\_CN(i,j,legNode)$}
\\ \hspace{4.7mm}\Return i, j+1, True, currentStitchRow + 1 \Comment{Go up 1 row}
\Else
\State $nextI, nextJ = SQUARE\_WAVE(i,j,legNode)$
\State $legNode = (nextJ == currentStitchRow)$ \\
\hspace{4.7mm}\Return nextI, nextJ, legNode, currentStitchRow
\EndIf
\end{algorithmic}
\label{alg:next_CN}
\end{algorithm}

\begin{algorithm}[t]
\small
\caption{$DRAW\_TOPOLOGY\_GRAPH()$}
\begin{algorithmic}[]
\State $CNList =  FOLLOW\_THE\_YARN()$
\State $n \gets len(CNList)$\\
\Comment{Draw the first disk\ \ \ \ \ \ \ \ \ \ \ \ \ \ \ \ \ \ \ \ \ \ \ \ \ \ \ \ \ \ \ \ \ \ \ \ \ \ \ \ \ \ \ \ \ \ \ \ \ \ \ \ \ \ \ \ \ \ \ \ \ \ \ \ \ \ \ \ \ \ }
\If {$DS[0][0].ST ==   K$}
	\State $DRAWDISK(0, 0, gray)$
\Else \Comment{Stitch type is P}
	 \State $DRAWDISK(0, 0, green)$
\EndIf

\hspace{-6mm}\Comment{Loop through the yarn list CNs\ \ \ \ \ \ \ \ \ \ \ \ \ \ \ \ \ \ \ \ \ \ \ \ \ \ \ \ \ \ \ \ \ \ \ \ \ \ \ \ \ \ \ \ \ \ }
\For {$	CNIndex  \gets 0 \hspace{1mm} to\hspace{1mm} n - 2$}
		 \State $currentI, currentJ, currentStitchRow  \gets \newline\mbox{\hspace{52mm}CNList[CNIndex]}$ 
     		\State $nextI, nextJ, nextYarnRow  \gets  CNList[CNIndex+1]$

\mbox{\hspace{-1.5mm}\Comment{Draw graph node\ \ \ \ \ \ \ \ \ \ \ \ \ \ \ \ \ \ \ \ \ \ \ \ \ \ \ \ \ \ \ \ \ \ \ \ \ \ \ \ \ \ \ \ \ \ \ \ \ \ \ \ \ \ \ \ \ \ \ \ \ \ \ \ \ \ }}
	\If {$DS[nextI][nextJ].AV ==  ``PCN"$} 
		\Comment{Last CN row}
              \State $DRAWDISK(nextI, nextJ, white)$
	\Else
		\Comment{Draw ACN disk}
            \If {$DS[nextI][nextJ].ST ==   K$}
              \State $DRAWDISK(nextI, nextJ, gray)$
           \Else \Comment{Stitch type is P}
              \State $DRAWDISK(nextI, nextJ, green)$
           \EndIf
    	\EndIf

\mbox{\hspace{-1mm}\Comment{Draw graph edge\ \ \ \ \ \ \ \ \ \ \ \ \ \ \ \ \ \ \ \ \ \ \ \ \ \ \ \ \ \ \ \ \ \ \ \ \ \ \ \ \ \ \ \ \ \ \ \ \ \ \ \ \ \ \ \ \ \ \ \ \ \ }}
     \State $legNode = (currentJ == currentStitchRow)$
	 \If  {$ODD(currentStitchRow)$}
			\State $ yarnColor \gets teal$
	\Else        
			\State $ yarnColor \gets magenta$
	\EndIf
     \If {$BORDER\_CN(currentI, currentJ, legNode)$}
			\State $DRAWLOOP(currentI, currentJ, nextI, nextJ,$\newline\mbox{\hspace{61mm} $yarnColor)$}
    \Else 
           \State $DRAWARROW(currentI, currentJ, nextI, nextJ,$ \newline\mbox{\hspace{61mm} $yarnColor)$}
	\EndIf

	\If{$ currentJ == nextJ$}
	\Comment{Checking for UACNs}
    \For {$testedI \gets currentI \hspace{1mm} to\hspace{1mm}  nextI$}
		  \If  {$DS[testedI][currentJ].AV ==  ``UACN"$}
               \State $DRAWSQUARESTROKE(testedI,$ \newline\mbox{\hspace{56.5mm} $currentJ, gray)$}
		 \EndIf
	\EndFor
	\EndIf

\EndFor
\end{algorithmic}
\label{alg:draw_topological_diagram}
\end{algorithm}

\subsubsection{Next-CN Algorithm}
After each CN is processed in the FOLLOW\_THE\_YARN algorithm, the
next CN in the square wave is generated by the NEXT\_CN algorithm
(Algorithm \ref{alg:next_CN}).
This algorithm usually invokes the SQUARE\_WAVE algorithm, which in turn uses
logic based on the values of $i$, $j$ and $legNode$ to compute
$nextI$ and $nextJ$. This logic was determined by studying the path
the yarn follows in an all-Knit pattern.

The new location is a leg node if $nextJ$ equals
$currentStitchRow$. If $(i,j)$ is on the border of the stitch pattern, the
$j$ index, which identifies the row of the CN, and $currentStitchRow$ are
incremented and the leg node status is set to True. This encodes that a yarn loops up to the next row to a leg node
position when it reaches the end of a row of stitches.

\subsection{Drawing the Topology Graph}
\label{subsection: drawing_the_topology_graph}

The topology graph information extracted from the algorithms described in
Section \ref{subsection:yarn_path_algorithms} can be visualized using
Algorithm \ref{alg:draw_topological_diagram}. This algorithm iterates
through the list of locations returned by Algorithm \ref{alg:follow_yarn}
and invokes different drawing routines to draw the nodes and edges of the
graph.
To more easily identify the path of the yarn through the graph,
two different line colors are used to represent the yarn. When knitting
an even row, when the yarn flows from left to right, the yarn/edge
color is magenta. When knitting an odd row, when the yarn flows from
right to left, the yarn/edge color is teal. 
The graph contains different types of CNs, and icons with different shapes
and colors are used to distinguish them from each other.
Colored disks are used for ACNs and PCNs, gray for knit ACNs, green for
purl ACNs and white for PCNs.
UACNs that lie along horizontal yarns are detected by Algorithm
\ref{alg:draw_topological_diagram} and gray squares are drawn to
represent them at their grid locations.
Currently, we only display a single disk at any location $(i,j)$,
even if more than one CN ends up at the location. However, the total
number of arrows going in and out of a node can provide
information about the number of CNs at a given location. 


Algorithm \ref{alg:draw_topological_diagram} loops through the locations
of the ACNs, generated by Algorithm \ref{alg:follow_yarn}, that are
sequentially ordered along the yarn that courses through
the fabric defined by the given stitch pattern.
At each iteration of the loop, the algorithm draws an arrow from the current
CN location to the next CN location in the list. The arrow represents the
yarn that connects the two CNs and the direction of the arrow corresponds
to the direction of the yarn during the knitting process. The arrow color
is determined by the yarnColor variable, which is based on the parity (and
direction) of
the current yarn row. When the current list element is on the border 
of the fabric, three arrows are drawn which loop the yarn around to the
next row being visualized.  The first two arrows have the same color as
the yarnColor value and the last arrow has the other color, to show that
a new row is beginning.

It is important to note when the current location and next location in the
yarn list have the same $j$ value there might be UACNs existing between
them.  The algorithm checks all CNs between the current CN location and the
next CN location and draws squares for CNs with actualization values equal
to UACN. A slightly modified version of Algorithm
\ref{alg:draw_topological_diagram} can be used to generate row-by-row
topological graphs for a given stitch pattern as seen in Figure
\ref{fig:pattern1stepbystep}.
This allows us to view the state of the topology graph as the fabric is
being constructed and helps to highlight how the CN actualization values
change during the knitting process.

\begin{figure}[t]
  \includegraphics[width=1\linewidth]{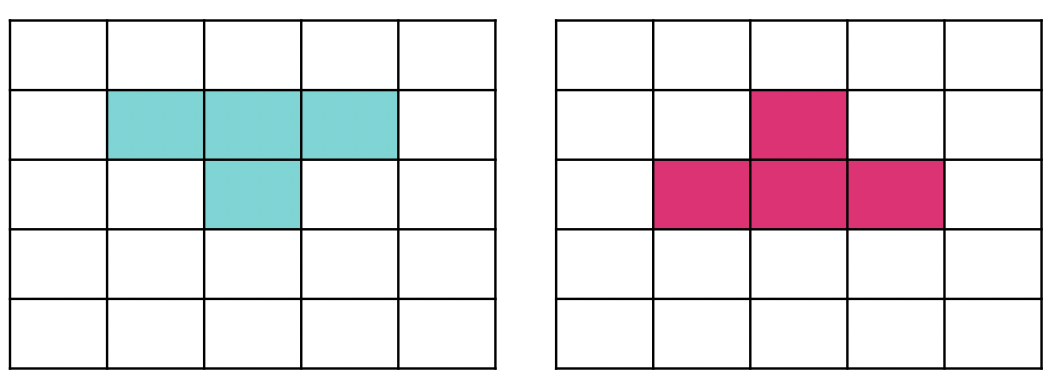}
  \caption{Teal and magenta stitch pattern templates.}
  \label{fig:testPatterns}
\vspace{5mm}
  \includegraphics[width=1\linewidth]{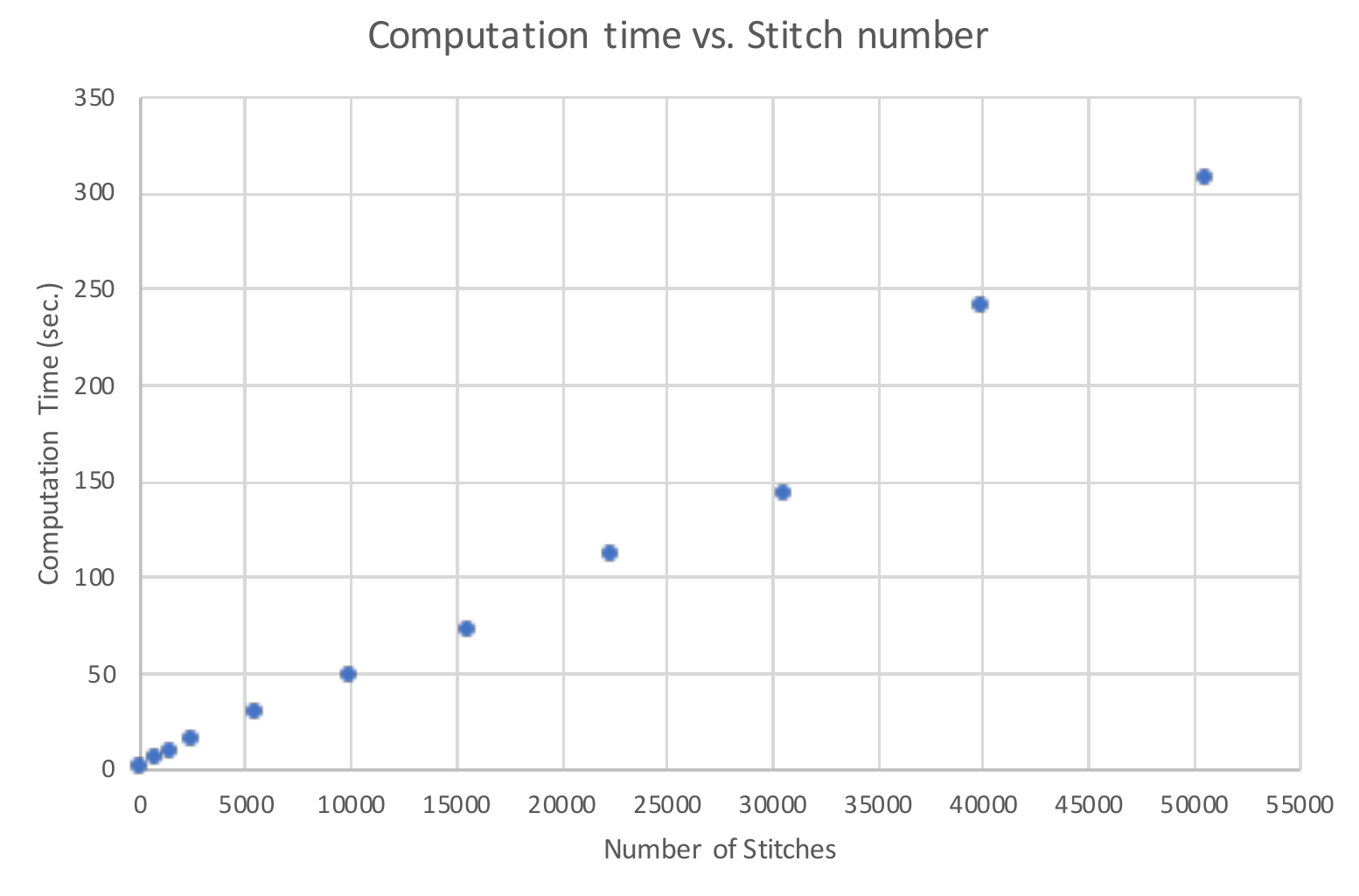}
  \caption{Plot of timing values from Table \ref{tab:timings} that demonstrates the linear time complexity, as a function of the
number of stitches, of \name\/'s evaluation algorithms.}
  \label{fig:timings}
\end{figure}

\begin{table}[t]
\centering
\begin{tabular}{|l|l|l|}
\hline
\multicolumn{1}{|c|}{Pattern Size} & \multicolumn{1}{c|}{\# Stitches} & \multicolumn{1}{c|}{Time (sec.)} \\ \hline
10 $\times$ 10                     & 100                              & 0.443                            \\ \hline
30 $\times$ 30                     & 900                              & 5.51                             \\ \hline
40 $\times$ 40                     & 1,600                            & 8.81                             \\ \hline
50 $\times$ 50                     & 2,500                            & 14.8                             \\ \hline
75 $\times$ 75                     & 5,625                            & 29.0                             \\ \hline
100 $\times$ 100                   & 10,000                           & 48.0                             \\ \hline
125 $\times$ 125                   & 15,625                           & 71.6                             \\ \hline
150 $\times$ 150                   & 22,500                           & 111                              \\ \hline
175 $\times$ 175                   & 30,625                           & 142                              \\ \hline
200 $\times$ 200                   & 40,000                           & 240                              \\ \hline
225 $\times$ 225                   & 50,625                           & 307                              \\ \hline
\end{tabular}
\caption{Computation times (in seconds) needed to generate and evaluate
repeating blocks of the stitch pattern in Figure \ref{fig:pattern1}. The first column contains the dimensions of the stitch pattern. The
second column contains the total number of stitches in the swatch and the
last column contains the time needed for the computation.}
\label{tab:timings}
\end{table}

\section{Results}

The \name\/ representation and associated algorithms were tested with 100
$5 \times 5$ stitch patterns in order to assess the robustness, accuracy
and effectiveness of our approach to evaluating the topology of knitted
fabrics.
The test patterns were generated by randomly selecting stitches for each
cell in the teal and magenta areas in the grids presented in
Figure \ref{fig:testPatterns}.
The set of selected stitches included Knit, Transfer right, Transfer left,
Tuck and Miss stitches. All of the white cells were defined as
Knit stitches. \name\/ generated 100 correct topology graphs from these
100 stitch patterns. The correctness of our results was
confirmed by comparisons with graphical outputs from the Shima Seiki
SDS-One APEX3 KnitDesign system.

\begin{figure*}[t]
\begin{tabular}{cc}
 \begin{minipage}{0.2\textwidth}
  \includegraphics[width=1\textwidth]{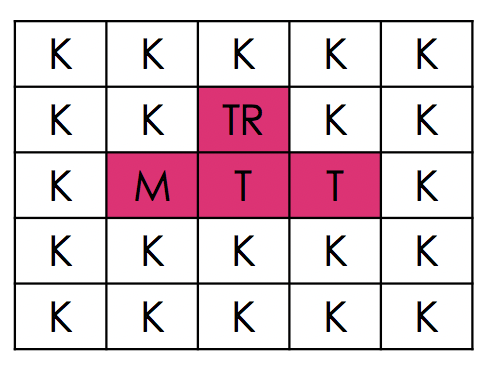} \\  	
      \mbox{\hspace{1.5cm}(a)}\\
    \includegraphics[width=1\textwidth]{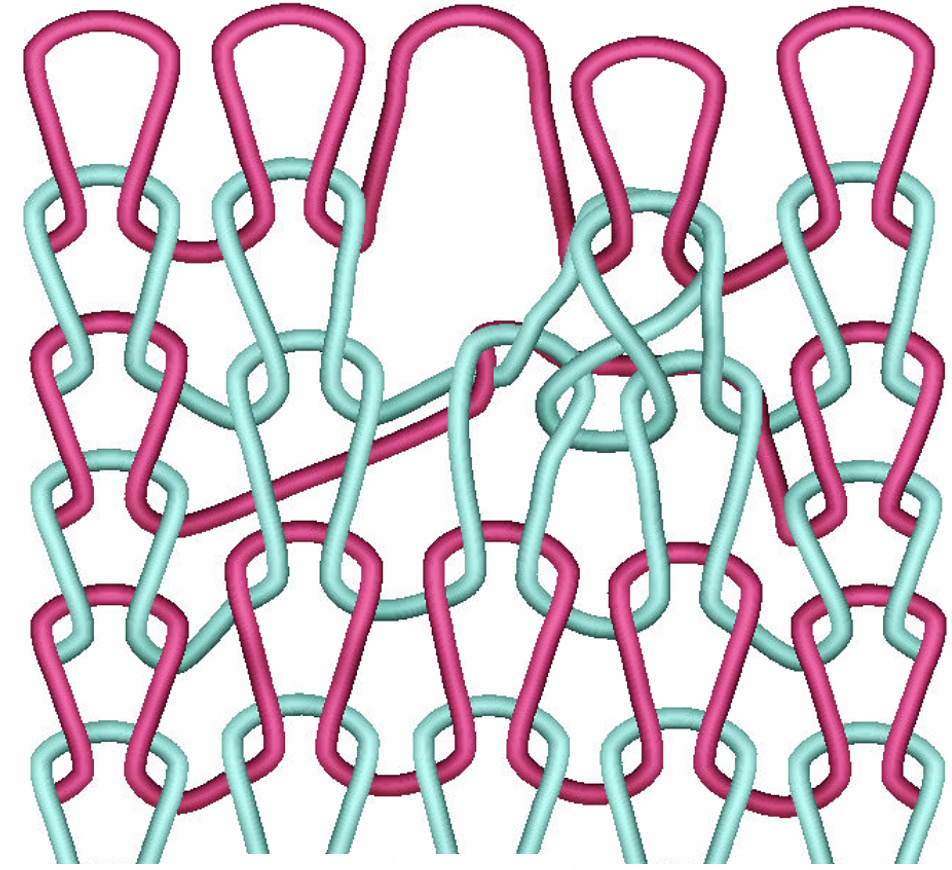} 
         \mbox{\hspace{1.5cm}(b)\hspace{7.9cm}(c)}
 \end{minipage}&
 \vspace{3mm}
 \begin{minipage}{0.8\textwidth} \includegraphics[width=0.9\textwidth]{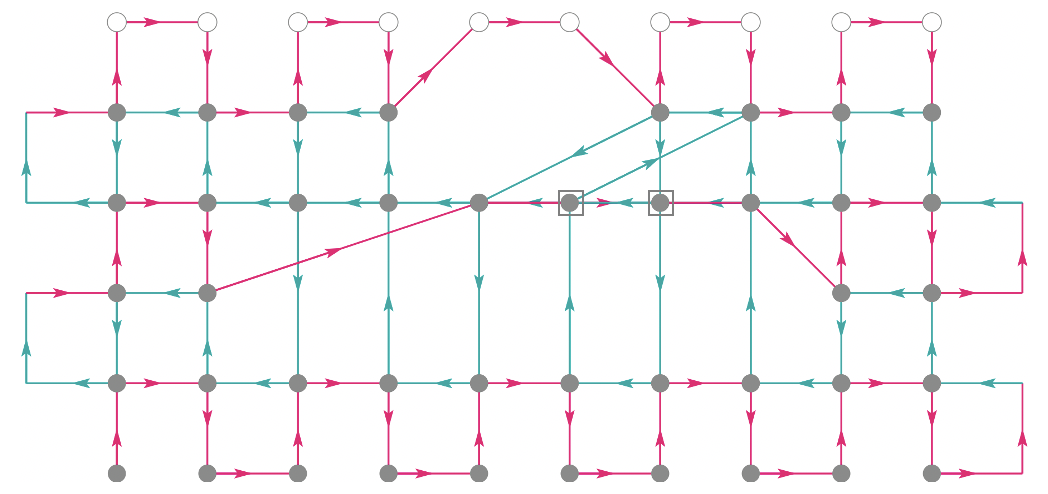} \end{minipage}
 \end{tabular}
  \includegraphics[width=1\linewidth]{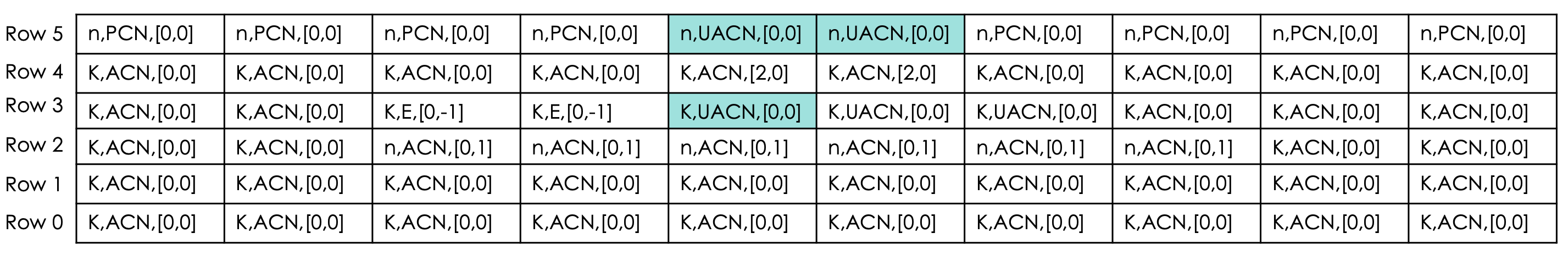}\\
      \mbox{\hspace{8.6cm}(d)}\\\\
  \includegraphics[width=1\linewidth]{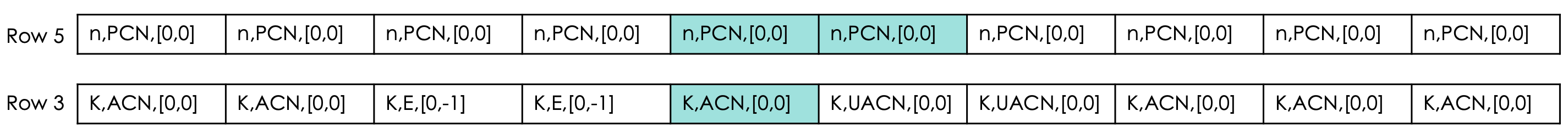}\\
    \mbox{\hspace{8.6cm}(e)}\\\\
  \begin{center}
  \vspace{-5mm}
   \includegraphics[width=0.70\linewidth]{legend.png}\\
  \end{center}
  \vspace{-2mm}
   \caption{Stitch pattern with a combination of Knit, Transfer, Miss and Tuck stitches: (a) Stitch instructions. (b) Simulation of stitch pattern. (c) Topology graph. (d) Corresponding data structure before evaluation.
   (e) Modified entries (highlighted in teal) in data structure after evaluation.}
   \label{fig:pattern1}
 \end{figure*}

\begin{figure*}[p]
  {\centering
  \includegraphics[width=1\linewidth]{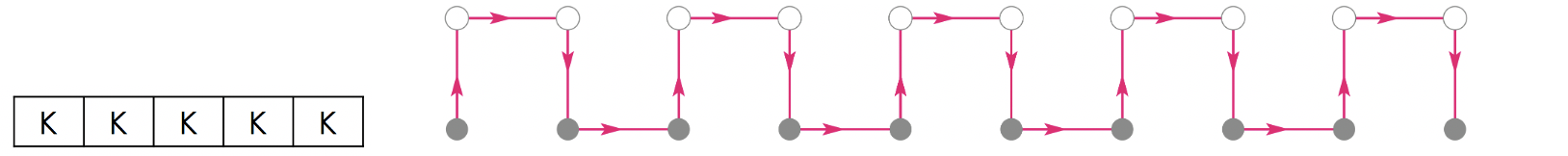}
  \hspace{3mm}
  \includegraphics[width=1\linewidth]{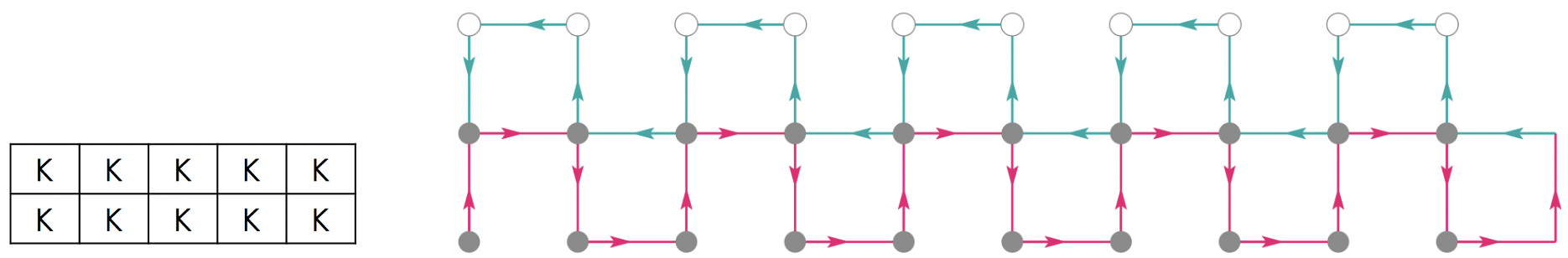}} \\
  \includegraphics[width=1\linewidth]{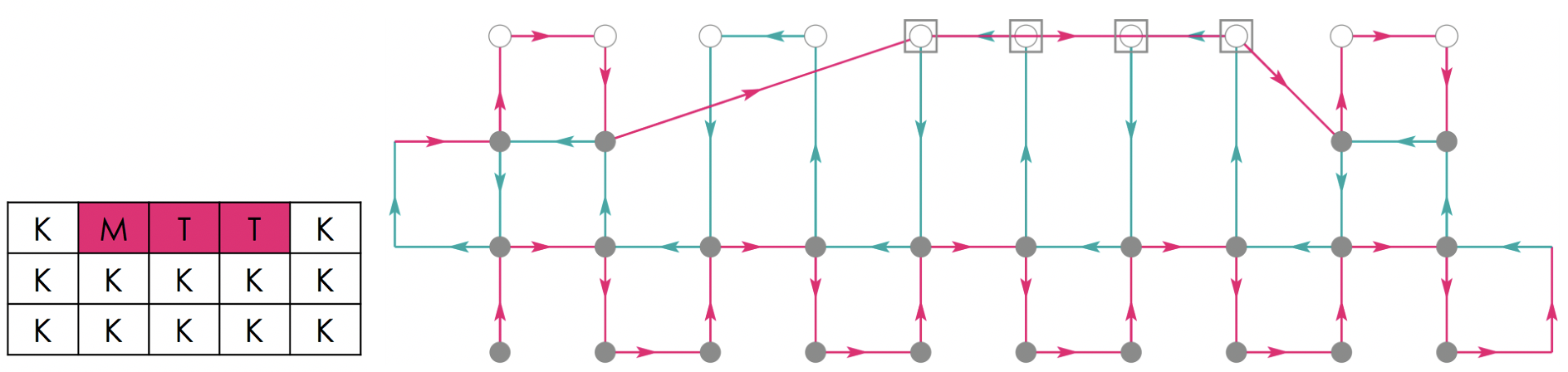}\\
  \includegraphics[width=1\linewidth]{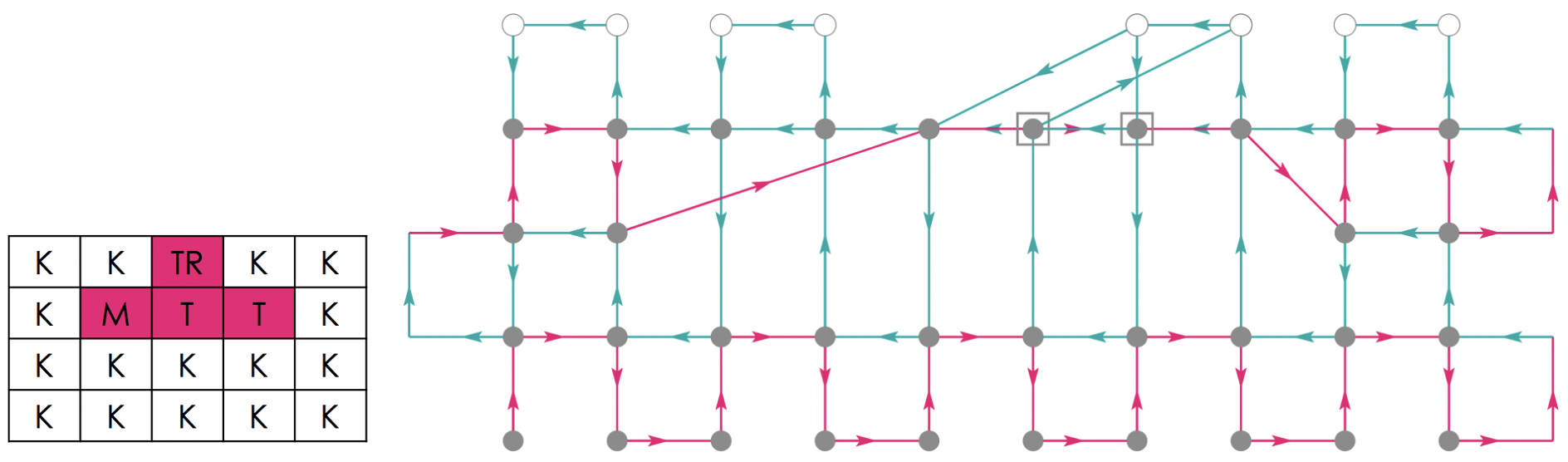}\\
   \includegraphics[width=1\linewidth]{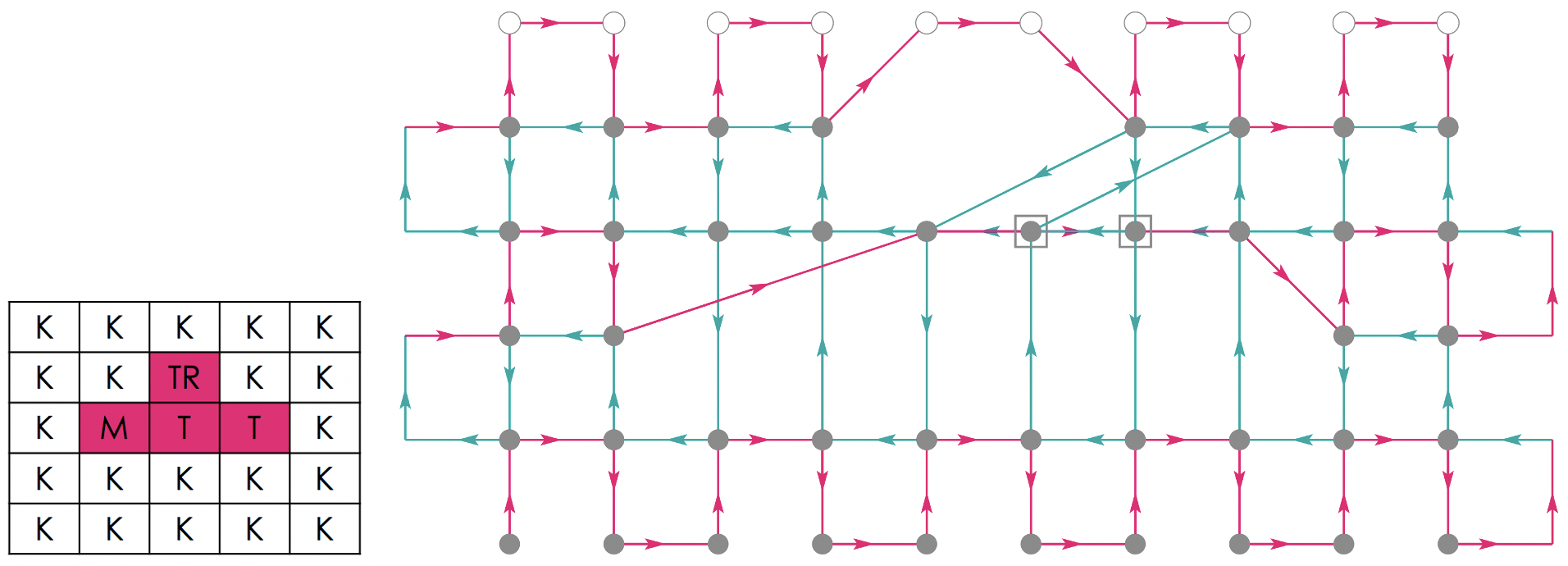}\\
  \begin{center}
   \includegraphics[width=0.70\linewidth]{legend.png}\\
   \end{center}
   \caption{The topology graph produced from a stitch pattern that
   includes Knit, Transfer, Miss and Tuck stitches. The descending
   diagrams present the topological state of the fabric as it is
   knitted row by row.}
   \label{fig:pattern1stepbystep}
\end{figure*}

\begin{figure*}[t]
\begin{tabular}{cc}
\begin{minipage}{0.2\textwidth}
\includegraphics[width=1\textwidth]{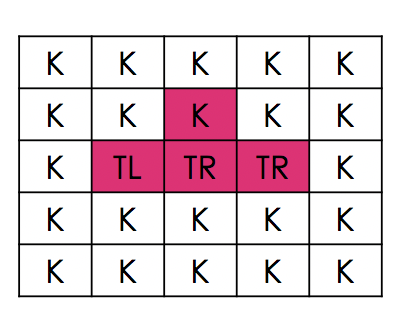} \\  	
  \mbox{\hspace{1.5cm}(a)}\\
 \includegraphics[width=1\textwidth]{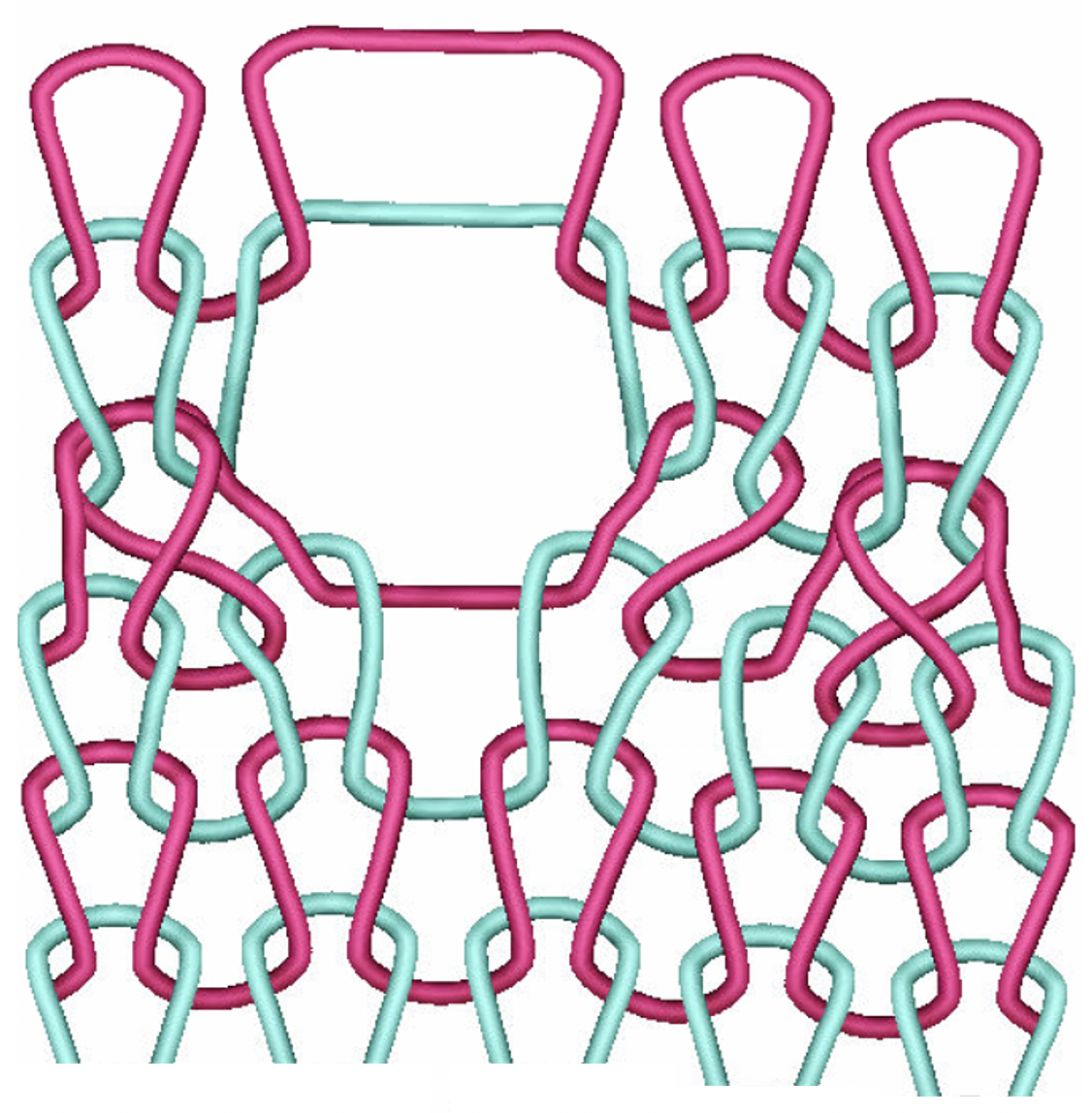} 
 \mbox{\hspace{1.5cm}(b)\hspace{7.9cm}(c)}
\end{minipage}&
\begin{minipage}{0.8\textwidth} 
\vspace{-0.5cm}
\includegraphics[width=0.9\textwidth]{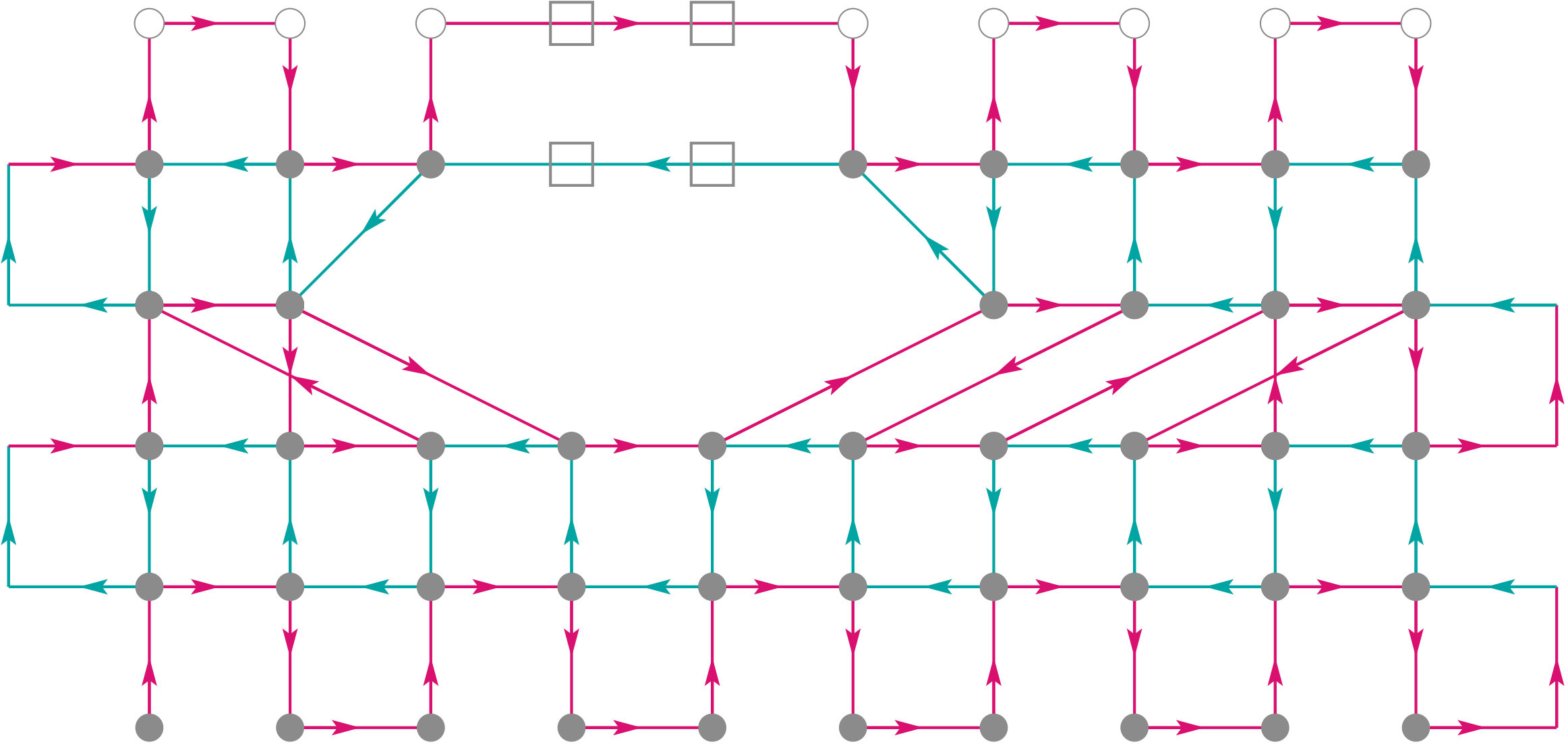} \end{minipage}
\end{tabular}
 \begin{center}
 \includegraphics[width=0.70\linewidth]{legend.png}\\
\end{center}
\vspace{-2mm}
 \caption{Stitch pattern with a combination of Knit and Transfer stitches: (a) Stitch instructions. (b) Simulation of the pattern. (c) Topology graph. The pattern creates a
``hole'' in the fabric and an ascending ``ladder'' of unanchored CNs.}
 \label{fig:pattern2}
 \end{figure*}

\begin{figure*}[p]
\begin{tabular}{cc}
\begin{minipage}{0.2\textwidth}
\includegraphics[width=1\textwidth]{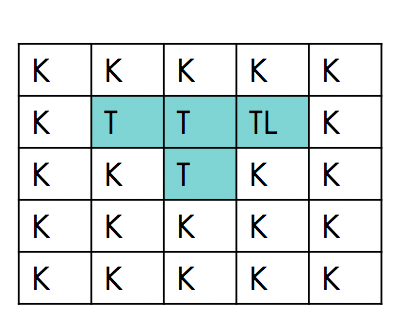} \\  	
  \mbox{\hspace{1.5cm}(a)}\\
 \includegraphics[width=1\textwidth]{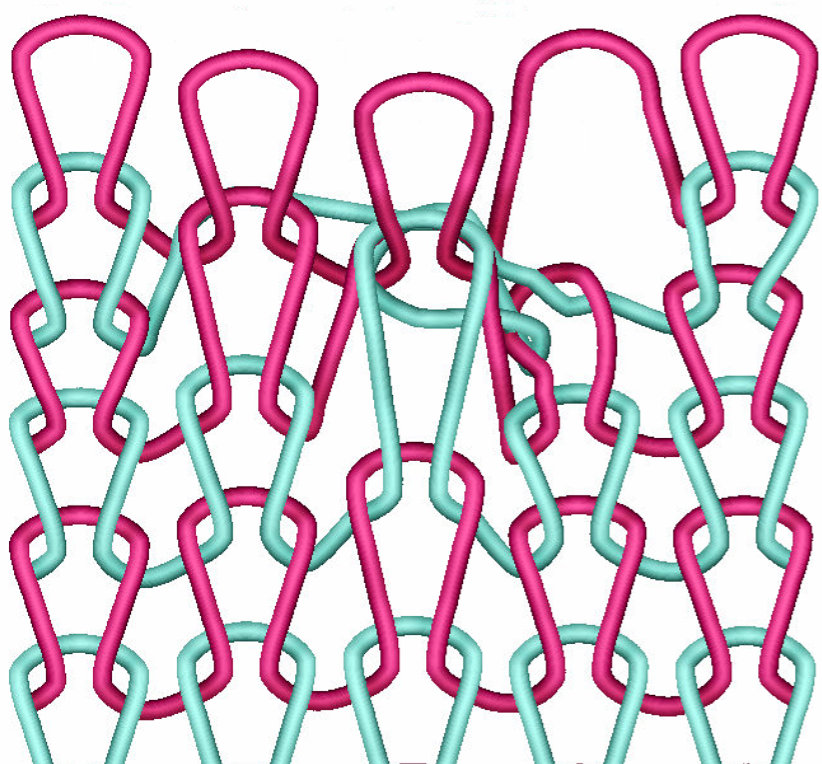} 
 \mbox{\hspace{1.5cm}(b)\hspace{7.9cm}(c)}
\end{minipage}&
\begin{minipage}{0.8\textwidth} 
\vspace{-0.5cm}
\includegraphics[width=0.9\textwidth]{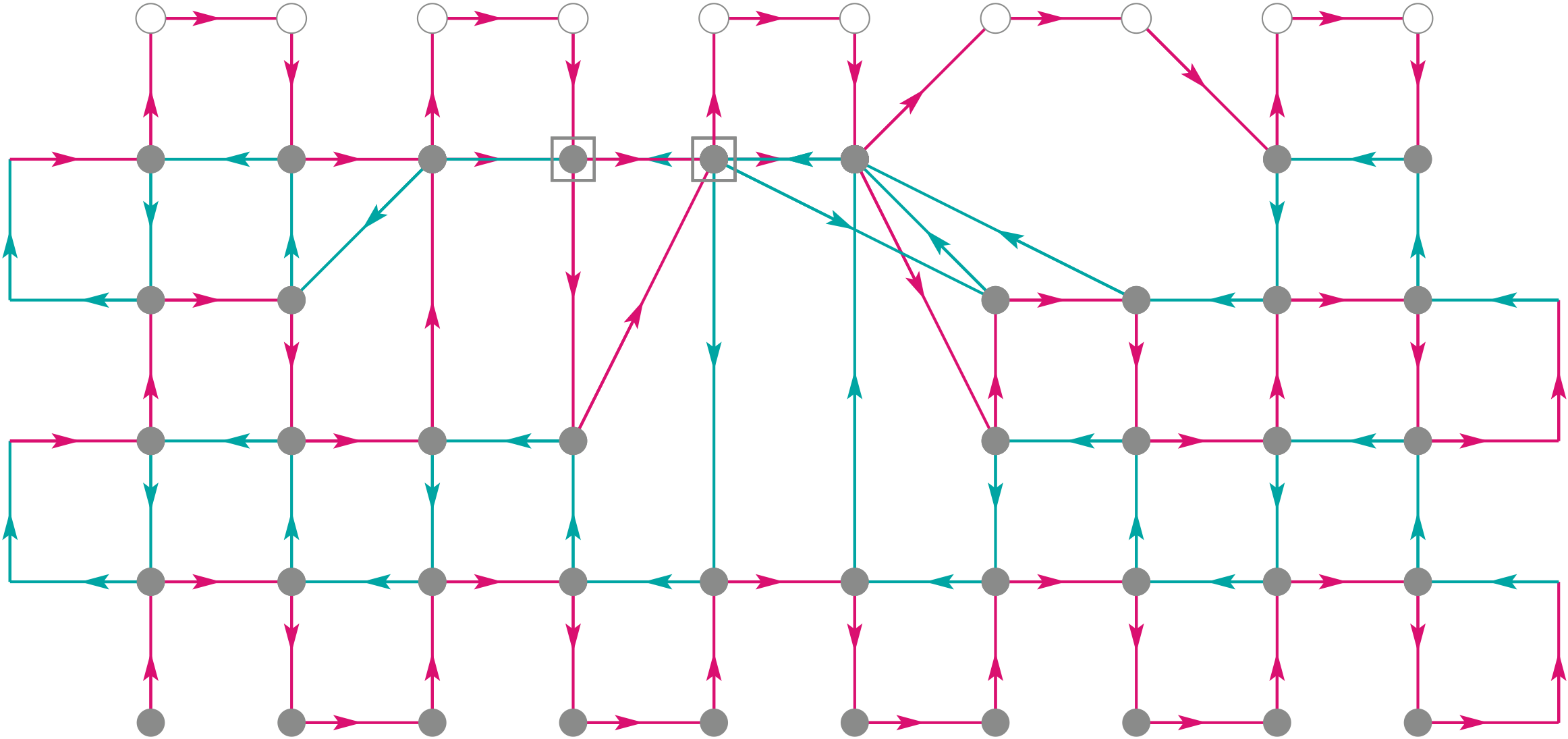} \end{minipage}
\end{tabular}
 \begin{center}
 \includegraphics[width=0.70\linewidth]{legend.png}\\
\end{center}
\vspace{-2mm}
 \caption{Stitch pattern with a combination of Knit, Tuck and Transfer stitches: (a) Stitch instructions. (b) Simulation of the pattern. (c) Topology graph.}
   \label{fig:pattern3}
\vspace{12mm}
\begin{tabular}{cc}
\begin{minipage}{0.2\textwidth}
\includegraphics[width=1\textwidth]{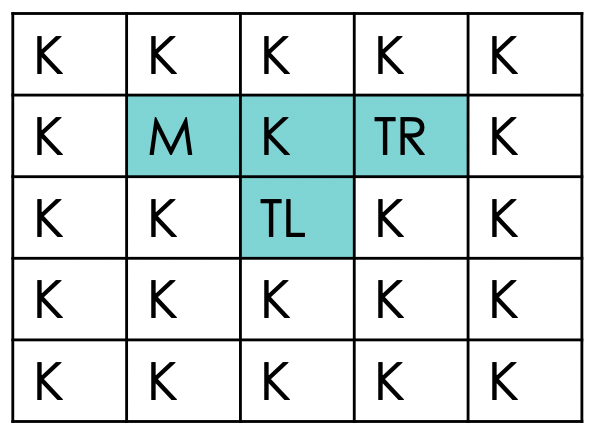} \\  	
  \mbox{\hspace{1.5cm}(a)}\\
 \includegraphics[width=1\textwidth]{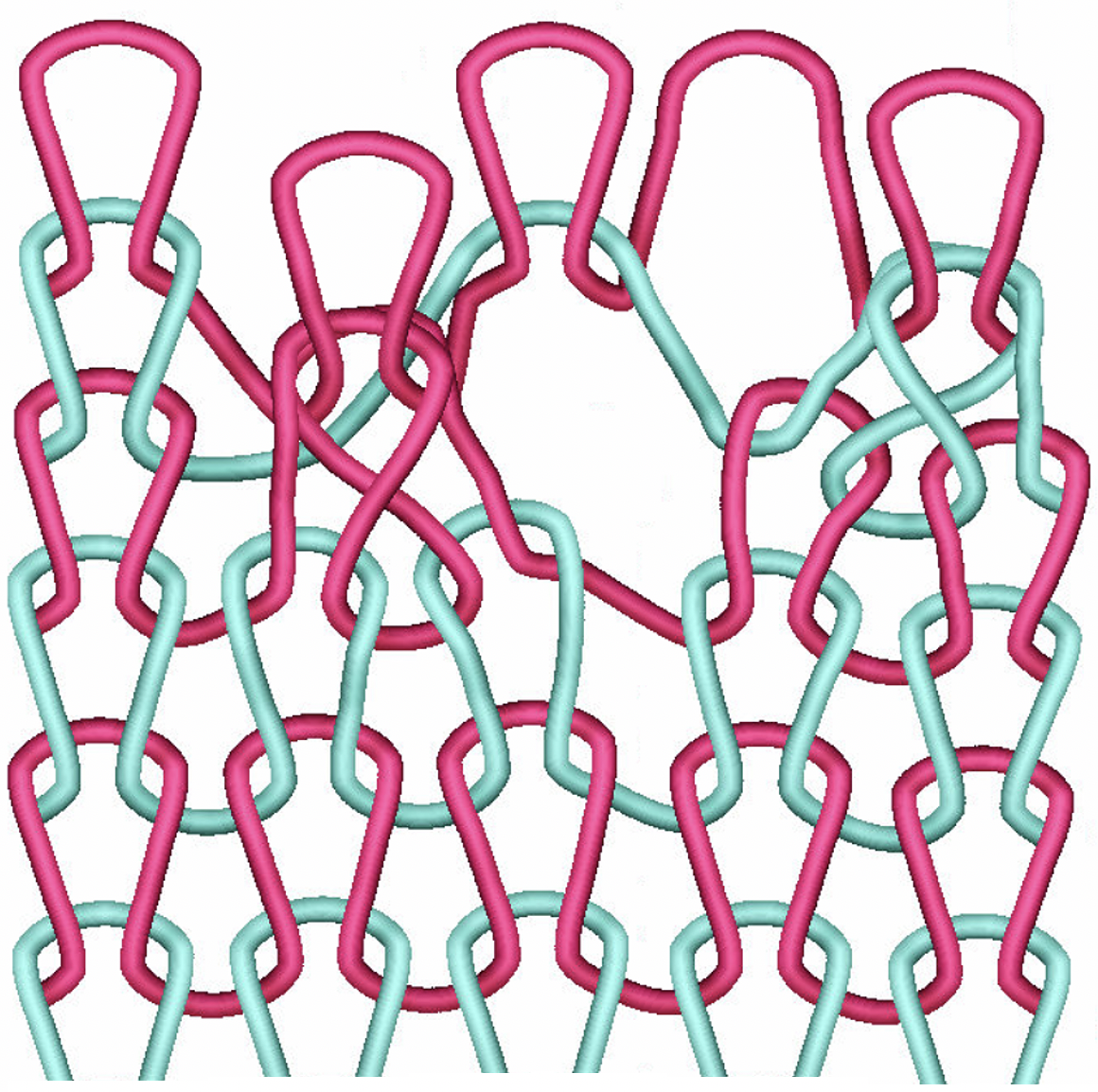} 
 \mbox{\hspace{1.5cm}(b)\hspace{7.9cm}(c)}
\end{minipage}&
\begin{minipage}{0.8\textwidth} 
\vspace{-0.5cm}
\includegraphics[width=0.9\textwidth]{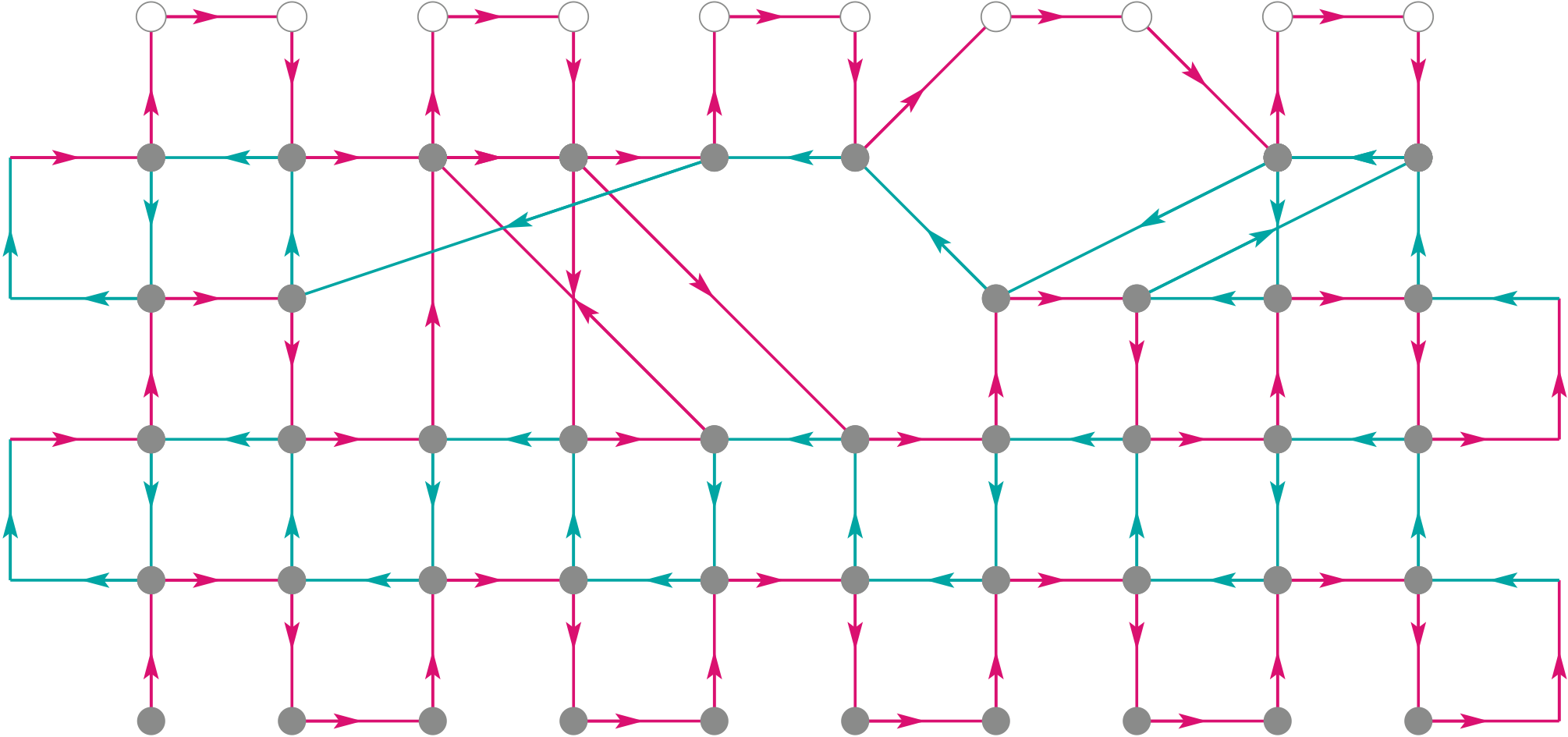} \end{minipage}
\end{tabular}
 \begin{center}
 \includegraphics[width=0.70\linewidth]{legend.png}\\
\end{center}
\vspace{-2mm}
 \caption{Stitch pattern with a combination of Knit, Miss and Transfer stitches: (a) Stitch instructions. (b) Simulation of the pattern. (c) Topology graph.}
   \label{fig:pattern8}
 \end{figure*}

\begin{figure*}[p]
\begin{tabular}{cc}
\begin{minipage}{0.2\textwidth}
\includegraphics[width=1\textwidth]{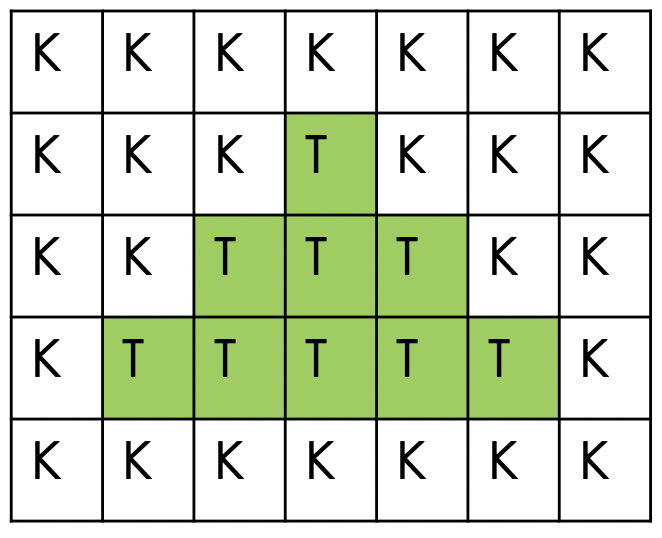} \\  	
  \mbox{\hspace{1.5cm}(a)}\\
 \includegraphics[width=1\textwidth]{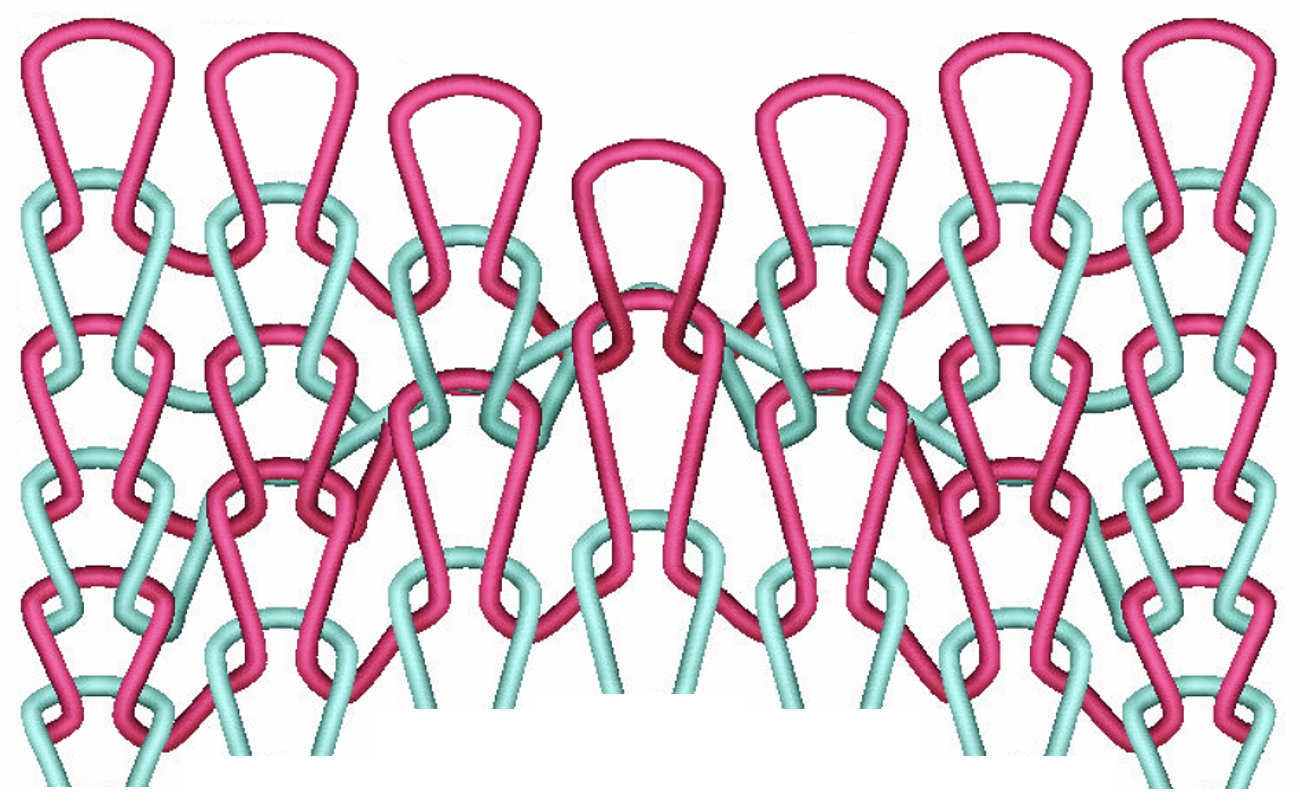} 
 \mbox{\hspace{1.5cm}(b)\hspace{7.9cm}(c)}
\end{minipage}&
\begin{minipage}{0.8\textwidth} 
\vspace{-0.5cm}
\includegraphics[width=0.9\textwidth]{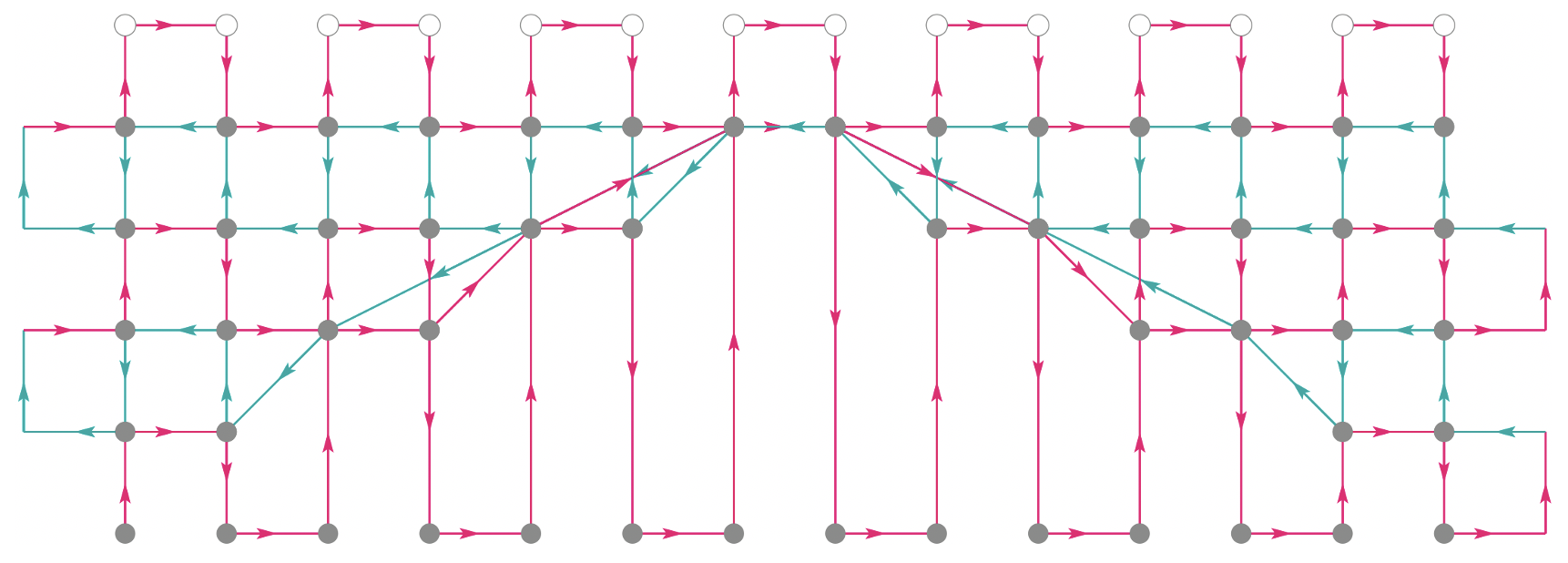} \end{minipage}
\end{tabular}
 \begin{center}
 \includegraphics[width=0.70\linewidth]{legend.png}\\
\end{center}
\vspace{-2mm}
  \caption{Pattern with a 3-level combination of Knit and Tuck stitches: (a) Stitch instructions. (b) Pattern simulation.  (c) Topology graph.}
 \label{fig:pattern4}
\vspace{8mm}
\begin{tabular}{cc}
\begin{minipage}{0.2\textwidth}
\includegraphics[width=1\textwidth]{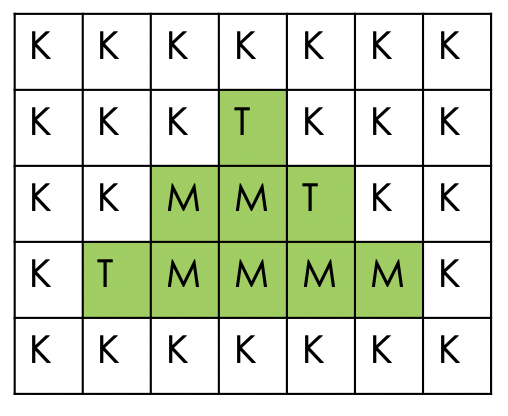} \\  	
  \mbox{\hspace{1.5cm}(a)}\\
 \includegraphics[width=1\textwidth]{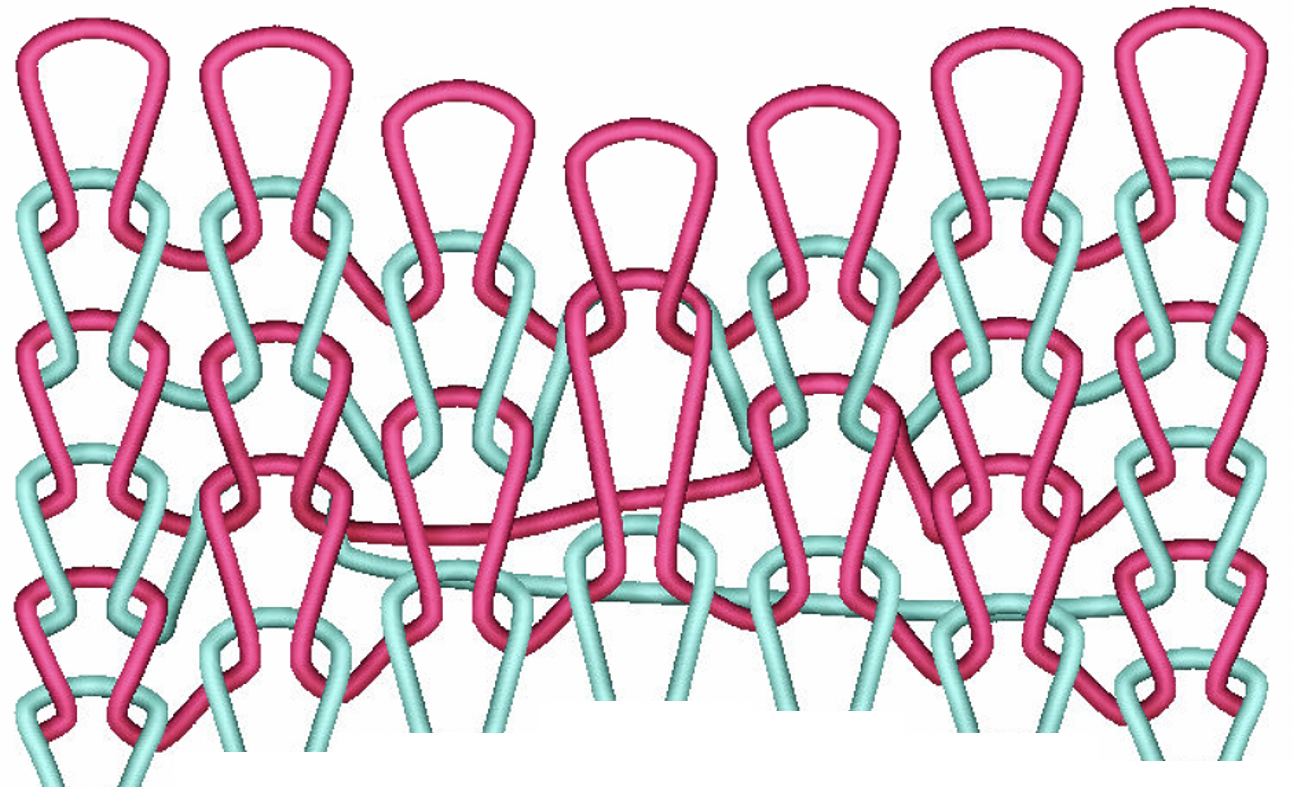} 
 \mbox{\hspace{1.5cm}(b)\hspace{7.9cm}(c)}
\end{minipage}&
\begin{minipage}{0.8\textwidth} 
\vspace{-0.5cm}
\includegraphics[width=0.9\textwidth]{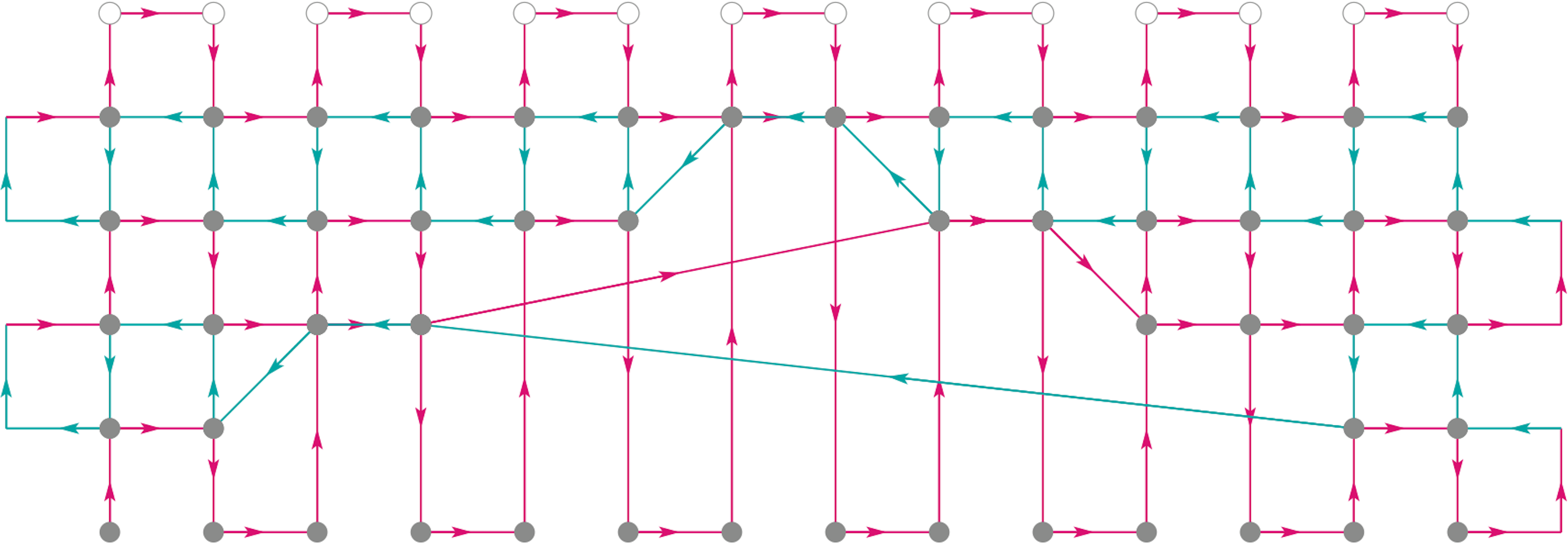}
\end{minipage}
\end{tabular}
 \begin{center}
 \includegraphics[width=0.70\linewidth]{legend.png}\\
\end{center}
\vspace{-2mm}
  \caption{Pattern with a 3-level combination of Knit, Miss and Tuck stitches: (a) Stitch instructions. (b) Pattern simulation.  (c) Topology graph.}
   \label{fig:pattern7}
\vspace{8mm}
\begin{tabular}{cc}
\begin{minipage}{0.2\textwidth}
\includegraphics[width=1\textwidth]{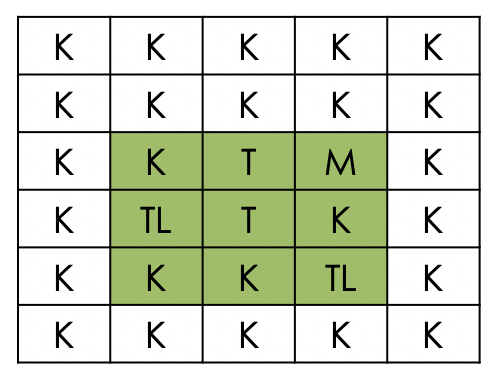} \\  	
  \mbox{\hspace{1.5cm}(a)}\\
 \includegraphics[width=1\textwidth]{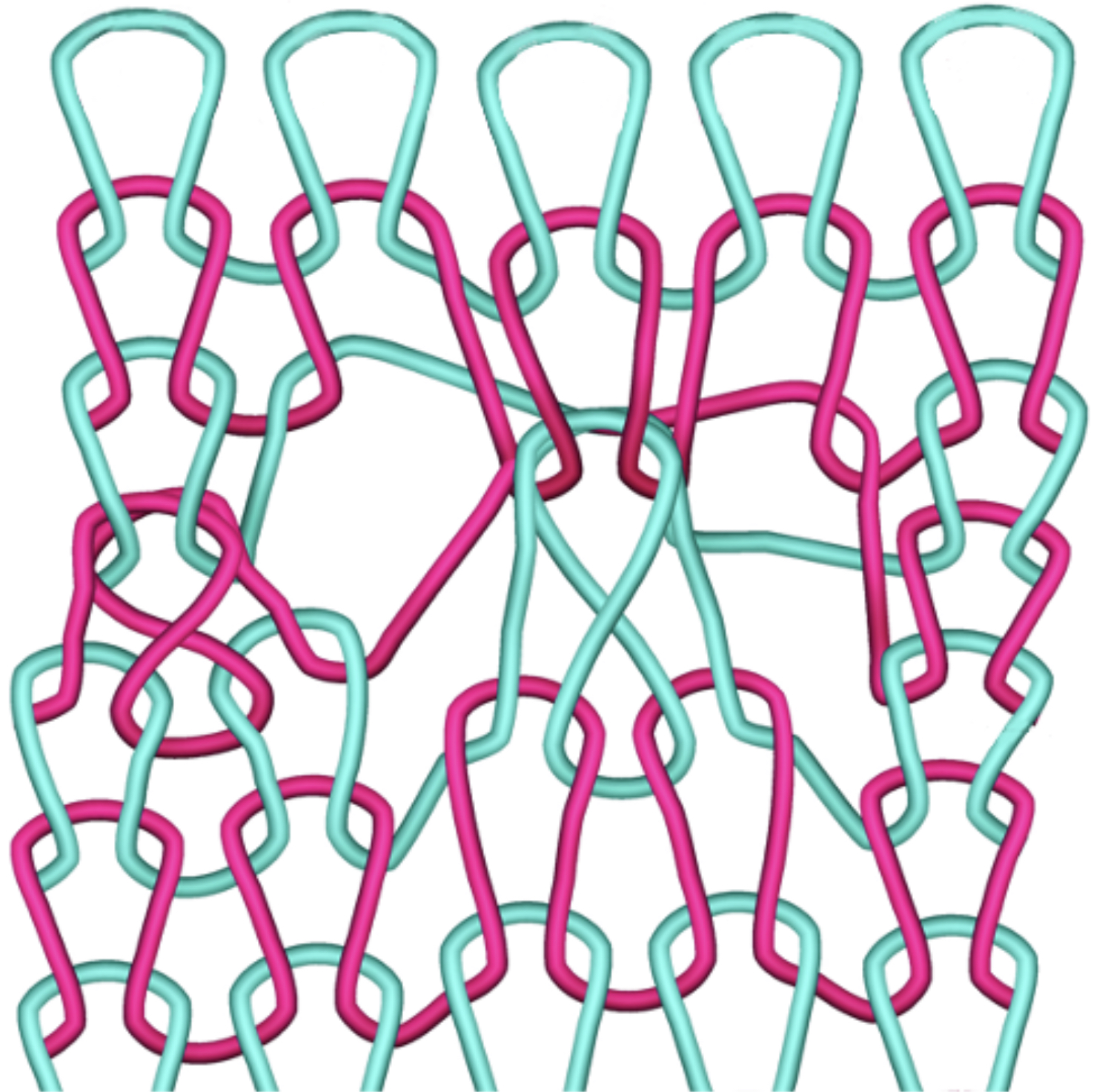} 
 \mbox{\hspace{1.5cm}(b)\hspace{7.9cm}(c)}
\end{minipage}&
\begin{minipage}{0.8\textwidth} 
\vspace{-0.5cm}
\includegraphics[width=0.9\textwidth]{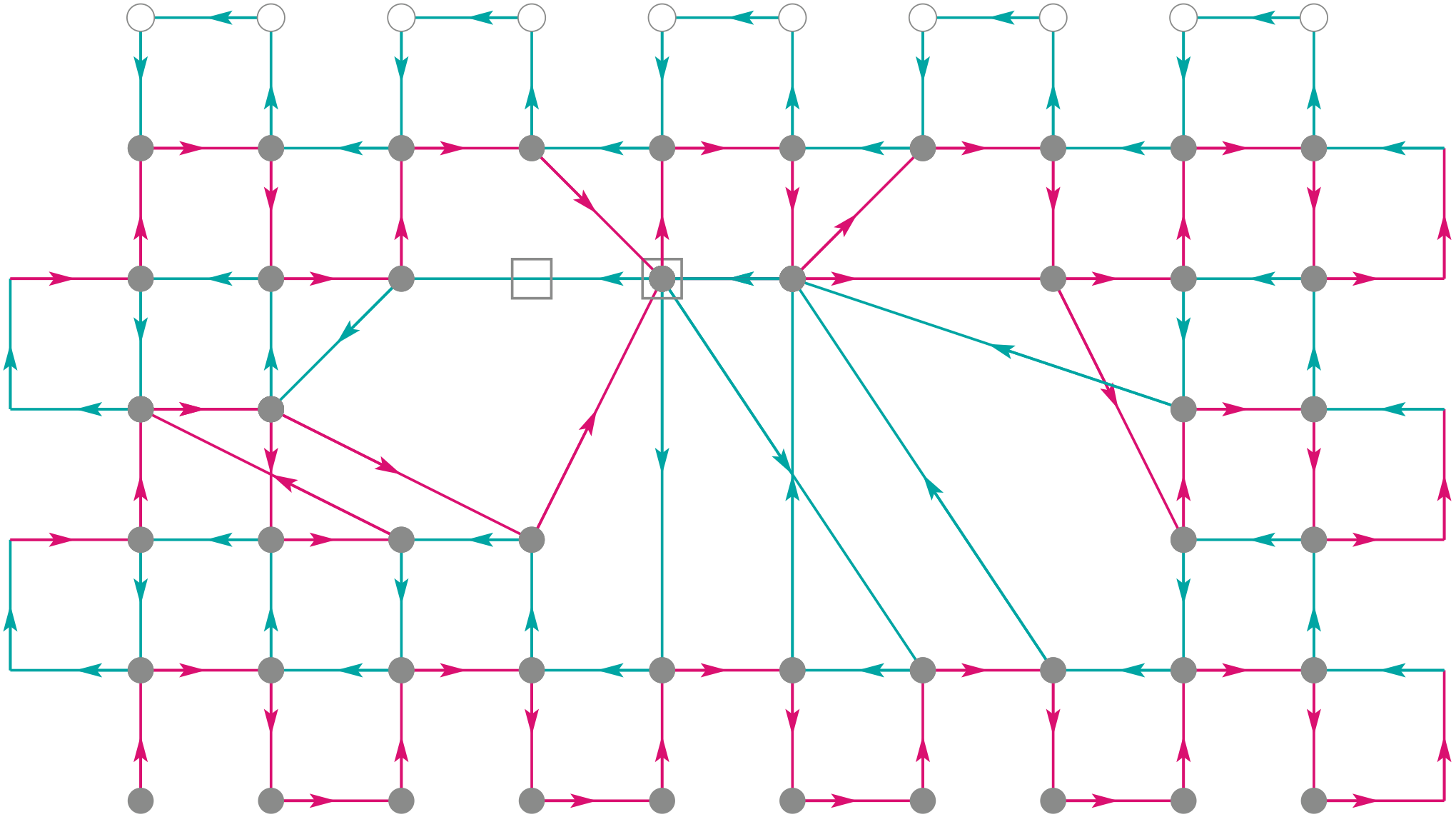} \end{minipage}
\end{tabular}
 \begin{center}
 \includegraphics[width=0.70\linewidth]{legend.png}\\
\end{center}
\vspace{-2mm}
 \caption{Pattern with a combination of Knit, Transfer, Tuck and Miss stitches: (a) Stitch instructions. (b) Pattern simulation. (c) Topology graph.}
 \label{fig:patternVertical}
 \end{figure*}

Figures \ref{fig:pattern1}, \ref{fig:pattern1stepbystep} and
\ref{fig:pattern2} use the
magenta template and
Figures \ref{fig:pattern3} and \ref{fig:pattern8} use the teal
template and produce complex yarn topologies from 
relatively simple combinations of Knit, Transfer, Tuck and Miss stitches.
Additionally a number of other patterns were tested
that included empty stitches around the pattern border to produce increases
and decreases in the fabric, as well as a set of (light green) 3-level
patterns, such as the ones in Figures \ref{fig:pattern4},
\ref{fig:pattern7} and \ref{fig:patternVertical}. These additional tests
also produced correct topology graphs of the resulting fabrics.

\begin{figure*}[p]
\begin{tabular}{cc}
\begin{minipage}{0.2\textwidth}
\includegraphics[width=1\textwidth]{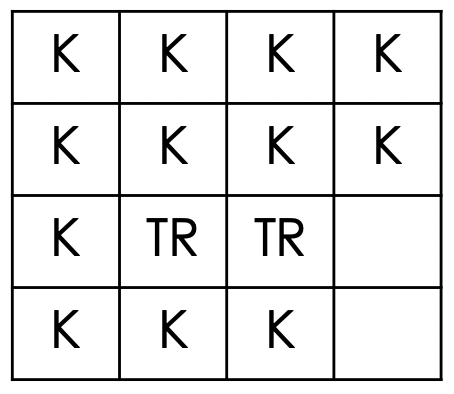} \\  	
  \mbox{\hspace{1.5cm}(a)}\\
 \includegraphics[width=1\textwidth]{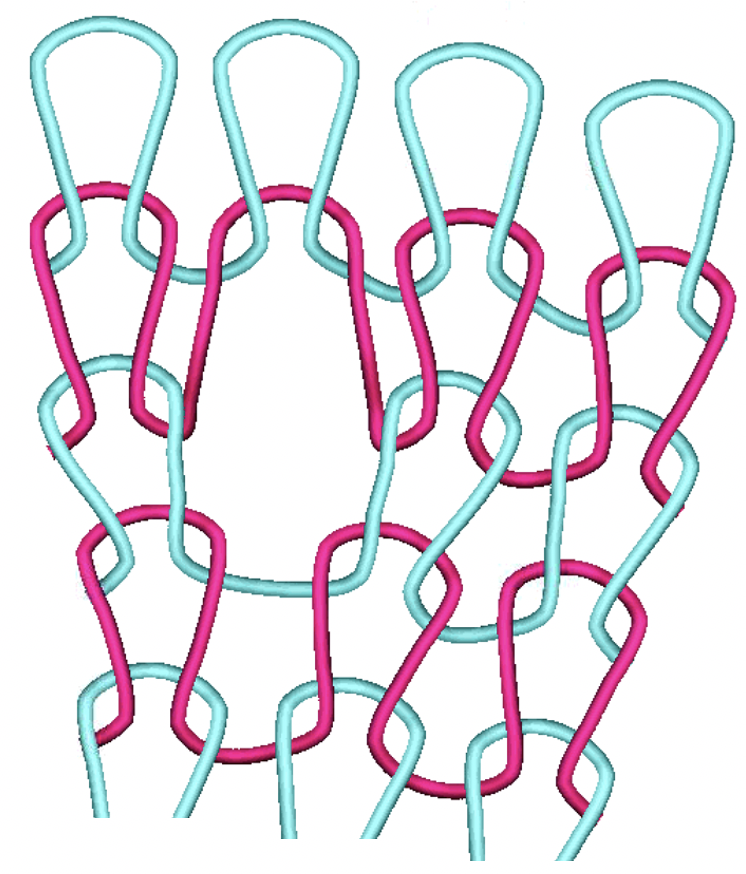} 
 \mbox{\hspace{1.5cm}(b)\hspace{7.9cm}(c)}
\end{minipage}&
\begin{minipage}{0.8\textwidth} 
\vspace{-1cm}
\includegraphics[width=1\textwidth]{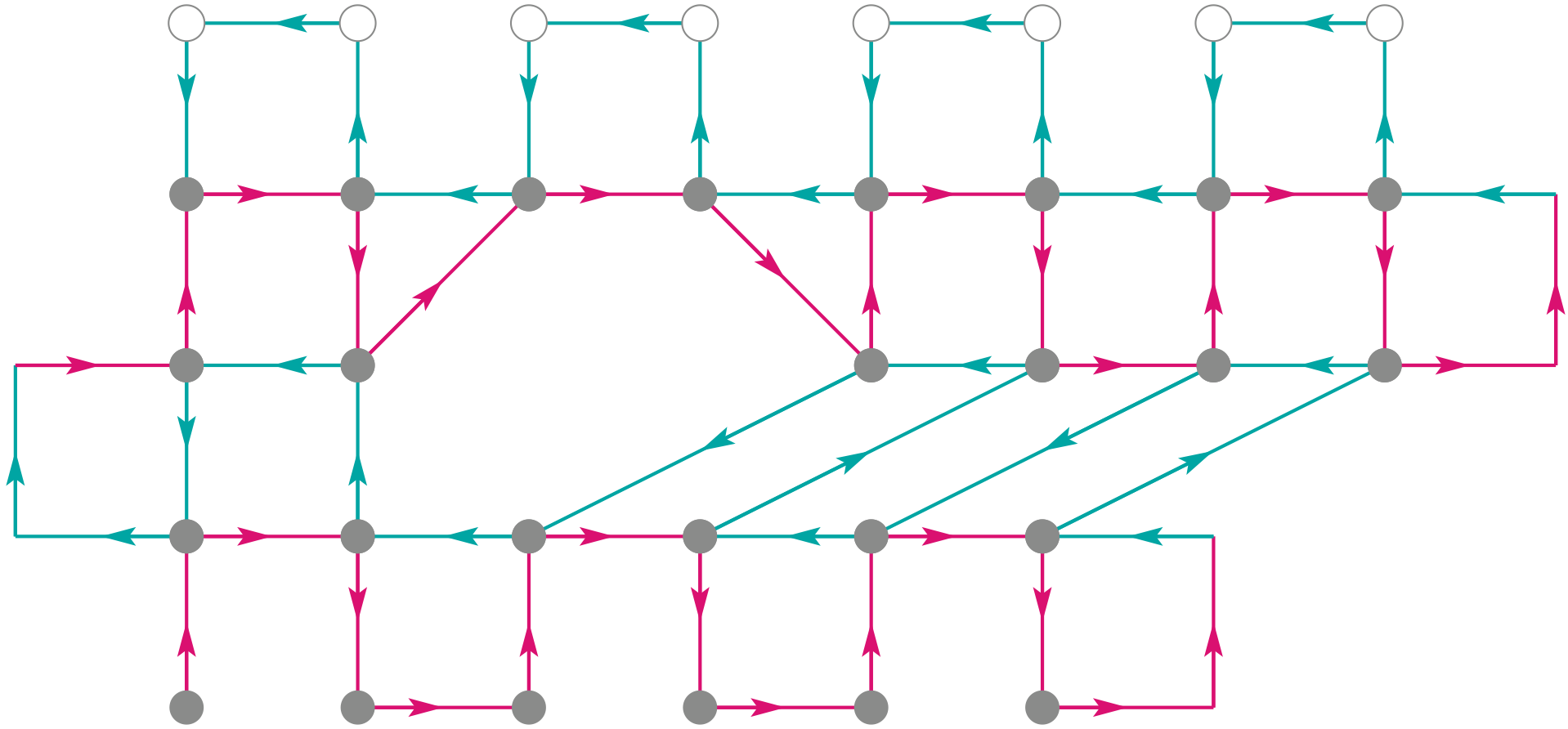} \end{minipage}
\end{tabular}
 \begin{center}
 \includegraphics[width=0.70\linewidth]{legend.png}\\
\end{center}
\vspace{-2mm}
 \caption{An increase of the fabric using Transfer and Empty stitches: (a) Stitch instructions. (b) Simulation of the pattern. (c) Topology graph.}
   \label{fig:increase}
\vspace{12mm}
\begin{tabular}{cc}
\begin{minipage}{0.2\textwidth}
\includegraphics[width=1\textwidth]{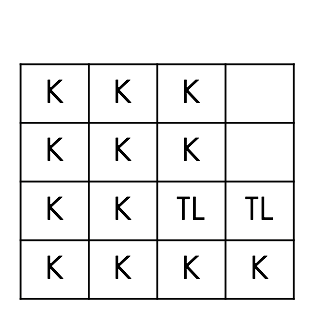} \\  	
  \mbox{\hspace{1.5cm}(a)}\\
 \includegraphics[width=1\textwidth]{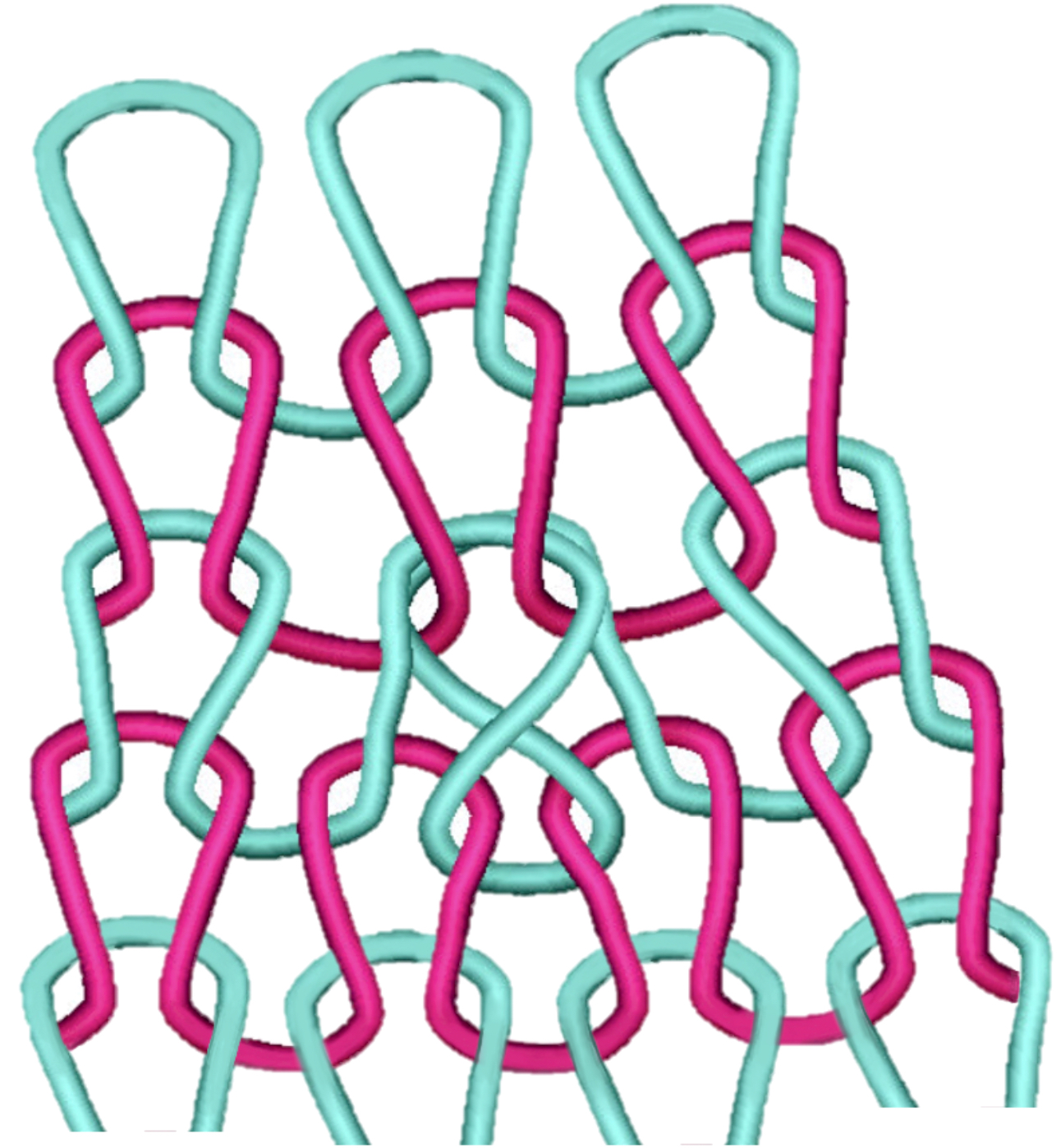} 
 \mbox{\hspace{1.5cm}(b)\hspace{7.9cm}(c)}
\end{minipage}&
\begin{minipage}{0.8\textwidth} 
\vspace{-1cm}
\includegraphics[width=1\textwidth]{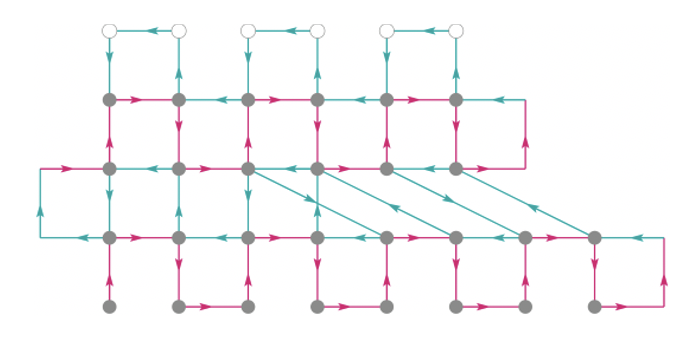} \end{minipage}
\end{tabular}
 \begin{center}
 \includegraphics[width=0.70\linewidth]{legend.png}\\
\end{center}
\vspace{-2mm}
 \caption{A decrease of the fabric using Transfer and Empty stitches: (a) Stitch instructions. (b) Simulation of the pattern. (c) Topology graph.}
   \label{fig:decrease}
 \end{figure*}

\begin{figure*}[t]
\begin{tabular}{cc}
\includegraphics[width=0.3\textwidth]{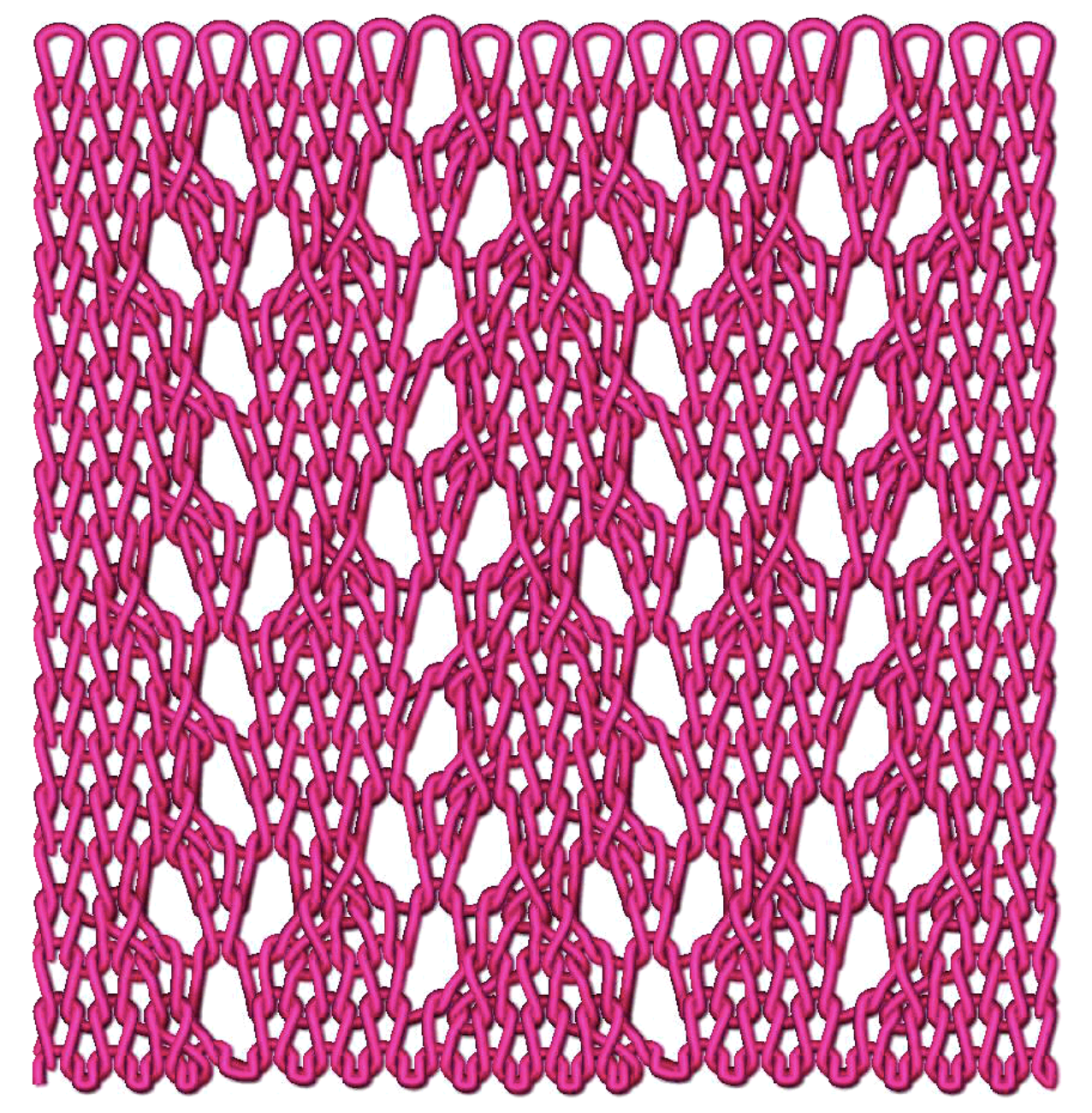} 
\includegraphics[width=0.7\textwidth]{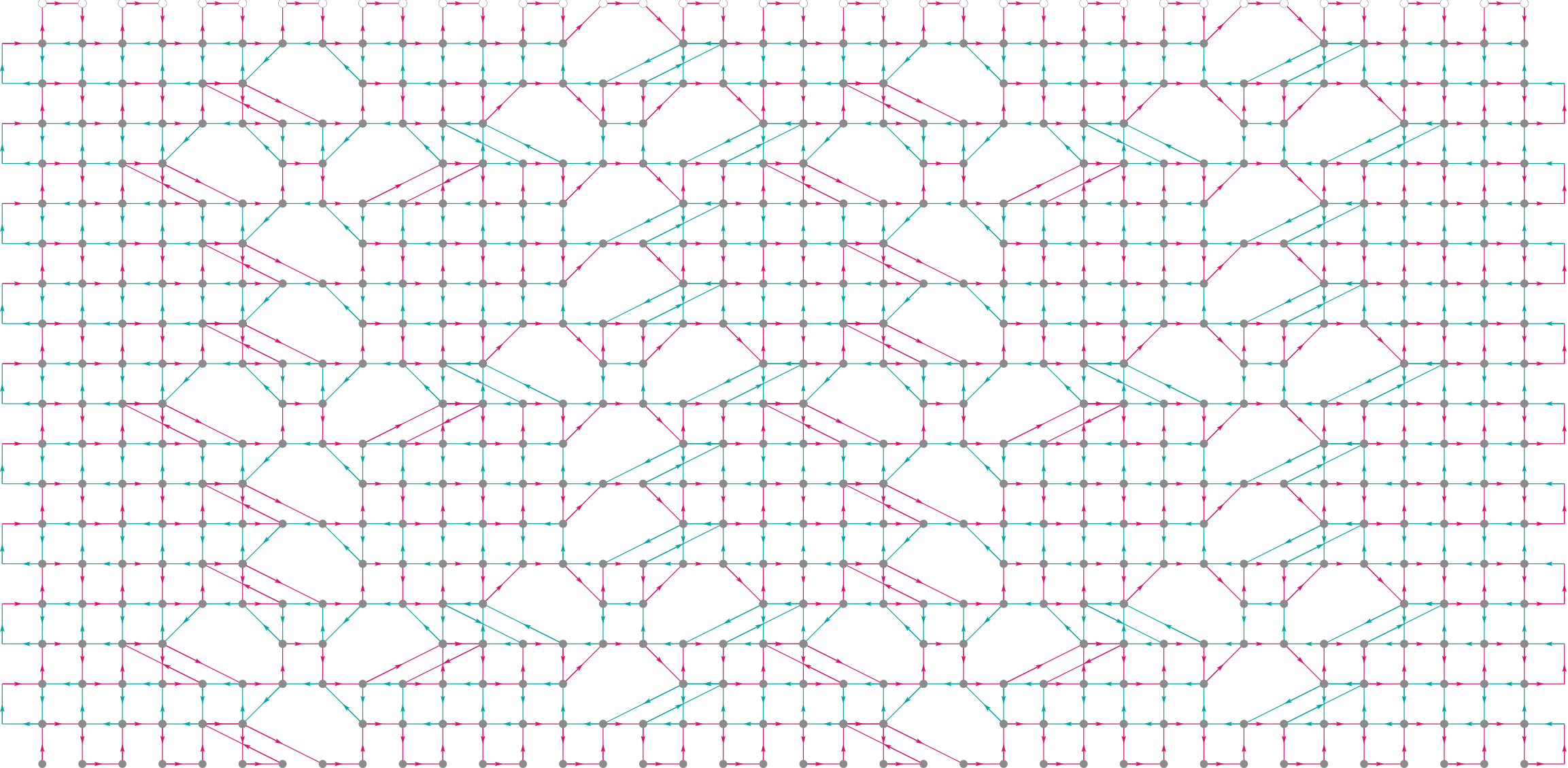} \\
\mbox{\hspace{-3.5cm}(a)\hspace{8.5cm} (b)}\\ \\
\mbox{\includegraphics[width=0.3\textwidth]{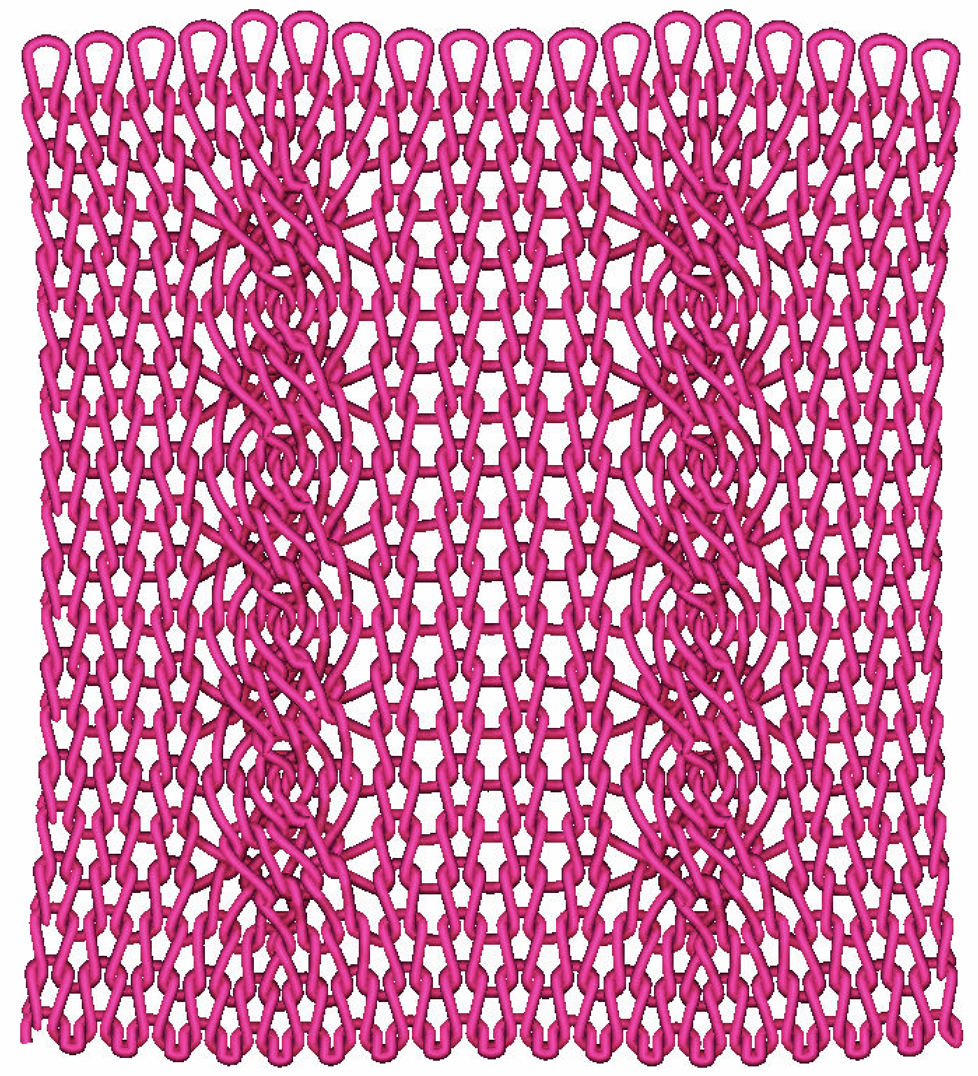} 
\includegraphics[width=0.7\textwidth]{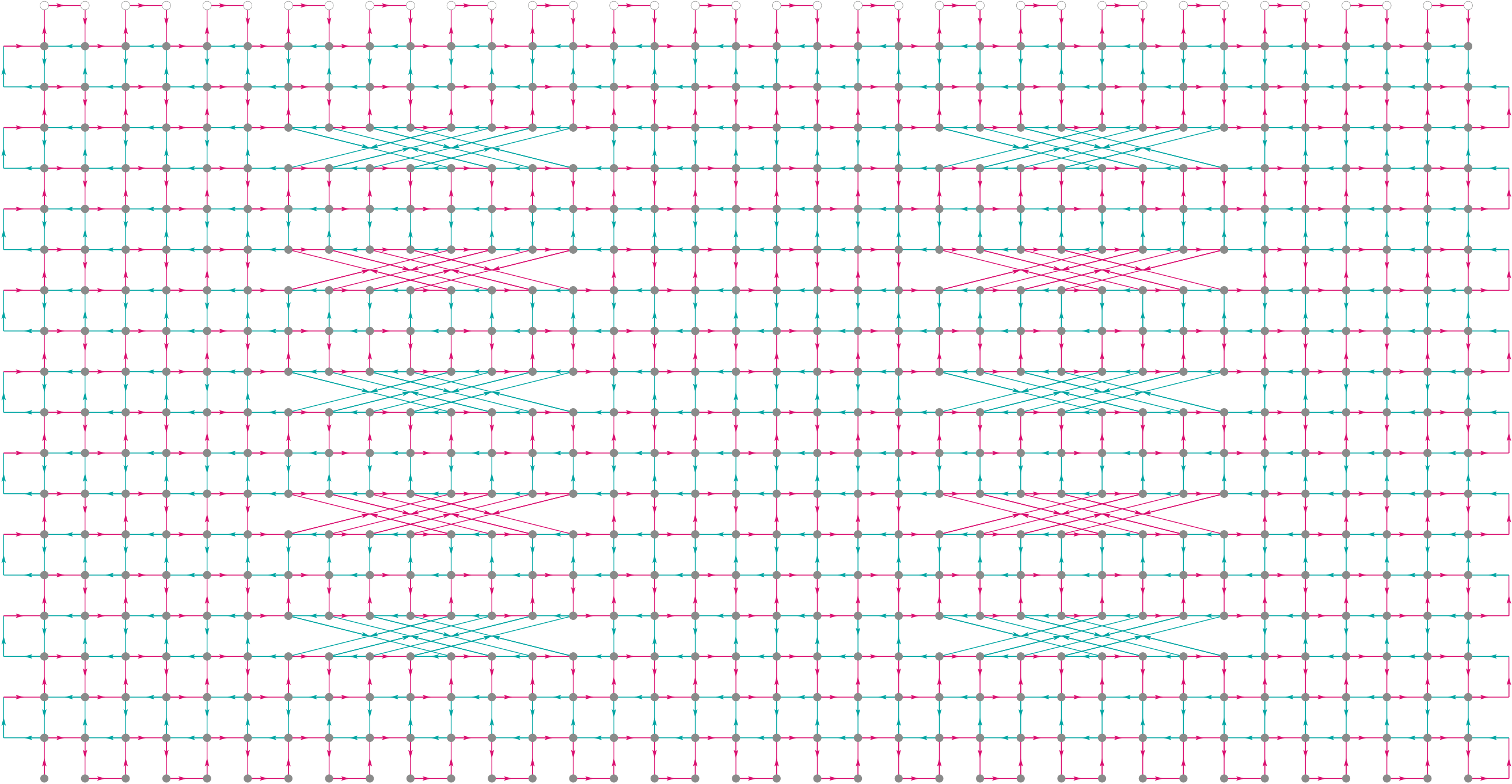}} \\
\mbox{\hspace{-3.5cm}(c)\hspace{8.5cm} (d)}\\
\end{tabular}
 \caption{Samples of larger stitch patterns and their corresponding topology graphs: (a) Simulation of a lace pattern. (b) Topology graph of the lace pattern. (c) Simulation for a cable pattern. (d) Topology graph of the cable pattern.}
\label{fig:bigPatches}
\end{figure*}

\begin{figure*}[p]
\begin{tabular}{cc}
\begin{minipage}{0.2\textwidth}
\includegraphics[width=1\textwidth]{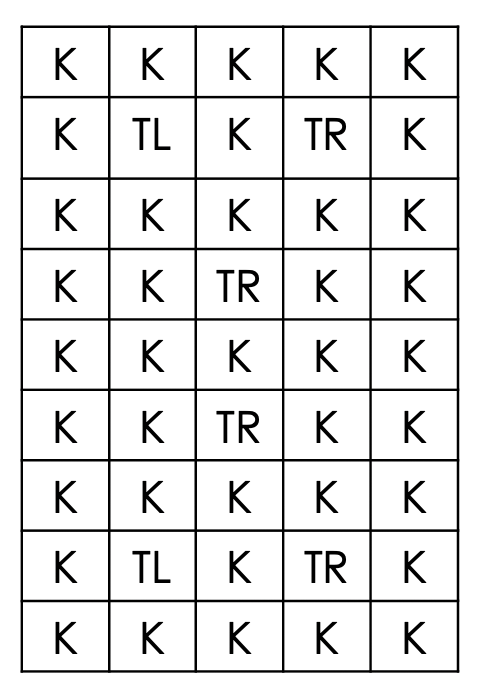} \\  	
  \mbox{\hspace{1.5cm}(a)}\\
 \includegraphics[width=1\textwidth]{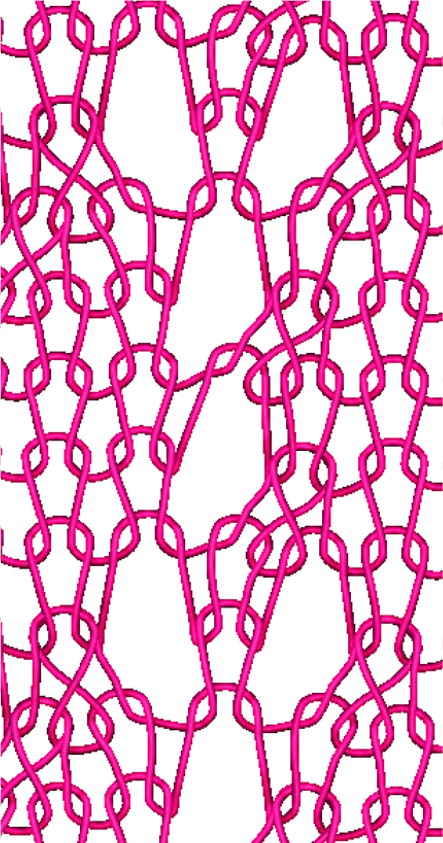} 
 \mbox{\hspace{1.5cm}(b)\hspace{7.9cm}(c)}
\end{minipage}&
\begin{minipage}{0.8\textwidth} 
\vspace{-0.5cm}
\includegraphics[width=0.9\textwidth]{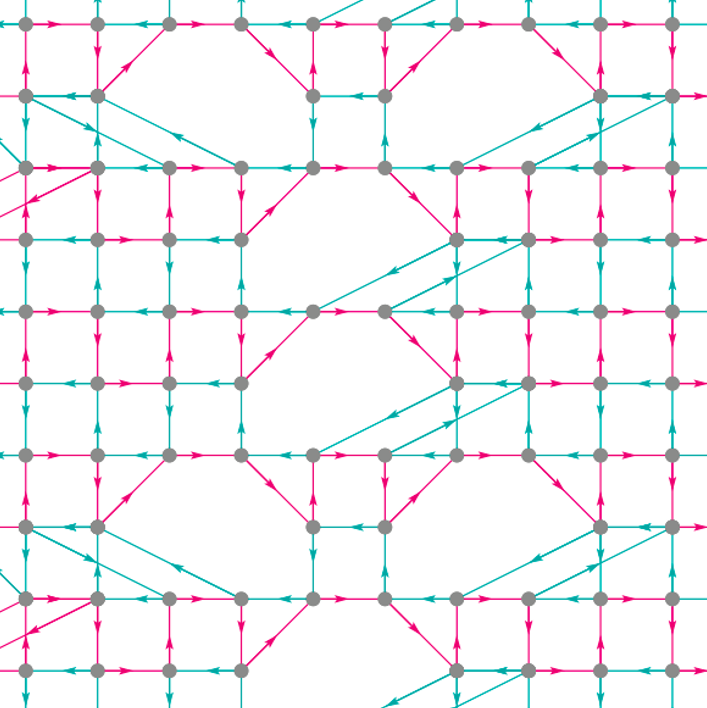}
\end{minipage}
\end{tabular}
 \begin{center}
 \includegraphics[width=0.70\linewidth]{legend.png}\\
\end{center}
\vspace{-2mm}
  \caption{A closer view of the lace pattern: (a) Stitch instructions. (b) Simulation of the pattern.  (c) Topology graph.}
   \label{fig:007AZoom}
\vspace{10mm}   
\begin{tabular}{cc}
\begin{minipage}{0.2\textwidth}
\includegraphics[width=1\textwidth]{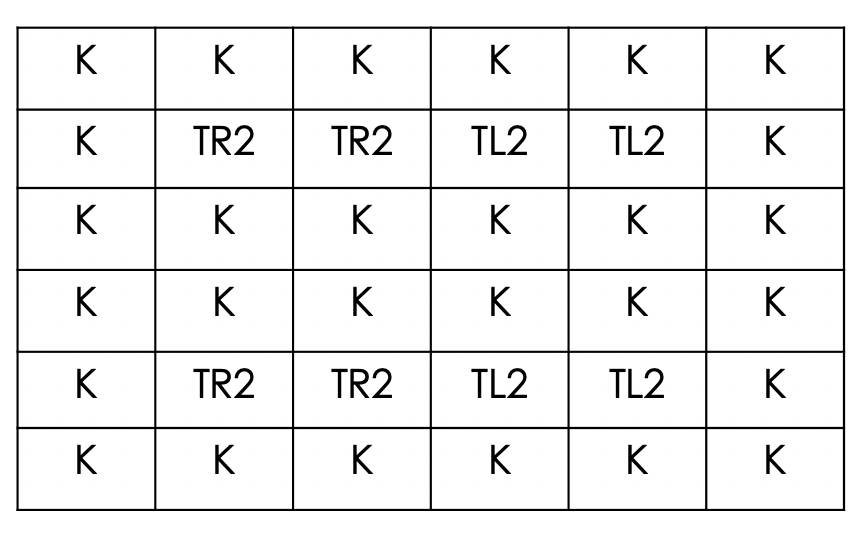} \\  	
  \mbox{\hspace{1.5cm}(a)}\\
 \includegraphics[width=1\textwidth]{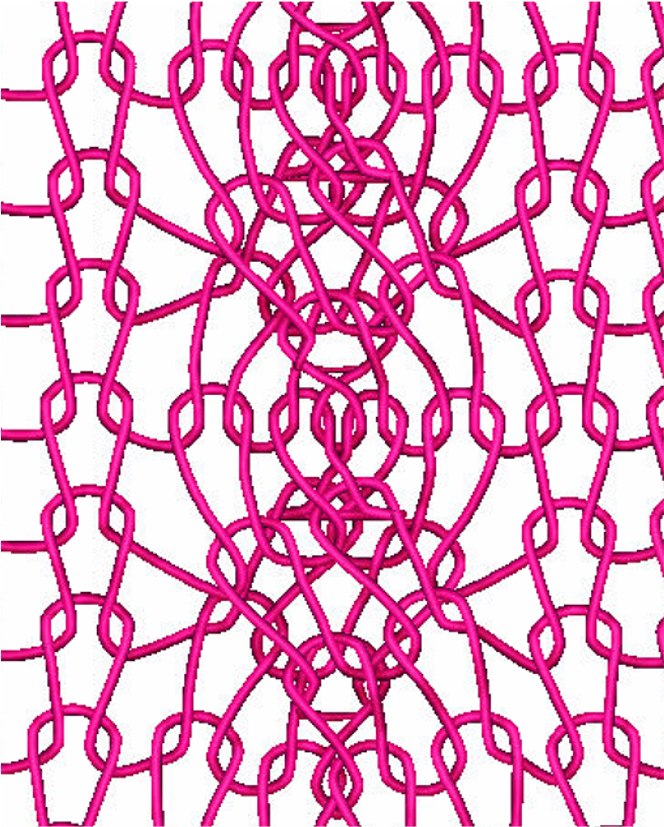} 
 \mbox{\hspace{1.5cm}(b)\hspace{7.9cm}(c)}
\end{minipage}&
\begin{minipage}{0.8\textwidth} 
\vspace{-0.5cm}
\includegraphics[width=0.9\textwidth]{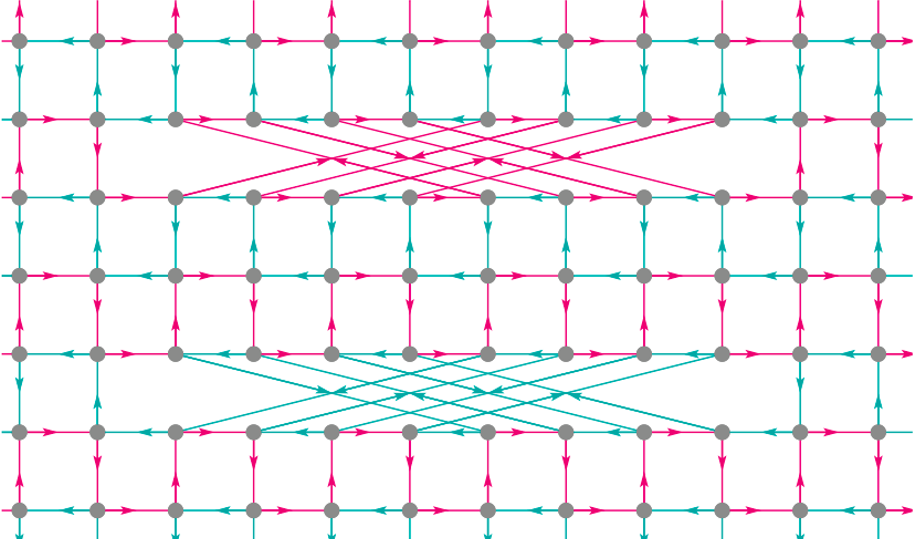}
\end{minipage}
\end{tabular}
 \begin{center}
 \includegraphics[width=0.70\linewidth]{legend.png}\\
\end{center}
\vspace{-2mm}
  \caption{A closer view of the cable pattern: (a) Stitch instructions. (b) Simulation of the pattern.  (c) Topology graph.}
   \label{fig:cableZoom}
\end{figure*}

Figure \ref{fig:pattern1} presents an example stitch pattern that includes
Knit, Transfer, Tuck and Miss stitches. The (a) figure is the stitch
pattern for this swatch. The (b) figure is graphical output from the
Shima Seiki SDS-One APEX3 KnitDesign system for the stitch
pattern. \footnote{The (b) portions of Figures \ref{fig:pattern1} to \ref{fig:cableZoom} were produced with the Shima Seiki SDS-One APEX3 KnitDesign system.}. 
The (c) figure is the topology graph of the
resulting fabric automatically generated by  our algorithms.

Figure \ref{fig:pattern1}(d) presents the initial populated data structure
and  \ref{fig:pattern1}(e) presents the state of the data structure after
the updates executed in the FOLLOW\_THE\_YARN algorithm.
Note that UACNs have been updated to PCNs in the top row and an ACN in
the third row by Algorithm \ref{alg:follow_yarn}.
Cells $(4,4)$ and $(5,4)$ in the data structure have a non-zero
${\Delta}i$, which means that these two CNs have been shifted to the right
as can be visually seen in the topology graph. Two interesting locations in
the diagram are locations $(5,3)$ and $(6,3)$, which in addition to the
ACNs shown with grey circles have at least one UACN represented by the
outlined square. The magenta  and teal directed edges are the yarns that
connect the CNs, with the arrows showing the direction the yarn was carried
during the knitting process. At least two yarns flow into and out of each
ACN and form the yarn contact at that location.

Figure \ref{fig:pattern1stepbystep}
presents a row-by-row visualization
of the Figure \ref{fig:pattern1} swatch. Given an $n \times m$
stitch pattern, we can evaluate and visualize the topology of the pattern
starting with the first row, followed by the first 2 rows, and so on.
This visualization allows one to see the intermediate states of the
data structure and the topology graph as the fabric is being knitted.

Figure \ref{fig:pattern2}
presents a stitch pattern that creates a
``hole'' in the swatch with transfer stitches that shift stitch
loops away from each other. As the head nodes of the stitches are shifted
to the left and right, the knit stitches above them are not
properly formed and produce a ``ladder'' of unanchored CNs.
These examples demonstrate that our approach is capable of generating
unusual and possibly challenging yarn topologies by representing and
aggregating local CN modifications.

Figures \ref{fig:increase} and \ref{fig:decrease} present stitch patterns
that include combinations of the Empty stitch (shown as an empty white
cell) and transfer stitches in order to realize increases (widening)
and decreases (narrowing) in the fabric. This allows \name\/ to
represent fabrics defined by more general, non-rectangular stitch patterns.
Figures \ref{fig:pattern4}, \ref{fig:pattern7} and
\ref{fig:patternVertical} present slightly larger
stitch patterns that include 3-level deep combinations of Knit, Tuck and
Miss stitches, which demonstrate \name\/'s ability to determine non-local
CN connectivity through the aggregation of local changes to CN
state information.
Figure \ref{fig:patternVertical} contains a pattern that transfers
CNs to the left and then shifts them upwards two rows with recursive calls
to FINAL\_LOC\_RECURSIVE.

Figure \ref{fig:bigPatches} presents two larger stitch patterns
($19 \times 19$ stitches). The top example is a lace pattern and the
bottom example is a cable. Figures \ref{fig:007AZoom} and 
\ref{fig:cableZoom} contain close-ups of
the repeating patterns and allow for better assessment of the results.
The topology graphs for these patterns
were computed in approximately two seconds and demonstrate that \name\/ is
capable of rapidly processing stitch patterns for larger fabric
swatches.  Additionally we performed timing tests on a swatch
that contains repeating blocks of the stitch pattern in
Figure \ref{fig:pattern1} in order to study the time complexity of
\name\/'s evaluation algorithms. The values in Table \ref{tab:timings}
and the plot in Figure \ref{fig:timings} demonstrate that the
computation time
needed to generate and evaluate the \name\/ data structure increases
linearly with the number of stitches in the modeled fabric. This
linear behavior verifies that the individual evaluation algorithms 
execute in near-constant time.
 
As for handling the edges of the fabric in all of the examples, the CNs on
the left and right boundaries are connected to the CNs either above or
below them, depending on the parities of their $(i,j)$ IDs.  For example,
boundary CNs whose $(i,j)$ parities are the same, e.g.~(odd,odd) or
(even,even), connect to the CN above them in the grid with an outgoing
yarn. CNs with differing $(i,j)$ parities are connected to the CN below
them by an incoming yarn. As for the top and bottom boundaries, when a
pattern is actually knitted \textit{cast-on} and \textit{ bind-off} stitch
combinations are
added to the fabric to prevent the top and bottom edges from
unraveling. \name\/ includes the topology for these specialized stitches,
but for visual simplicity they are not included in our graph
visualizations.

\section{Discussion}
\label{sect:Discussion}

It should be noted that  \name\/'s algorithms are based on several
assumptions about the stitch patterns that define the analyzed fabric. Applying a few simple rules/constraints on a fabric's stitch pattern
guarantees the structural integrity (soundness) of the resulting fabric,
and additionally ensures that our algorithms execute correctly.
A structurally sound knitted fabric is one where all of
its loops have been intertwined with another yarn.  This property
defines a knitted fabric that will not unravel.
Using \name\/ terminology, all CNs that define the head of a loop must eventually be
actualized in a structurally sound fabric.

The restrictions imposed on the stitch patterns processed by
\name\/ are the following. 1) The stitches along
the boundary of the pattern should be Knit or Purl stitches, or the
increase and decrease combinations in Figures \ref{fig:increase} and
\ref{fig:decrease}.
2) Empty stitches should only be defined on the ``outside'' of the stitch
pattern, i.e.~an Empty stitch should not be surrounded by non-Empty
stitches. Note that an Empty stitch is different from an empty CN
state, which can be surrounded by non-empty CNs, for example from a Miss
stitch.
3) Transfer stitches should not shift their
head nodes (their loops) outside of the boundary of the stitch pattern,
i.e.~to an Empty stitch. An example of a stitch pattern that is structurally unsound, violates
the stitch rules and leads to an incorrectly evaluated topology is one
with a Tuck stitch along the side boundary. This creates an unanchored CN 
along a side edge of the fabric, a situation that leads to stitch
unraveling. While the rules limit which stitch patterns may be 
represented and evaluated by \name\/, they still allow for the modeling
and analysis of a broad variety of structurally sound knitted fabrics.

\setcounter{page}{14}

We have described the query algorithms FINAL\_LO-CATION
(Algorithm \ref{alg:final_loc}) and ACNS\_AT (Algorithm
\ref{alg:acns_at}). These queries are central to the new \name\/ query
algorithms currently under development, which include determining the
topological connections between CNs, the connections between locations,
as well as the
ordering of yarns at a certain $(i,j)$ location and at arbitrary spatial
yarn crossings. CN and location connectivity is important to all 
flow simulations (heat, water and electrical current) conducted on a
fabric substrate. Yarn stacking
information would be additionally informative
to electric circuit simulations over a knitted fabric, assuming that
different yarn sections could have different conductivity / resistance.

For example, to compute the flow of water, heat or current through
the yarns of a fabric, one needs to know how the yarn is in contact with
itself. The exchange of heat, water, and current occurs where
the yarn intertwines. The algorithm ACNS\_AT and FINAL\_LOCATION together
provide the list of locations of these yarn intertwinings.
The under-development CONNECTED\_TO algorithm would provide the
connections between the contacts needed to compute flow over the
whole fabric.
We also intend to
integrate \name\/'s topology and yarn ordering information with the work
of Wadekar et al.~\cite{Wadekar2020GMK} to extend their
yarn-level geometric models to complex knitted fabrics. These models
support rapid evaluation of the mechanical properties of knitted fabrics.

It is important to note that the data structure and algorithms
presented in this work have been based on high-level stitch commands that
encapsulate multiple lower-level machine commands, thus, the space of
knitted structures supported by \name\/ is smaller than the space of
structures generated from an arbitrary set of knitting machine commands
(e.g.~rack a needle bed, move a loop between beds, knit or drop a loop).
These actions could be added to the data structure, but would
require modifications of the algorithms. Also, initially we focused on
stitch commands that create CN pairs only in one of the needle beds,
thus, leaving out stitches that require CNs to be defined on both beds,
such as the split stitch.

\section{Conclusions and Future Work}

In this paper we have proposed a process-oriented representation, \name\/,
that defines a foundational data structure for representing the topology
of the yarns in weft-knitted textiles.
We have  defined a process space that includes commonly
used yarn-level operations, abstracted as mappings on yarn contact
neighborhoods, produced by a weft-knitting knitting process. This space captures the
essence of knitting processes in terms of their actions on yarns,  but is
independent of a particular machine architecture.  In contrast to fabric
space, the process space represents the structure of the fabric
implicitly, eliminating the need for an explicit, fully-evaluated
representation of its topology.  Process space supports a concise,
computationally efficient evaluation approach based on on-demand, near
constant-time queries.
We have defined the important features of this process space and designed
a data structure to represent it and algorithms to evaluate it.  Through
the testing of more than 100 stitch patterns, we
demonstrate the robustness and effectiveness of this representation scheme.

Work is underway to expand the scope, utility and applicability
of \name\/.
Given the information in the data structure, algorithms are being
developed that can determine if a stitch pattern defines a structurally
sound fabric. This feature would be useful for textile designers in that it
would provide important information about the stitch pattern before
attempting to manufacture it.
Another test that could be easily implemented is related to 
manufacturability of a particular fabric.  As noted in Section
\ref{subsection:yarn_path_algorithms}, most loops of yarn can only be stretched
up to three needle locations away
before either
the yarn or the needle break. Flagging stitch combinations that produce
excessive yarn stretching, and therefore threaten the integrity of
the fabric or the knitting machine itself, is straightforward to
realize.

As mentioned in Section \ref{sect:Discussion}, \name\/ 
currently only supports the creation of CNs at a specific $(i,j)$ location
on a single needle bed. However, the data structure can be extended to
allow for creation of CNs on both beds at a specific $(i,j)$ location by
including a separate CN array for each of the beds.
This would also allow us to extend the set of stitch commands supported
by \name\/. Our future work will also include algorithms to determine the stacking order of
CNs at a specific $(i,j)$ location, as well, as the ordering of yarns where
they cross in the fabric away from CN locations.

\section*{Acknowledgements}
This work was partially supported by a National Science Foundation
Graduate Research Fellowship under Grant No.~DGE-10028090/DGE-1104459 and
NSF grants CMMI-1344205 and CMMI-1537720.

\section*{References}

\bibliography{bibfile}

\end{document}